\pdfoutput=1
\RequirePackage{ifpdf}
\ifpdf 
\documentclass[pdftex]{sigma}
\else
\documentclass{sigma}
\fi

\numberwithin{equation}{section}

\usepackage{esint}
\usepackage{scalerel,stackengine}
\usepackage{enumitem}
\usepackage{slashed}
\usepackage[mathscr]{eucal}

{\theoremstyle{definition}
\newtheorem{Def}{Definition}[section]

\newtheorem{Remark}[Def]{Remark}}

\newtheorem{Thm}[Def]{Theorem}
\newtheorem{Prp}[Def]{Proposition}
\newtheorem{Lemma}[Def]{Lemma}
\newtheorem{Corollary}[Def]{Corollary}

\newcommand{\la}{\langle}
\newcommand{\ra}{\rangle}

\newcommand{\Sl}{\mathopen{\prec}}
\newcommand{\Sr}{\mathclose{\succ}}
\newcommand{\C}{\mathbb{C}}
\newcommand{\R}{\mathbb{R}}
\newcommand{\1}{\mbox{\rm 1 \hspace{-1.05 em} 1}}

\newcommand{\N}{\mathbb{N}}
\newcommand{\Pdd}{\mbox{$\partial$ \hspace{-1.2 em} $/$}}

\newcommand{\Hil}{\mathscr{H}}
\newcommand{\J}{\mathfrak{J}}

\newcommand{\Jtest}{\mathfrak{J}^\text{\rm{\tiny{test}}}}
\newcommand{\Jlin}{\mathfrak{J}^\text{\rm{\tiny{lin}}}}

\newcommand{\bep}{\begin{pmatrix}}
\newcommand{\enp}{\end{pmatrix}}

\renewcommand{\O}{\mathscr{O}}

\newcommand{\F}{{\mathscr{F}}}

\renewcommand{\P}{{\mathcal{P}}}

\newcommand{\B}{{\mathscr{B}}}
\renewcommand{\O}{{\mathscr{O}}}
\renewcommand{\L}{{\mathcal{L}}}
\newcommand{\Sact}{{\mathcal{S}}}
\newcommand{\s}{{\mathfrak{s}}}

\newcommand{\Lin}{\text{\rm{L}}}

\newcommand{\scrM}{\myscr M}

\newcommand{\reg}{\text{\rm{reg}}}
\newcommand{\PP}{\text{\rm{PP}}}

\DeclareFontFamily{OT1}{rsfso}{}
\DeclareFontShape{OT1}{rsfso}{m}{n}{ <-7> rsfso5 <7-10> rsfso7 <10-> rsfso10}{}
\DeclareMathAlphabet{\myscr}{OT1}{rsfso}{m}{n}

\DeclareMathOperator{\re}{Re}
\DeclareMathOperator{\im}{Im}

\DeclareMathOperator{\Tr}{Tr}
\DeclareMathOperator{\tr}{tr}

\DeclareMathOperator{\supp}{supp}

\renewcommand{\u}{\mathfrak{u}}
\renewcommand{\v}{\mathfrak{v}}

\begin{document}

\allowdisplaybreaks

\newcommand{\arXivNumber}{1711.07058}

\renewcommand{\PaperNumber}{091}

\FirstPageHeading

\ShortArticleName{The Causal Action in Minkowski Space and Surface Layer Integrals}

\ArticleName{The Causal Action in Minkowski Space\\ and Surface Layer Integrals}

\Author{Felix FINSTER}

\AuthorNameForHeading{F.~Finster}

\Address{Fakult\"at f\"ur Mathematik, Universit\"at Regensburg, D-93040 Regensburg, Germany}
\Email{\href{mailto:finster@ur.de}{finster@ur.de}}

\ArticleDates{Received September 19, 2019, in final form September 11, 2020; Published online September 27, 2020}

\Abstract{The Lagrangian of the causal action principle is computed in Minkowski space for Dirac wave functions interacting with classical electromagnetism and linearized gravity in the limiting case when the ultraviolet cutoff is removed. Various surface layer integrals are computed in this limiting case.}


\Classification{49S05; 83A05; 35Q75; 81T27}

\tableofcontents

\section{Introduction} \label{secintro}
The theory of causal fermion systems is a recent approach to fundamental physics
(see the basics in Section~\ref{secbackground}, the reviews~\cite{ nrstg, review, dice2014}, the textbook~\cite{cfs}
or the website~\cite{cfsweblink}).
In this approach, the physical equations are formulated
via a variational principle, the {\em{causal action principle}}.
In order to rewrite the resulting dynamics in a familiar way in terms of Dirac
particles interacting with classical gauge fields, one considers the so-called
{\em continuum limit}.
In the analysis of the continuum limit, one studies the Euler--Lagrange (EL) equations corresponding to
the causal action principle in the limit~$\varepsilon \searrow 0$
when an ultraviolet regularization is removed.
In the present paper, we proceed differently and study instead the Lagrangian itself asymptotically
as~$\varepsilon \searrow 0$. This serves two different aims: First,
we thus obtain an ``effective action'' in Minkowski space.
Second, it becomes possible to compute various surface layer integrals
as derived abstractly in~\cite{positive, noether, jet, osi}.

There are several reasons why the asymptotics~$\varepsilon \searrow 0$
of the Lagrangian and the causal action were not computed
earlier. Foremost, it is unclear how
to make mathematical sense of the causal action principle in the limit~$\varepsilon \searrow 0$,
because the constraints (the so-called trace and boundedness constraints;
for details see Section~\ref{seccfs}) do not have an obvious correspondence in this limit.
Moreover, the fact that the causal action has a different mathematical structure than usual
actions in spacetime discouraged the author from taking the ``naive'' limit~$\varepsilon \searrow 0$
of the Lagrangian seriously.
It was only after the discovery of conserved surface layer integrals in~\cite{noether}
that the successful computation of these surface layer integrals
gave a hint that the asymptotics as~$\varepsilon \searrow 0$ of the Lagrangian
should indeed have a direct physical significance.

The present paper is the first work in which the Lagrangian of the causal action
is computed asymptotically as~$\varepsilon \searrow 0$.
Our results give new insight into the nature of the interaction as described by the
causal action principle.
One important finding is that there are contributions
to the Lagrangian which vanish if and only if the coupled
Einstein--Dirac--Maxwell equations are satisfied (see Sections~\ref{secJJ}, \ref{secdFFT}
and~\ref{secFFT}). For example, there is a contribution of the form
\[ \L(x,y) \sim \big( (J_i + c j_i) (y-x)^i \big) \cdot \big( (J_k + c j_k ) (y-x)^k \big)
{\mathfrak{K}}(x,y), \]
where~$J$ \looseness=-1 and~$j$ are the Dirac and Maxwell currents, respectively,
and~${\mathfrak{K}}$ is a distributional kernel supported on the light cone (for details see~\eqref{Jjcontrib}).
When varying the electromagnetic potential and the wave functions
naively (``naive'' in the sense that the above-mentioned constraints
in the causal action principle are disregarded), the above contribution to the action is
critical if and only if the Maxwell equations $J_i + c j_i =0$ hold.
This ``naive'' derivation of the Maxwell equations from the causal action principle is quite
different from the derivation in~\cite[Chapters~3--5]{cfs}, where the EL-equations
corresponding to the causal action principle were analyzed in a limiting case
referred to as the continuum limit (for an outline see Section~\ref{seccl}).
This difference is not only technical, but it even involves other contributions
to the Lagrangian with a form of the singularity on the light cone,
as made precise by the notion of the {\em degree}
(see~\eqref{degdef} in Section~\ref{seccl}).
In more general words, the Lagrangian has the remarkable property that it gives rise to the
classical field equations several times to different degrees on the light cone.
Clearly, in an interacting system, the resulting hierarchy of equations must all be satisfied
for the same values of the coupling constants. This implies that minimizing the causal action
gives rise to a specific class of regularizations for which the regularization parameters satisfy
all the resulting consistency conditions.

The above finding that the Lagrangian gives rise to the classical field equations several times to different
degrees on the light cone is not as surprising as it might seem at first sight.
Indeed, this phenomenon is closely related to the freedom in testing the EL equations.
In the formalism of the continuum limit, testing is performed by smooth variations which
vanish on the diagonal (see~\cite[Section~3.5.2]{cfs}).
These variations have the advantage that the resulting EL equations are well-defined in the
continuum limit and can therefore be analyzed in detail with mathematical rigor.
But from an abstract point of view,
given a minimizer of the causal action principle, the EL equations must hold for
much more general variations.
This becomes clear in the jet formalism introduced in~\cite{jet},
where the continuum limit analysis corresponds to a very special choice of
the jet space~$\Jtest$ (see~\cite[Section~7.2]{perturb}).
In particular, it should be allowed to test by varying the bosonic potentials
(more precisely in direction of bosonic jets; see~\eqref{bosonicjet}
in Section~\ref{seccl} or again~\cite[Section~7.2]{perturb}).
Such variations yield EL equations with a different singular behavior on the light cone
and, of course, these equations must again reduce to the classical field equations
in a suitable limiting case.
In this way, our findings can be regarded as
a consistency check that the causal action principle makes both mathematical and physical sense.

The contributions to the Lagrangian to lower degree on the light cone can no longer be interpreted
as classical field equations. Instead, they are needed in order to obtain
non-trivial conserved surface layer integrals.
{\em{Surface layer integrals}} are a generalization of surface integrals to the
setting of causal fermion systems (for a general introduction see~\cite[Section~2.3]{noether}
or Section~\ref{secosi} below). As shown in~\cite{noether}, symmetries give
rise to corresponding conservation laws for surface layer integrals.
Moreover, in~\cite{jet, osi} other conserved surface layer integrals were discovered.
In the present paper, we compute various surface layer integrals in Minkowski space:
To begin with, we complete the analysis in~\cite{noether} by showing that the calculations
results for the conservation law describing current conservation obtained in~\cite[Section~5]{noether}
for one sector also apply to more realistic systems including neutrinos (Section~\ref{seccurrent}).
Next, we analyze the {\em{symplectic form}}~$\sigma$
as found in~\cite[Sections~3 and~4.3]{jet} as well as the conserved
{\em{surface layer inner product}} as first described in~\cite[Theorem~1.1 and Corollary~3.11]{osi}.
In our setting of Minkowski space, these surface layer integrals take the form
\begin{gather}
\sigma(\u, \v) = \int_{-\infty}^{t_0} {\rm d}t \int_{\R^3} {\rm d}^3x \int_{t_0}^\infty {\rm d}t' \int_{\R^3} {\rm d}^3y\,
\big( \nabla_{1,\u} \nabla_{2,\v} \L(x,y) - \nabla_{1,\v} \nabla_{2,\u} \L(x,y) \big), \label{siggen} \\
 (\u, \v ) = \int_{-\infty}^{t_0} {\rm d}t \int_{\R^3} {\rm d}^3x \int_{t_0}^\infty {\rm d}t' \int_{\R^3} {\rm d}^3y\,
\big( \nabla_{1,\u} \nabla_{1,\v} \L(x,y) - \nabla_{2,\u} \nabla_{2,\v} \L(x,y) \big) , \label{osispgen}
\end{gather}
where~$\u$ and~$\v$ are jets describing first variations of the vacuum
which preserve the EL equations (for details see~\cite{jet} of Section~\ref{secosi} below).
The technical core of the present paper is to
analyze how the electromagnetic field and the Dirac wave functions
contribute to these surface layer integrals. In order to keep the calculations as simple as possible, we
restrict attention to regularizations which are static and spherically symmetric.
As shown in Section~\ref{secosib}, the electromagnetic field gives the contributions
(see Theorem~\ref{thmbose})
\begin{gather}
\sigma(\u, \v) = \frac{c_1}{\delta^4}
\int_{\R^3} \big( A_\u^i (F_\v)_{i0} - A_\v^i (F_\u)_{i0}\big) {\rm d}^3x, \label{symplecticbose} \\
(\u, \v) = \frac{c_2}{\delta^4} \int_{\R^3} \frac{1}{ | \vec{k} |} \big( \big(\hat{F}_\u\big)_{0i} \big(\hat{F}_\u\big)_0^{ i}
- \frac{1}{4} \big(\hat{F}_\u\big)_{ij} \big(\hat{F}_\u\big)^{ij} \big) {\rm d}^3k \label{bose}
\end{gather}
(where~$\vec{k} \in \R^3$ is the spatial momentum).
Here~$\u$ and~$\v$ are the jets generated by the electromagnetic potentials~$A_\u$ and~$A_\v$
(for details see~\eqref{varyPsi} and~\eqref{bosonicjet} in Section~\ref{seccl}),
and~$F_\u$ and~$F_\v$ are the corresponding field tensors.
Moreover, $\delta$ is a length scale describing the
regularization (for details see~\cite[Sections~1.2.1 and~4.2.5]{cfs}), and $c_{1\!/\!2}$ are two real-valued constants.
The integral in~\eqref{symplecticbose} is the well-known symplectic form of classical electrodynamics
(see for example~\cite[Section~2.3]{deligne+freed}).
Due to the absolute value in the denominator in~\eqref{bose}, the surface layer inner product is
semi-definite on the bosonic potentials (but it is of course degenerate on gauge orbits).
The inner product~\eqref{bose} is commonly used in quantum field theory as giving rise to
the scalar product on the bosonic Fock space (for example, it is used implicitly
when introducing the creation and annihilation operators in~\cite[Sections~8.3 and~8.4]{weinberg}).
Compared to~\eqref{symplecticbose}, the surface layer inner product~\eqref{bose}
contains an additional derivative and an additional factor~$1/|\vec{k}|$.
Combining these two factors gives rise to a factor plus one on the upper and minus one
on the lower mass shell, thereby implementing the {\em frequency splitting}
currently used in quantum field theory (more precisely, in the bosonic quantization procedure
the scalar product~\eqref{bose} is constructed from the symplectic form by flipping the sign
for the negative-frequency solutions; this frequency splitting also becomes apparent in
the Feynman propagator, defined by the condition that ``positive frequencies propagate to
the future and negative frequencies propagate to the past'').
It is remarkable that in the setting of causal fermion systems, the frequency splitting does not
need to be put in by hand, but it follows from the theory simply by computing
a surface layer integral which, by the structure of the EL equations corresponding to the
causal action, is known to be conserved in time.

For the contributions by the Dirac wave functions, first variations of the vacuum
are described in the jet formalism by a variation~$\delta \psi$
of a Dirac wave function~$\psi$, where~$\psi$ is a solution on the lower mass shell
(describing a state of the Dirac sea), whereas~$\delta \psi$ is a solution on the upper
mass shell (describing a particle state).
Furthermore, we restrict attention to the components of the fermionic jets which preserve the
chiral symmetry (see Definition~\ref{defchiralsymm}). Moreover, we only consider the situation where both~$\psi$ and~$\delta \psi$ are solutions of the
Dirac equation corresponding to the same mass~$m$.
This is motivated by the fact that we restrict attention to electromagnetic and gravitational interactions
which leave the flavor unchanged. The superscripts~$\u$ and~$\v$ indicate which
Dirac wave function is varied. Then, as shown in Section~\ref{secosif},
the contributions to the conserved surface layer integrals
are given by (see Theorem~\ref{thmfermi} and Proposition~\ref{prpex})
\begin{gather}
\sigma\big( (\delta \psi^\u, \psi^\u), (\delta \psi^\v, \psi^\v) \big)
 = \frac{c_3}{\delta^4} \int_{\R^3} \frac{{\rm d}^3k}{(2 \pi)^3} \int_{\R^3} \frac{{\rm d}^3q}{(2 \pi)^3}
 \frac{1}{m^2} \big( \omega(\vec{q})^2 + \omega(\vec{k})^2 \big) \notag \\
\hphantom{\sigma\big( (\delta \psi^\u, \psi^\u), (\delta \psi^\v, \psi^\v) \big) =}{}
\times \sum_{c=L,R} \im \big( \Sl \delta \hat{\psi}^\u(\vec{k}) | \chi_c \hat{\psi}^\u(-\vec{q}) \Sr
\Sl \hat{\psi}^\v(-\vec{k}) | \chi_{\bar{c}} \delta \hat{\psi}^\v(\vec{q}) \Sr \notag \\
\qquad \hphantom{\sigma\big( (\delta \psi^\u, \psi^\u), (\delta \psi^\v, \psi^\v) \big) =}{}
- \Sl \delta \hat{\psi}^\u(\vec{k}) | \gamma^\alpha \chi_c \hat{\psi}^\u(-\vec{q}) \Sr
\Sl \hat{\psi}^\v(-\vec{k}) | \gamma_\alpha \chi_c \delta \hat{\psi}^\v(\vec{q}) \Sr \big), \label{symplecticfermi} \\
\big( (\delta \psi^\u, \psi^\u) , (\delta \psi^\v, \psi^\v) \big)
 = \frac{c_4}{\delta^4} \int_{\R^3} \frac{{\rm d}^3k}{(2 \pi)^3} \int_{\R^3} \frac{{\rm d}^3q}{(2 \pi)^3}
\re \big( \Sl \delta \hat{\psi}^\u(\vec{k}) | \delta \hat{\psi}^\v(\vec{k}) \Sr \Sl \hat{\psi}^\v(\vec{q}) | \hat{\psi}^\u(\vec{q}) \Sr \big) \notag \\
\hphantom{\big( (\delta \psi^\u, \psi^\u) , (\delta \psi^\v, \psi^\v) \big) =}{}
 \times \frac{1}{m^3} \big\{ \vec{k} \cdot \vec{q} \big( \omega(\vec{q}) + \omega(\vec{k}) \big) + |\vec{q}|^2 \omega(\vec{q}) + |\vec{k}|^2 \omega(\vec{k}) \big\} \label{fermi}
\end{gather}
with real constants~$c_3$ and~$c_4$. Here~$\omega(\vec{k}):=\sqrt{|\vec{k}|^2+m^2}$
is the absolute value of the corresponding frequency,
$\chi_{L/R} := \big(1 \mp \gamma^5\big)/2$ are the chiral projections,
and~$\Sl \cdot|\cdot \Sr$ is the usual inner product on Dirac spinors (being indefinite of
signature $(2,2)$).
The conservation of these surface layer integrals gives rise to pointwise conditions
for the spinors (see~\eqref{conscond} in Theorem~\ref{thmfermiconserve}).
The inner product~\eqref{fermi} is again semi-definite (see Proposition~\ref{prpdefinite}).
Combining~\eqref{bose} and~\eqref{fermi}, we obtain a scalar product
on the jet space spanned by the fermionic and bosonic jets.

Our computations also reveal that the
structure of the above surface layer integrals is quite different from that of the
surface layer integrals describing current and energy conservation
in~\cite[Sections~5.2 and~6.2]{noether}. Namely, while the surface layer integrals
in~\cite{noether} are of short range (meaning that the main contribution comes from
points~$x$ and~$y$ whose distance is small on the Compton scale),
the symplectic form and the surface layer inner product are essentially {\em nonlocal} and
of {\em{long range}} in the sense that the main contributions come from points~$y$ lying on
the light cone centered at~$x$ but which may be far apart from~$x$
(again on the Compton scale as measured in the reference frame distinguished by the
regularization). This does not pose any problems when computing the surface
layer integrals for the asymptotic incoming or outgoing states in a scattering process,
and in this case the above formulas~\eqref{symplecticbose}--\eqref{fermi} again hold.
But when choosing~$t_0$ as an intermediate time while the interaction takes place,
then the surface layer integral will also depend on the past and future of~$t_0$.
We also point out that in the interacting case, the
bosonic and fermionic parts of the surface layer inner product
should no longer be conserved separately,
because the general conservation law in~\cite{osi} only applies to the whole
jet space spanned by the fermionic and bosonic jets.
For the symplectic form, however, there are indications that the bosonic
and fermionic parts should even be conserved separately
(see Proposition~\ref{prpsympdiv} and Remark~\ref{remsympdiv}).
As is worked out in detail in~\cite{fockbosonic, fockfermionic}, the symplectic form and
the surface layer inner product indeed give rise to a conserved
scalar product on the complex Fock space of the interacting theory.

In Section~\ref{secpositive}, we consider the surface layer integral
\[ -\int_{-\infty}^{t_0} {\rm d}t \int_{\R^3} {\rm d}^3x \int_{t_0}^\infty {\rm d}t' \int_{\R^3} {\rm d}^3y\,
\nabla_{1,\v} \nabla_{2,\v} \L(x,y) . \]
This surface layer integral is not necessarily conserved, but it is shown in~\cite[Section~7]{positive}
that it is non-negative if one varies about a local minimizer of the causal action.
Therefore, as a~consistency check and in order to verify that the regularized Dirac sea configuration
is indeed a minimizer, we compute this surface layer integral and find that, due to contributions
by the Maxwell current, it is indeed non-negative.
The paper concludes with a few remarks and an outlook (Section~\ref{secdiscuss}).

Taken together, our results show in a surprising and compelling way that the
different contributions to the Lagrangian in Minkowski space fit together consistently both with
the general conservation laws of causal fermion systems and to
structures present in classical field theory and quantum field theory.

We finally outline our methods. As in~\cite[Chapter~5]{cfs} we describe the vacuum by
a system of Dirac seas including leptons and quarks.
Nevertheless, for simplicity we restrict attention to an interaction via electromagnetic
fields and linearized gravity. Since these fields do not describe
changes of flavor, in most parts of the analysis it suffices to consider
a single Dirac sea of mass~$m$. In the first step of our analysis we
apply the formalism of the continuum limit (for an introduction see~\cite[Section~2.4]{cfs})
to obtain contributions which are distributions in spacetime with a $\delta$-singularity on the light cone
and which have a pole in~$\varepsilon$.
More precisely, all relevant contributions to the Lagrangian can be written as
\begin{gather} \label{Lcontrib}
\L(x,y) \asymp \text{(smooth functions in~$x$ and~$y$)} \cdot
\frac{1}{\varepsilon^p t^q} \delta(|t|-r) \log^s\big(t^2-r^2\big) \epsilon(t)^{s'}
\end{gather}
with~$s, s' \in \{0,1\}$, where we set~$t=\xi^0$, $r=|\vec{\xi}|$ with~$\xi:=y-x$
(here and in what follows, the symbol~$\asymp$ indicates that we
restrict attention to a specific contribution).
These formulas involve regularization parameters which we simply treat as
effective empirical parameters.
We also point out that the formalism of the continuum limit gives rise to the formulas~\eqref{Lcontrib}
only away from the diagonal (i.e., for~$x \neq y$).
Here we simply extend these formulas in the distributional sense.
Clearly, this extension is unique only up to singular contributions supported on the diagonal
(like contributions ${\sim} \varepsilon^{-q} \delta^4(x-y)$ or distributional derivatives thereof).
It turns out that this simple method gives physically sensible results.

For the computation of the conserved surface layer integrals, in Sections~\ref{seccurrent}--\ref{secpositive}
we analyze integrals involving~\eqref{Lcontrib} with Fourier methods.
Indeed, the distributions supported on the light cone~\eqref{Lcontrib}
have a nice structure in momentum space (see for example Fig.~\ref{figK12} on p.~\pageref{figK12}).
Rewriting multiplication in position space as convolution in momentum space, our task is to
compute certain convolution integrals.
More specifically, for the computation of the symplectic form and the surface layer inner product,
we consider the combination given in~\cite[Theorem~3.1]{osi}
\begin{gather} \label{osicombined}
\int_{-\infty}^{t_0} {\rm d}t \int_{\R^3} {\rm d}^3x \int_{t_0}^\infty {\rm d}t' \int_{\R^3} {\rm d}^3y\,
 ( \nabla_{1, \u} - \nabla_{2, \u} )
 ( \nabla_{1, \v} + \nabla_{2, \v} ) \L(x,y) .
\end{gather}
By anti-symmetrizing and symmetrizing in~$\u$ and~$\v$, one gets the
symplectic form~\eqref{siggen} and the surface layer inner product~\eqref{osispgen}, respectively.
In order to compute the fermionic surface layer integrals, we make use of the
specific support properties of the Dirac wave functions and the convolution kernels
in momentum space (see Fig.~\ref{figsupport} on p.~\pageref{figsupport}).
When computing the bosonic surface layer integrals, the main difficulty is that
the light-cone expansion involves unbounded line integrals (see for example Lemma~\ref{lemmaunbounded}).
Using the causal structure of the Lagrangian, we show that these unbounded line integrals vanish.
These arguments implicitly pose conditions on the admissible class of regularizations,
which can be understood intuitively that the regularized objects are ``supported mainly near
the light cone'' and ``vanish approximately for spacelike distances''
(see Section~\ref{secosiconserve}).
Similar as worked out in~\cite{reg} in the vacuum, one could analyze in detail what
these conditions mean and how they can be satisfied.
But this analysis goes beyond the scope of the present paper.
Here we are content with showing that the highly singular contributions can be
given a mathematical meaning using certain computation rules which are
motivated and introduced.

\section{Physical background and motivation} \label{secbackground}
In this section, we give a brief introduction to causal fermion systems and
outline all the concepts needed later on.
Our presentation has similarities to other introductions (for example in~\cite[Section~2]{dice2014}, \cite[Section~1]{nrstg} or~\cite[Section~1.2]{cfs}, \cite[Section~4]{review}),
but it is streamlined towards the causal action principle in Minkowski space.

\subsection{From relativistic quantum mechanics to causal fermion systems} \label{secquantumcfs}
We begin in the setting of relativistic quantum mechanics in the presence of an external
classical electromagnetic field.
Let~$\scrM$ be Minkowski space and $\mu$ the natural volume measure thereon, i.e., ${\rm d}\mu = {\rm d}^4x$ if $x=\big(x^0,x^1,x^2,x^3\big)$ is an inertial frame. We also denote time by~$t=x^0$
and write spatial vectors as~$\vec{x}=\big(x^1,x^2,x^3\big)$.
We consider Dirac wave functions in the presence of an external electromagnetic potential~$A$,
which satisfy the Dirac equation
\begin{gather*}
\big( i \Pdd + \slashed{A} - m \big) \psi = 0 ,
\end{gather*}
where~$m$ is the rest mass, and the slash denotes contraction with
the Dirac matrices~$\gamma^j$ in the Dirac representation.
On the Dirac solutions, we consider the usual scalar product
\begin{gather}\label{ScalProd}
( \psi | \phi )_t := 2 \pi \int_{t=\textrm{const}} \Sl \psi | \gamma^0 \phi\Sr (t,\vec x) \, {\rm d}^3 x
\end{gather}
(here~$\Sl \cdot |\cdot \Sr$ is the indefinite inner product on Dirac spinors, also written
as~$\Sl \psi | \phi \Sr = \overline{\psi} \phi$, where~$\overline{\psi} := \psi^\dagger \gamma^0$ is the adjoint spinor,
and the dagger denotes complex conjugation and transposition). If one evaluates~\eqref{ScalProd} for~$\phi=\psi$,
the integrand can be written as~$\big(\overline{\psi}\gamma^0\psi\big)(t,\vec{x}) = \big(\psi^\dagger \psi\big)(t,\vec{x})$,
having the interpretation as the probability density of the Dirac particle corresponding to~$\psi$
to be at time~$t$ at the position~$\vec{x}$. Due to current conservation, the integral in~\eqref{ScalProd} is time independent.

Next, we choose an ensemble of Dirac solutions~$\psi_1, \ldots, \psi_f$.
For simplicity in presentation, we restrict attention to the case~$f<\infty$ of a finite number of
Dirac wave functions, which we assume to be continuous.
It is a central idea behind causal fermion systems to describe the physical system and
to formulate its dynamical equations purely in terms of the ensemble of wave
functions~$\psi_1, \ldots, \psi_f$.
Another idea is that the causal fermion system should encode the form of the wave functions
in a gauge-invariant way. To this end, we denote the complex vector space spanned by the
wave functions~$\psi_1, \ldots, \psi_f$ by~$\Hil$. On~$\Hil$ we consider the restriction of the
scalar product~\eqref{ScalProd}, i.e., $\la \cdot|\cdot \ra_\Hil := ( \cdot|\cdot )_t|_{\Hil \times \Hil}$.
Thus~$(\Hil, \la \cdot|\cdot \ra_\Hil)$ is an $f$-dimensional complex vector space
formed of wave functions.
For any spacetime point~$x \in \scrM$, we now introduce the sesquilinear form
\begin{gather} \label{bxdef}
b_x \colon \ \Hil \times \Hil \rightarrow \C ,\qquad b_x(\psi, \phi) = -\Sl \psi(x) | \phi(x)\Sr ,
\end{gather}
which maps two solutions of the Dirac equation to their inner product at~$x$.
The sesquilinear form $b_x$ can be represented by a self-adjoint operator $F(x)$ on $\Hil$,
which is uniquely defined by the relations
\[ \la \psi | F(x) \phi \ra_\Hil =b_x(\psi,\phi) \qquad \text{for all~$\psi, \phi \in \Hil$}. \]
More concretely, in the basis~$(\psi_k)_{k = 1, \ldots,f}$ of~$\Hil$, the last relation can be written as
\begin{gather} \label{Fdef}
\la \psi_i | F(x) \psi_j \ra_\Hil = - \Sl\psi_i(x) | \psi_j(x) \Sr .
\end{gather}
If the basis is orthonormal, the calculation
\[ F(x) \psi_j = \sum_{i=1}^f \la \psi_i | F(x) \psi_j \ra_\Hil \psi_i
= - \sum_{i=1}^f \Sl\psi_i(x) | \psi_j(x) \Sr \psi_i \]
(where we used the completeness relation~$\phi = \sum_i \la \psi_i | \phi \ra \psi_i$)
shows that the operator~$F(x)$ has the matrix representation
\[ (F(x) )^i_j = - \Sl\psi_i(x) | \psi_j(x) \Sr . \]
In physical terms, the matrix elements give information on the correlation of the
wave functions~$\psi_i$ and~$\psi_j$ at the spacetime point~$x$.
Therefore, we refer to~$F(x)$ as the {\em local correlation operator} at~$x$.

Let us analyze the properties of $F(x)$. First of all, the calculation
\[ \la F(x) \psi | \phi \ra_\Hil = \overline{ \la \phi | F(x) \psi \ra_\Hil}
= \overline{ -\Sl\phi(x) | \psi(x) \Sr} = -\Sl\psi(x) | \phi(x) \Sr = \la \psi | F(x) \phi \ra_\Hil \]
shows that the operator~$F(x)$ is self-adjoint
(where we denoted complex conjugation by a bar).
Furthermore, since the pointwise inner product $\Sl\psi(x) | \phi(x) \Sr$ has signature $(2,2)$,
we know that~$b_x$ has signature $(p,q)$ with $p,q \leq 2$.
As a consequence, counting multiplicities, the opera\-tor~$F(x)$ has at most two positive and at most two negative eigenvalues. It is useful to denote the set of all symmetric linear operators on~$\Hil$ which have rank at most
four and (counting multiplicities) have at most two positive and at most two negative eigenvalues by~$\F
\subset \Lin(\Hil)$. Then the local correlation operator~$F(x)$ is an element of~$\F$.

Constructing the operator $F(x) \in \F$ for every spacetime point $x \in \scrM$, we
obtain the {\em{local correlation map}}
\begin{gather} \label{lcm}
F \colon \ \scrM \rightarrow \F ,\qquad x \mapsto F(x) .
\end{gather}
This allows us to introduce a measure $\rho$ on $\F$ as follows. For any~$\Omega \subset \F$,
one takes the pre-image $F^{-1}(\Omega) \subset \scrM$ and computes its spacetime volume,
\[ \rho(\Omega) := \mu \big( F^{-1}(\Omega) \big) . \]
This gives rise to the so-called {\em{push-forward measure}} which in mathematics
is denoted by~$\rho = F_\ast \mu$ (for details see for example~\cite[Section~3.6]{bogachev}).
The $\rho$-measurable sets are defined as
the $\sigma$-algebra of all subsets of~$\F$ whose pre-image~$F^{-1}(\Omega)$
is $\mu$-measurable.

The resulting structures of a measure~$\rho$ on the set~$\F$ of linear
operators on a Hilbert space~$\Hil$ form a causal fermion system.
For clarity, we now introduce a few general concepts
(Sections~\ref{seccfs} and~\ref{secosi}), and then return to systems in Minkowski space
(Section~\ref{seccapmink}).

\subsection{Causal fermion systems and the causal action principle} \label{seccfs}
We now give the abstract definitions (for more details see for example~\cite[Section~1.1]{cfs}).

\begin{Def}[causal fermion system] \label{defcfs} Given a separable complex Hilbert space~$\Hil$ with scalar product~$\la \cdot|\cdot \ra_\Hil$
and a parameter~$n \in \N$ (the {\em{``spin dimension''}}), we let~$\F \subset \Lin(\Hil)$ be the set of all
self-adjoint operators on~$\Hil$ of finite rank, which (counting multiplicities) have
at most~$n$ positive and at most~$n$ negative eigenvalues. On~$\F$ we are given
a positive measure~$\rho$ (defined on a $\sigma$-algebra of subsets of~$\F$), the so-called
{\em universal measure}. We refer to~$(\Hil, \F, \rho)$ as a {\em causal fermion system}.
\end{Def}

A causal fermion system describes a spacetime together
with all structures and objects therein.
In order to single out the physically admissible
causal fermion systems, one must formulate physical equations. To this end, we impose that
the universal measure should be a minimizer of the causal action principle,
which we now introduce. For any~$x, y \in \F$, the product~$x y$ is an operator of rank at most~$2n$.
However, in general it is no longer a selfadjoint operator because~$(xy)^* = yx$,
and this is different from~$xy$ unless~$x$ and~$y$ commute.
As a consequence, the eigenvalues of the operator~$xy$ are in general complex.
We denote these eigenvalues counting algebraic multiplicities
by~$\lambda^{xy}_1, \ldots, \lambda^{xy}_{2n} \in \C$
(more specifically,
denoting the rank of~$xy$ by~$k \leq 2n$, we choose~$\lambda^{xy}_1, \ldots, \lambda^{xy}_{k}$ as all
the non-zero eigenvalues and set~$\lambda^{xy}_{k+1}, \ldots, \lambda^{xy}_{2n}=0$).
We introduce the Lagrangian and the causal action by
\begin{alignat}{3}
& \text{Lagrangian:} \quad && \L(x,y) = \frac{1}{4n} \sum_{i,j=1}^{2n} \big( \big|\lambda^{xy}_i \big|
- \big|\lambda^{xy}_j \big| \big)^2, & \label{Lagrange} \\
& \text{causal action:}\quad && \Sact(\rho)= \iint_{\F \times \F} \L(x,y) \, {\rm d}\rho(x) {\rm d}\rho(y) . &\label{Sdef}
\end{alignat}
The {\em causal action principle} is to minimize~$\Sact$ by varying the measure~$\rho$
under the following constraints:
\begin{alignat}{3}
& \text{volume constraint:}\quad && \rho(\F) = \text{const}, & \label{volconstraint} \\
& \text{trace constraint:} \quad && \int_\F \tr(x) \, {\rm d}\rho(x) = \text{const},& \label{trconstraint} \\
& \text{boundedness constraint:} \quad && \iint_{\F \times \F} |xy|^2 \,{\rm d}\rho(x) {\rm d}\rho(y)\leq C , &\label{Tdef}
\end{alignat}
where~$C$ is a given parameter, $\tr$ denotes the trace of a linear operator on~$\Hil$, and
the absolute value of~$xy$ is the so-called spectral weight,
\[ |xy| := \sum_{j=1}^{2n} \big|\lambda^{xy}_j \big| . \]
This variational principle is mathematically well-posed if~$\Hil$ is finite-dimensional.
For the existence theory and the analysis of general properties of minimizing measures
we refer to~\cite{lagrange, discrete, continuum}.
In the existence theory one varies in the class of regular Borel measures
(with respect to the topology on~$\Lin(\Hil)$ induced by the operator norm),
and the minimizing measure is again in this class. With this in mind, here we always assume that
\begin{gather*} 
\text{$\rho$ is a regular Borel measure} .
\end{gather*}

Let~$\rho$ be a {\em{minimizing}} measure. {\em{Spacetime}}
is defined as the support of this measure,
\[
M := \supp \rho . \]
Thus the spacetime points are selfadjoint linear operators on~$\Hil$.
These operators contain a lot of additional information which, if interpreted correctly,
gives rise to spacetime structures like causal and metric structures, spinors
and interacting fields. We refer the interested reader to~\cite[Chapter~1]{cfs}.

The only results on the structure of minimizing measures
which will be needed in what follows concern the treatment of the
trace constraint and the boundedness constraint.
As a consequence of the trace constraint, for any minimizing measure~$\rho$
the local trace is constant in spacetime, i.e.,
there is a real constant~$c \neq 0$ such that (see~\cite[Proposition~1.4.1]{cfs})
\begin{gather} \label{trc}
\tr x = c \qquad \text{for all~$x \in M$} .
\end{gather}
Restricting attention to operators with fixed trace, the trace constraint~\eqref{trconstraint}
is equivalent to the volume constraint~\eqref{volconstraint} and may be disregarded.
The boundedness constraint, on the other hand, can be treated with a Lagrange multiplier.
More precisely, in~\cite[Theorem~1.3]{lagrange} it is shown that for every minimizing measure~$\rho$,
there is a Lagrange multiplier~$\kappa>0$ such that~$\rho$ is a local minimizer of the causal action
with the Lagrangian replaced by
\begin{gather} \label{Lkappa}
\L_\kappa(x,y) := \L(x,y) + \kappa |xy|^2 .
\end{gather}

\subsection{The Euler--Lagrange equations and surface layer integrals} \label{secosi}
Surface layer integrals generalize surface integrals to the setting of causal fermion systems.
It is a major objective of this paper to compute these surface layer integrals in Minkowski space.
In preparation of introducing the concept of a surface layer integrals, we
need to state the Euler--Lagrange (EL) equations, explain the notion of jets
and explain the linearized field equations.
Let~$\rho$ be a minimizer of the causal action principle.
We introduce the function~$\ell_\kappa$ by
\begin{gather} \label{ldef}
\ell_\kappa(x) = \int_M \L_\kappa(x,y) \,{\rm d}\rho(y) - \s ,
\end{gather}
where~$\L_\kappa$ is the Lagrangian incorporating the boundedness constraint~\eqref{Lkappa}
and~$\s$ is a real parameter.
The {\em{Euler--Lagrange}} (EL) {\em{equations}} state that this function vanishes and is minimal
on the support of~$\rho$, i.e., for a suitable choice of~$\s$,
\begin{gather*} 
\ell_\kappa |_M \equiv \inf_\F \ell_\kappa = 0 .
\end{gather*}
For the derivation and technical details we refer to~\cite{jet}.
These EL equations are nonlocal in the sense that
they make a statement on~$\ell_\kappa$ even for points~$x \in \F$ which
are far away from spacetime~$M$.
It turns out that for the applications we have in mind, it is preferable to
evaluate the EL equations locally in a neighborhood of~$M$.
This concept leads to the {\em{weak EL equations}} introduced in~\cite[Section~4]{jet}.
We here give a slightly less general version of these equations which
is sufficient for our purposes. In order to explain how the weak EL equations come about,
we begin with the simplified situation that~$\F$ has a smooth manifold structure and~$\ell_\kappa$ is
a smooth function on~$\F$.
In this case, the minimality of~$\ell_\kappa$ implies that the derivative of~$\ell_\kappa$
vanishes on~$M$, i.e.,
\begin{gather} \label{ELweak}
\ell_\kappa|_M \equiv 0 \qquad \text{and} \qquad D \ell_\kappa|_M \equiv 0 .
\end{gather}
In order to combine these two equations in a compact form,
it is convenient to consider a pair~$\u := (a, u)$
consisting of a real-valued function~$a$ on~$M$ and a vector field~$u$,
and to denote the combination of multiplication and directional derivative by
\begin{gather} \label{Djet}
\nabla_{\u} \ell_\kappa(x) := a(x) \ell_\kappa(x) + \big(D_u \ell_\kappa \big)(x) .
\end{gather}
The equations~\eqref{ELweak} imply that~$\nabla_{\u} \ell_\kappa(x)$
vanishes for all~$\u$ and for all~$x \in M$.
The pair~$\u=(a,u)$ is referred to as a {\em{jet}}.
The real vector space of all jets is denoted by~$\J$.
One advantage of working with jets is that the two equations in~\eqref{Djet}
can be combined to one equation
\begin{gather} \label{ELtest}
\nabla_{\u} \ell_\kappa|_M = 0 \qquad \text{for all~$\u \in \Jtest$} ,
\end{gather}
referred to as the {\em{weak EL equations}}.
Before going on, we point out that~$\F$ does in general not have a smooth manifold structure,
nor is the Lagrangian~$\L_\kappa$ smooth. In order to treat this situation in a convincing way,
one introduces suitable subspaces of~$\J$ formed of jets for which the above directional derivatives are
mathematically well-defined.
Since the details will not be needed here, we do not give the definitions but refer instead
to~\cite[Section~4]{jet} or~\cite[Section~2.2]{fockbosonic}.

The EL equations are nonlinear because changing the measure~$\rho$
has two effects: First the function~$\ell_\kappa$ in~\eqref{ldef} is modified, and moreover
this function must be evaluated at different point~$x \in M$.
Since such nonlinear equations are difficult to analyze, it is a useful tool to linearize.
Usually, linearized fields are obtained by considering a family of
nonlinear solutions and linearizing with respect to a parameter~$\tau$
describing the field strength.
The analogous notion in the setting of causal fermion systems
is a linearization of a family of measures~$(\tilde{\rho}_\tau)$ which all satisfy the weak EL equations~\eqref{ELtest}.
It turns out to be fruitful to construct this family of measures by multiplying
a given critical measure~$\rho$ by a weight function~$f_\tau$ and then
``transporting'' the resulting measure with a mapping~$F_\tau$. More precisely, one considers the ansatz
\begin{gather} \label{rhoFf}
\tilde{\rho}_\tau = (F_\tau)_* \big( f_\tau \rho \big) ,
\end{gather}
where~$f_\tau \in C^\infty(M, \R^+)$ and~$F_\tau \in C^\infty(M, \F)$ are smooth mappings,
and~$(F_\tau)_*\mu$ denotes again the push-forward a measure.

The property of the family of measures~$\tilde{\rho}_\tau$ of the form~\eqref{rhoFf}
to satisfy the weak EL equation for all~$\tau$
means infinitesimally in~$\tau$ that the jet~$\v$ defined by
\begin{gather} \label{vinfdef}
\v = (b,v) := \frac{{\rm d}}{{\rm d}\tau} (f_\tau, F_\tau) \big|_{\tau=0}
\end{gather}
satisfies the {\em{linearized field equations}} (for the derivation see~\cite[Section~3.3]{perturb}
or, in the simplified smooth setting, the textbook~\cite[Chapter~6]{intro})
\begin{gather*} 
\la \u, \Delta \v \ra|_M = 0 \qquad \text{for all test jets~$\u$},
\end{gather*}
where for any~$x \in M$,
\begin{gather*} 
\la \u, \Delta \v \ra(x) := \nabla_{\u} \left( \int_M ( \nabla_{1, \v} + \nabla_{2, \v} ) \L_\kappa(x,y) \,{\rm d}\rho(y) - \nabla_\v \s \right)
\end{gather*}
(and~$\nabla_1$ and~$\nabla_2$ act on the arguments~$x$ and~$y$ of the
Lagrangian, respectively).
We denote the vector space of all solutions of the linearized field equations by~$\Jlin \subset \J$.

After the above preparations, we are now ready to introduce {\em{surface layer integrals}}.

\begin{figure}[t]\centering
\includegraphics{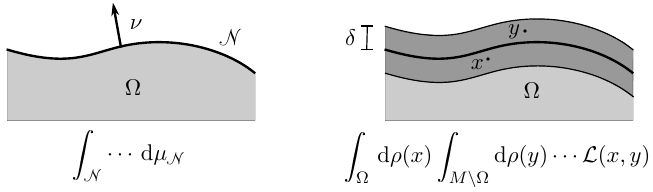}

\caption{A surface integral and a corresponding surface layer integral.}\label{fignoether1}
\end{figure}

In the setting of causal fermion systems, the usual integrals over hypersurfaces in spacetime are undefined.
Instead, one considers so-called {\em{surface layer integrals}}, being double integrals of the form
\begin{gather} \label{IntrOSI}
\int_\Omega {\rm d}\rho(x) \int_{M \setminus \Omega} {\rm d}\rho(y) (\cdots) \L_\kappa(x,y) ,
\end{gather}
where~$\Omega$ is a subset of~$M$ and~$(\cdots)$ stands for a differential operator
acting on the Lagrangian. The structure of such surface layer integrals can be understood most easily
in the special situation that the Lagrangian is of short range
in the sense that~$\L_\kappa(x,y)$ vanishes unless~$x$ and~$y$ are close together.
In this situation, we get a contribution to the double integral~\eqref{IntrOSI} only
if both~$x$ and~$y$ are close to the boundary~$\partial \Omega$.
With this in mind, surface layer integrals can be understood as an adaptation
of surface integrals to the setting of causal variational principles.
This consideration is illustrated in Fig.~\ref{fignoether1}, where the range of
the Lagrangian is denoted by~$\delta$ (for a more detailed explanation see~\cite[Section~2.3]{noether}).
Therefore, in the setting of causal variational principles, they take the role of surface integrals
in Lorentzian geometry.
We remark that in applications in Minkowski space or on a Lorentzian manifold,
the Lagrangian typically decays on the Compton scale $1/m$ (where~$m$ denotes the
rest mass of the Dirac particles).

Surface layer integrals were first introduced in~\cite{noether} in order to
formulate Noether-like theorems for causal variational principles.
In particular, it was shown that there is a conserved
surface layer integral which generalizes the Dirac current
in relativistic quantum mechanics (see~\cite[Section~5]{noether}).
More recently, in~\cite{jet} another conserved
surface layer integral was discovered which gives rise to a symplectic form on the
solutions of the linearized field equations (see~\cite[Sections~3.3 and~4.3]{jet}).
A systematic study of conservation laws for surface layer integrals is given in~\cite{osi}.
Finally, in~\cite[Proposition~7.1]{positive} a surface layer integral was introduced
which is not conserved, but positive. We now collect all the
surface layer integrals to be considered in this paper:

\begin{Def} \label{defosi} For any jets~$\u$, $\v$ and a subset~$\Omega \subset M$,
we define the following surface layer integrals:
\begin{itemize}\itemsep=0pt
\item[$\blacktriangleright$] conserved one-form
\begin{gather*}
\gamma(\v) = \int_{\Omega} {\rm d}\rho(x) \int_{M \setminus \Omega} {\rm d}\rho(y)
 ( \nabla_{1,\v} - \nabla_{2,\v} ) \L(x,y), 
\end{gather*}
\item[$\blacktriangleright$] symplectic form
\begin{gather*}
\sigma(\u, \v) = \int_{\Omega} {\rm d}\rho(x) \int_{M \setminus \Omega} {\rm d}\rho(y)
 ( \nabla_{1,\u} \nabla_{2,\v} - \nabla_{2,\u} \nabla_{1,\v} ) \L(x,y), 
\end{gather*}
\item[$\blacktriangleright$] surface layer inner product
\begin{gather*}
(\u, \v) = \int_{\Omega} {\rm d}\rho(x) \int_{M \setminus \Omega} {\rm d}\rho(y)
 ( \nabla_{1,\u} \nabla_{1,\v} - \nabla_{2,\u} \nabla_{2,\v} ) \L(x,y), 
\end{gather*}
\item[$\blacktriangleright$] positive surface layer integral
\begin{gather}
 -\int_\Omega {\rm d}\rho(x) \int_{M \setminus \Omega} {\rm d}\rho(y) \nabla_{1,\v} \nabla_{2,\v} \L(x,y). \label{osipos}
\end{gather}
\end{itemize}
\end{Def}

\subsection{The causal action principle in Minkowski space} \label{seccapmink}
In the present paper, we are concerned with causal fermion systems in Minkowski
space. This means more precisely that the measure~$\rho = F_* \mu$ will always
be the push-forward of a local correlation map~\eqref{lcm}, i.e.,
\begin{gather} \label{rhoF}
\rho = (F_\tau)_* \mu
\end{gather}
(where~${\rm d}\mu={\rm d}^4x$ is again the volume measure of Minkowski space).
Indeed, later on the ensemble of Dirac wave functions will be a bit more complicated, because
we will work with several generations of lepton and quarks (for details see Section~\ref{secsetup}),
but this extension will be straightforward and without altering the following considerations.
The goal of this section is to explain how for causal fermion systems in Minkowski space,
the causal action principle can be rewritten as a variational principle for
wave functions in spacetime, similar as originally formulated in~\cite{pfp}.

The first step of the construction is to rewrite $\rho$-integrals as integrals over
Minkowski space. Indeed, by definition of the push-forward measure,
the action can be written as
\[ \Sact = \iint_{\scrM \times \scrM} \L ( F(x), F(y) ) \,{\rm d}\rho(x) {\rm d}\rho(y) . \]
In order to express the Lagrangian in terms of the wave functions, it is useful to
introduce the {\em{wave evaluation map}}~$\Psi(x)$ by
\begin{gather} \label{wepdef}
\Psi(x) \colon \ \Hil \rightarrow \C^4 ,\qquad \Psi(x) \psi_k = \psi_k(x) .
\end{gather}
Working with the wave evaluation operator gives a clear distinction between
vectors in the Hilbert space~$(\Hil, \la \cdot |\cdot \ra_\Hil)$ and their representation
by wave functions in spacetime. This is important because when varying the system,
the Hilbert space~$(\Hil, \la \cdot |\cdot \ra_\Hil)$ remains fixed, whereas the corresponding
wave functions may change. More specifically, a vector~$u \in \Hil$ is represented
by its corresponding {\em{physical wave function}}~$\psi^u$ given by
\begin{gather} \label{psiudef}
\psi^u(x) := \Psi(x) u \in \C^4 .
\end{gather}
The wave evaluation operator makes it possible to express
the local correlation operator~\eqref{Fdef} as
\begin{gather} \label{FPsi}
F(x) = -\Psi(x)^* \Psi(x) ,
\end{gather}
where the adjoint is taken with respect to the spin inner product, i.e.,
\[ \Sl \phi | \Psi(x) u \Sr = \la \Psi(x)^* \phi | u \ra_\Hil \qquad \text{for all~$\phi \in \C^4$ and~$u \in \Hil$} . \]
Using that trace of an operator product is invariant under cyclic permutations, we find that for any~$p \in \N$,
\begin{gather*}
 \tr \big( ( F(x) F(y) )^p \big) =
\tr \big( \big( \Psi(x)^* \Psi(x) F(y) ( F(x) F(y) )^{p-1} \big) \\
\hphantom{\tr \big( ( F(x) F(y) )^p \big)}{} = -\Tr \big( \big( \Psi(x) F(y) ( F(x) F(y) )^{p-1} \Psi(x)^*\big)\\
\hphantom{\tr \big( ( F(x) F(y) )^p \big)}{}
= \Tr \big( \big( (\Psi(x) \Psi(y)^* ) (\Psi(y) \Psi(x)^* ) \big)^p \big) ,
\end{gather*}
where~$\tr$ denotes the trace on~$\Hil$, whereas~$\Tr$ denotes the trace of a $(4 \times 4)$-matrix.
Introducing the {\em{kernel of the fermionic projector}}~$P(x,y)$ and the {\em{closed chain}}~$A_{xy}$ by
\begin{gather} \label{Pxydef}
P(x,y) := -\Psi(x) \Psi(y)^* ,\qquad A_{xy} := P(x,y) P(y,x) ,
\end{gather}
we obtain the simple relation
\[ \tr \big( ( F(x) F(y) )^p \big) = \Tr \big( ( A_{xy} )^p \big) \qquad \text{for all~$p \in \N$} . \]
Since the eigenvalues of an operator of finite rank can be expressed in terms of traces of
powers of the operator, we conclude that the operator~$F(x) F(y)$
has the same non-zero eigenvalues as the matrix~$A_{xy}$ (counting algebraic multiplicities).
This makes it possible to compute the eigenvalues~$\lambda^{xy}_1, \ldots, \lambda^{xy}_4$
in the Lagrangian~\eqref{Lagrange} and the boundedness constraint as the eigenvalues of the $(4 \times 4)$-matrix~$A_{xy}$. Moreover, the trace in the trace constraint (see~\eqref{trconstraint} and~\eqref{trc})
can be written as
\begin{gather} \label{trP}
\tr ( F(x) ) = -\tr ( \Psi(x)^* \Psi(x) ) = -\Tr ( \Psi(x) \Psi(x)^* )
= \Tr ( P(x,x) ) .
\end{gather}
Apart from the computational benefits, this argument explains why the action
and the constraints can be expressed in terms of the kernel of the fermionic projector.
This is the reason why the kernel of the fermionic projector will play a central role in
the subsequent analysis.

In order to understand how the wave functions come into play, we let~$\psi_1, \ldots, \psi_f$
be an orthonormal basis of~$\Hil$. Inserting a completeness relation, we obtain
for any~$\phi \in \C^4$,
\begin{gather*}
P(x,y) \phi \overset{\eqref{Pxydef}}{=}- \Psi(x) \Psi(y)^* \phi = -\sum_{k=1}^f \big(\Psi(x) \psi_k \big) \la\psi_k | \Psi(y)^* \phi \ra_\Hil \\
\hphantom{P(x,y) \phi}{} \ \,\, = -\sum_{k=1}^f \big(\Psi(x) \psi_k \big) \Sl \Psi(y) \psi_k | \phi \Sr
\overset{\eqref{wepdef}}{=} -\sum_{k=1}^f \psi_k(x) \Sl \psi_k(y) | \phi \Sr .
\end{gather*}
This relation can be written in the shorter form with bra/ket-notation as
\[ P(x,y) = -\sum_{k=1}^f |\psi_k(x)\Sr \Sl \psi_k(y) . \]
This shows that the kernel of the fermionic projector is composed of all the
wave functions~$\psi_1,\allowbreak \ldots, \psi_f$ of the system. Since, as noted above,
the action and the constraints can be expressed in~$P(x,y)$,
the causal action principle can be regarded as a variational principle
where one varies the ensemble of wave functions~$\psi_1, \ldots, \psi_f$.
Restricting attention to variations of this form, the volume constraint~\eqref{volconstraint}
is automatically satisfied (because we are working in Minkowski space with a fixed
spacetime volume~${\rm d}\mu = {\rm d}^4x$). Treating the boundedness constraint
by a Lagrange multiplier~\eqref{Lkappa}, our task is to minimize the action
\[ \Sact[P] := \iint_{\scrM \times \scrM} \L_\kappa [A_{xy} ] \, {\rm d}^4x {\rm d}^4y \]
under variations of the wave functions~$\psi_1, \ldots, \psi_f$ which respect
the trace constraint (see~\eqref{trc} and~\eqref{trP})
\[ \Tr ( P(x,x) ) = c \qquad \text{for all~$x \in \scrM$} . \]

\subsection{Regularized Dirac sea configurations} \label{secsea}
We now specify how to choose the ensemble of wave functions,
which in Section~\ref{secquantumcfs} was denoted by~$\psi_1, \ldots, \psi_f$ and which spanned
the Hilbert space of the causal fermion system.
In order to describe the Minkowski vacuum, we choose
the Hilbert space~$\Hil$ as the completion of the subspace of all negative-energy solutions
of the Dirac equation $({\rm i} \Pdd - m) \psi = 0$
(for simplicity of presentation, we here consider only one type of Dirac particles;
in Section~\ref{secPgen} we shall generalize the setting in a straightforward way to include all
leptons and quarks of the standard model).
The restriction of the scalar product to $\Hil$ is denoted by~$\la \cdot |\cdot \ra := (\cdot|\cdot)|_{\Hil \times \Hil}$.
For this choice of Hilbert space, the above construction of the local correlation
operators~\eqref{Fdef} does not apply because~$\Hil$ is infinite-dimensional,
and the wave functions in~$\Hil$ are defined only up to sets of measure zero.
As a consequence, the sesquilinear form in~\eqref{bxdef} is ill-defined.
In order to cure this problem, one needs to introduce an ultraviolet regularization by setting
\begin{equation*} 
\psi_\varepsilon={\mathfrak{R}}_\varepsilon(\psi) ,
\end{equation*}
where the {\em{regularization operators}} ${\mathfrak{R}}_\varepsilon \colon \Hil \rightarrow C^0\big(\R^{1,3}, \C^4\big)$
are linear operators whose range consists of continuous wave functions, and which converge to the identity, as $\varepsilon \searrow 0$, i.e.,
\[ \psi=\lim_{\varepsilon \searrow 0}{\mathfrak{R}}_\varepsilon(\psi) . \]
A simple example of a regularization operator is given by convolution with a suitable mollifier.
Working with the regularized wave functions, the sesquilinear form in~\eqref{bxdef}
is well-defined and bounded. Therefore, we can introduce the
{\em{regularized local correlation operator}}~$F^\varepsilon(x)$ by
\[ 
 \la \psi | {F^\varepsilon(x)} \phi \ra := -\overline{\psi_\varepsilon(x)} \phi_\varepsilon(x) \qquad \forall\, \psi, \phi \in \Hil , \]
and applying the above construction to~$F^\varepsilon$ gives a causal fermion system~$(\Hil, \F, \rho)$.

This construction requires some explanations. First, we note for clarity that the support of the
resulting measure~$\rho$ is given by the image of~$F^\varepsilon$,
\[ M := \supp \rho = \overline{F^\varepsilon\big(\R^{1,3}\big)} . \]
Thus, although~$\F$ is infinite-dimensional (more precisely, the operators in~$\F$ which
have exactly two positive and exactly two negative eigenvalues form an infinite-dimensional Banach manifold),
the support of~$\rho$ is four-dimensional.
It is a general concept that the causal action principle should give rise to measures which ``concentrate''
on low-dimensional subsets of~$\F$. This effect has been studied and proven in simple examples
in~\cite{sphere, support}. This analysis also reveals the underlying mechanisms.

The choice of $\Hil$ as the space of all negative-energy solutions of the Dirac equation realizes Dirac's original proposal that, in the vacuum state of the theory, all the states of negative energy must be occupied ({\em{Dirac sea}}). In the theory of causal fermion systems, this concept is taken seriously. However, the original problems inherent in this concept (like the infinite negative energy density of the sea) do not arise, because the measure describing the Dirac sea introduced above
is a critical point of the causal action principle in the above-mentioned
continuum limit. In simple terms, this means that
the ``states of the Dirac sea drop out of the Euler--Lagrange equations''. As a consequence, measures representing interacting
systems are realized by finite perturbations relative to the sea, giving rise to the usual description
in terms of particles and anti-particles.
The regularization operator $\mathfrak{R}_{\varepsilon}$ has the effect of ``smoothing'' the wave functions on a microscopic scale.
The length scale $\varepsilon$ involved in its definition can be thought of as the Planck length. In
the Theory of Causal Fermion Systems, the regularization is not merely a~technical tool in order to make ill-defined expressions meaningful, but it realizes the idea that, on microscopic length scales, the structure of spacetime must be modified. With this in mind, we consider the {\em{regularized objects}} as the \textit{physical objects}.

\subsection{The formalism of the continuum limit} \label{seccl}
As explained in Section~\ref{secquantumcfs}, it is a central idea
behind causal fermion systems to describe the physical system
purely in terms of the ensemble of wave functions.
Implementing this idea mathematically leads to the definition
of causal fermion systems (see Definition~\ref{defcfs}).
The dynamics of a causal fermion systems is described by
the causal action principle as introduced abstractly in Section~\ref{seccfs}.
In Section~\ref{seccapmink}, the causal action principle was rewritten
as a variational principle for an ensemble of wave functions in Minkowski
space. This action principle can be understood
as describing an interaction of all the physical wave functions of the system.
In order to write this interaction in a more tractable form, it is very helpful to
describe the collective behavior of all the physical wave functions by
bosonic potentials. This procedure has been carried out systematically in~\cite{cfs},
leading to the so-called {\em{continuum limit}} analysis where the interaction is described
effectively by classical bosonic gauge fields coupled to fermionic wave functions.

More specifically, in order to describe systems involving particles and/or anti-par\-ticles, following Dirac's hole theory one extends $\Hil$ by solutions of the Dirac equation of positive energy and/or removes states of negative energy. Bosonic fields
like the electromagnetic or gravitational fields correspond to collective ``excitations'' of the Dirac sea and spinorial wave functions described by a Dirac equation modified by a potential~$\B$,
\[ \big({\rm i} \gamma^j \partial_j + \B - m \big) \psi = 0 . \]
In order to make this picture precise, one makes use of the fact that
in a subtle analysis of the asymptotics~$\varepsilon \searrow 0$,
referred to as the {\em{continuum limit}}
(for details see~\cite[Chapter~4]{pfp} and~\cite[Section~2.4]{cfs}),
the measure $\rho$ describing the Minkowski vacuum turns out to be a critical point of the causal action.
If particles and/or anti-particles are present, this is no longer the case.
The measures, $\rho$, that are critical points of the causal action functional defined in~\eqref{Sdef}
are then perturbations of the measure describing the Minkowski vacuum state. They lead to a description of interactions among the Dirac particles, which,
asymptotically as~$\varepsilon \searrow 0$, can be described by bosonic fields. As already mentioned above, the analysis sketched here enables one to derive classical field equations, in particular the Maxwell equations and the Einstein field equations, from the causal action principle. Details are presented in~\cite[Chapters~3--5]{cfs}.
Here we only review a few constructions, also making the connection to the abstract setting
of causal fermion systems as outlined in Sections~\ref{seccfs} and~\ref{secosi}.

Before beginning, we note that, in our setting, the jet formalism in Section~\ref{secosi} simplifies because we always
restrict attention to measures of the form~\eqref{rhoF} which are the push-forward
of the volume measure of Minkowski space. Likewise, when varying this
measure according to~\eqref{rhoFf}, the function~$f$ is identically equal to one.
As a consequence, the scalar component of the resulting jet in~\eqref{vinfdef} vanishes, i.e.,
$\v=(0,v)$ with a vector field~$v$ on~$\F$. Another equivalent way of describing the variation
is to replace the mapping~$F$ in~\eqref{rhoF} by the family~$F_\tau$ with~$F_0=F$.
Using~\eqref{FPsi}, the variation can be described by a family~$\Psi_\tau$ of
wave evaluation operators, i.e.,
\[ F_\tau(x) = -\Psi_\tau(x)^* \Psi_\tau(x) , \]
where~$\Psi_\tau(x)\colon \Hil \rightarrow \C^4$ again gives a representation of the
vectors of~$\Hil$ as wave functions in Minkowski space (but these wave functions do not
need to satisfy the Dirac equation). Therefore, the vector field describing infinitesimal variations
is given by
\begin{gather} \label{varyPsi}
v\big(F(x)\big) = \frac{{\rm d}}{{\rm d}\tau} v (F_\tau(x) ) \Big|_{\tau=0} =
- (\delta \Psi )(x)^* \Psi(x) -\Psi(x)^* (\delta \Psi )(x) ,
\end{gather}
where~$(\delta \Psi)(x):=\frac{{\rm d}}{{\rm d}\tau} \Psi_\tau(x)|_{\tau=0}$ is the first variation of the
wave evaluation operator.
In this way, the jets can be associated to first variations of the wave evaluation operator.
For notational convenience, in what follows we denote the jets by the wave evaluation
operator and its first variation,
\begin{gather*} 
\v = ( \delta \Psi, \Psi) .
\end{gather*}
Using~\eqref{Pxydef}, the first variation of the kernel of the fermionic projector is given by
\begin{gather*} 
\delta P(x,y) := - (\delta \Psi )(x) \Psi(y)^* - \Psi(x) ( \delta \Psi )(y)^* .
\end{gather*}
We also use the notation
\[ \nabla_{1,\v} P(x,y) = - (\delta \Psi )(x) \Psi(y)^* \qquad \text{and} \qquad
\nabla_{2,\v} P(x,y) = - \Psi(x) ( \delta \Psi )(y)^* . \]
Thus here the jet derivative simply is a variational derivative where
all the wave functions are varied.
More specifically, one can consider different types of such variations:
\begin{enumerate}\itemsep=0pt
\item[(a)] One can vary individual physical wave functions. This gives rise to
the so-called {\em{fermionic jets}}.
\item[(b)] Alternatively, one can vary all physical wave functions collectively.
In physical applications when the physical wave functions satisfy the Dirac equation,
such variations can be described by bosonic fields, like for example
\begin{gather} \label{bosonicjet}
\delta \Psi(x) = -(s_m \slashed{A} \Psi)(x) ,
\end{gather}
where~$s_m$ is a Dirac Green's operator and~$A$ is the electromagnetic potential.
The corresponding jets are referred to as {\em{bosonic jets}}.
\end{enumerate}

Working with such variations, it is more convenient to implement the trace constraint~\eqref{trconstraint}
not by the pointwise condition~\eqref{trc} but instead by a Lagrange multiplier term.
Then the EL equations~\eqref{ELtest} can be written as (see~\cite[Proposition~1.4.3]{cfs})
\begin{gather} \label{Qrel}
\int_M Q(x,y) \psi^u(y) \,{\rm d}\rho(y) = \frac{\lambda}{2} \psi^u(x) \qquad \text{for all~$u \in \Hil$ and~$x \in M$} ,
\end{gather}
where~$\psi^u$ is the physical wave function~\eqref{psiudef}, $\lambda$ is the Lagrange multiplier corresponding to the trace constraint,
and the kernel~$Q(x,y)$ describes the first variation of the Lagrangian by
\begin{gather*} 
\delta \L_\kappa(x,y) =
\Tr_{S_y} ( Q(y,x) \delta P(x,y) ) + \Tr_{S_x} ( Q(x,y) \delta P(x,y)^* ) .
\end{gather*}
In the {\em continuum limit analysis} (for details see~\cite[Section~3.5.2]{cfs}), these EL equations are
evaluated for~$u$ a Dirac wave function of the form of an
{\em{ultrarelativistic wave packet}} of negative energy which is very small in a neighborhood
of~$x$. Evaluating in this specific way, one can evaluate the EL equations
independent of the detailed form of the regularization. The dependence on the
regularization is captured by the {\em{formalism of the continuum limit}}, which we now outline.
The starting point is the light-cone expansion of the unregularized kernel of the fermionic in Minkowski space
in the presence of an external potential, which has the form
\begin{gather} \label{fprep}
P(x,y) = \sum_{n=-1}^\infty
{\mbox{(smooth line integrals)}} \times T^{(n)}(x,y) + \text{(smooth contributions)} .
\end{gather}
Here the left side is a by-distribution in Minkowski space, and the right side
is its Hadamard expansion. The ``smooth line integrals'' involve the bosonic potentials
and their derivatives.
In order to regularize the light-cone expansion on the length scale~$\varepsilon$, we proceed as follows.
The smooth contributions are all left unchanged. For the regularization
of the factors~$T^{(n)}$, we employ the replacement rule
\begin{gather*} 
m^p T^{(n)} \rightarrow m^p T^{(n)}_{[p]} ,
\end{gather*}
where the factors~$T^{(n)}_{[p]}$ are smooth functions of~$\xi$.
These factors can be treated symbolically using the
following simple calculation rules. In computations one may treat the~$T^{(n)}_{[p]}$
like complex functions. However, one must be careful when tensor indices of factors~$\slashed{\xi}$
are contracted with each other. Naively, this gives a factor~$\xi^2$ which vanishes on the
light cone and thus changes the singular behavior on the light cone. In order to describe this
effect correctly, we first write every summand of the light cone expansion~\eqref{fprep}
such that it involves at most one factor~$\slashed{\xi}$ (this can always be arranged using
the anti-commutation relations of the Dirac matrices).
We now associate every factor~$\slashed{\xi}$ to the corresponding factor~$T^{(n)}_{[p]}$.
In short calculations, this can be indicated by putting brackets around the two factors,
whereas in the general situation we add corresponding indices
to the factor~$\slashed{\xi}$, giving rise to the replacement rule
\begin{gather*} 
m^p \slashed{\xi} T^{(n)} \rightarrow m^p \slashed{\xi}^{(n)}_{[p]} T^{(n)}_{[p]} .
\end{gather*}
For example, we write the regularized fermionic projector of the vacuum as
\[ P^\varepsilon(x,y) = \frac{{\rm i}}{2} \sum_{n=0}^\infty \frac{m^{2n}}{n!} \slashed{\xi}^{(-1+n)}_{[2n]} T^{(-1+n)}_{[2n]}
+ \sum_{n=0}^\infty \frac{m^{2n+1}}{n!} T^{(n)}_{[2n+1]} . \]

The kernel~$P^\varepsilon(y,x)$ is obtained by taking the conjugate,
\begin{gather*} 
 ( P^\varepsilon(x,y) )^* = P^\varepsilon(y,x)
\end{gather*}
(where the star denotes the adjoint with respect to the inner product~$\Sl \cdot|\cdot \Sr$).
The conjugates of the factors~$T^{(n)}_{[p]}$ and~$\xi^{(n)}_{[p]}$ are the complex conjugates,
\[ \overline{T^{(n)}_{[p]}} := \big(T^{(n)}_{[p]} \big)^* \qquad \text{and} \qquad
\overline{\xi^{(n)}_{[p]}} := \big(\xi^{(n)}_{[p]} \big)^* . \]
One must carefully distinguish between these factors with and without complex conjugation.
In particular, the factors~$\slashed{\xi}^{(n)}_{[p]}$ need not be symmetric,
\[ \big( \slashed{\xi}^{(n)}_{[p]} \big)^* \neq \slashed{\xi}^{(n)}_{[p]} \qquad \text{in general} . \]

When forming composite expressions, the tensor indices of the factors~$\xi$ are
contracted to other tensor indices.
The factors~$\xi$ which are contracted to other factors~$\xi$ are called {\em{inner factors}}.
The contractions of the inner factors are handled with the so-called {\em{contraction rules}}
\begin{gather}
\big(\xi^{(n)}_{[p]}\big)^j \big(\xi^{(n')}_{[p']}\big)_j =
\frac{1}{2} \Big( z^{(n)}_{[p]} + z^{(n')}_{[p']} \Big), \label{eq52} \\
\big(\xi^{(n)}_{[p]}\big)^j \overline{\big(\xi^{(n')}_{[p']}\big)_j} =
\frac{1}{2} \Big( z^{(n)}_{[p]} + \overline{z^{(n')}_{[p']}} \Big), \label{eq53} \\
z^{(n)}_{[p]} T^{(n)}_{[p]} = -4 \Big( n T^{(n+1)}_{[p]}
+ T^{(n+2)}_{\{p \}} \Big) , \label{eq54}
\end{gather}
which are to be complemented by the complex conjugates of these equations.
Here the factors~$z^{(n)}_{[p]}$
can be regarded simply as a book-keeping device
to ensure the correct application of the rule~\eqref{eq54}.
The factors~$T^{(n)}_{\{p\}}$
have the same scaling behavior as the~$T^{(n)}_{[p]}$,
but their detailed form is somewhat different; we simply treat them as a new class of symbols.
In cases where the lower index does not need to be specified we write~$T^{(n)}_\circ$.
After applying the contraction rules, all inner factors~$\xi$ have disappeared.
The remaining so-called {\em{outer factors}}~$\xi$
need no special attention and are treated
like smooth functions.

Next, to any factor~$T^{(n)}_\circ$ we associate the {\em{degree}} $\deg T^{(n)}_\circ$
on the light cone by
\begin{gather} \label{degdef}
\deg T^{(n)}_\circ = 1-n .
\end{gather}
The degree is additive in products, whereas the degree of a quotient is defined as the
difference of the degrees of numerator and denominator. The degree of an expression
can be thought of as describing the order of its singularity on the light cone, in the sense that a
larger degree corresponds to a stronger singularity (for example, the
contraction rule~\eqref{eq54} increments~$n$ and thus decrements the degree, in
agreement with the naive observation that the function~$z=\xi^2$ vanishes on the light cone).
Using formal Taylor series, we can expand in the degree. In all our applications, this will
give rise to terms of the form{\samepage
\begin{gather} \label{sfr}
\eta(x,y)
\frac{ T^{(a_1)}_\circ \cdots T^{(a_\alpha)}_\circ
\overline{T^{(b_1)}_\circ \cdots T^{(b_\beta)}_\circ} }
{ T^{(c_1)}_\circ \cdots T^{(c_\gamma)}_\circ
\overline{T^{(d_1)}_\circ \cdots T^{(d_\delta)}_\circ} } \qquad \text{with~$\eta(x,y)$ smooth} .
\end{gather}
The quotient of the two monomials in this equation is referred to as a {\em{simple fraction}}.}

A simple fraction can be given a quantitative meaning by considering one-dimensional integrals
along curves which cross the light cone transversely away from the origin~$\xi=0$.
This procedure is called {\em{weak evaluation on the light cone}}.
For our purpose, it suffices to integrate over the time coordinate~$t=\xi^0$ for fixed~$\vec{\xi} \neq 0$.
Moreover, using the symmetry under reflections~$\xi \rightarrow -\xi$, it suffices to consider the upper
light cone~$t \approx |\vec{\xi}|$. The resulting integrals diverge if the regularization
is removed. The leading contribution for small~$\varepsilon$ can be written as
\begin{gather*}
\int_{|\vec{\xi}|-\varepsilon}^{|\vec{\xi}|+\varepsilon} {\rm d}t\, \eta(t,\vec{\xi})
\frac{ T^{(a_1)}_\circ \cdots T^{(a_\alpha)}_\circ
\overline{T^{(b_1)}_\circ \cdots T^{(b_\beta)}_\circ} }
{ T^{(c_1)}_\circ \cdots T^{(c_\gamma)}_\circ
\overline{T^{(d_1)}_\circ \cdots T^{(d_\delta)}_\circ} }
 \approx \eta(|\vec{\xi}|,\vec{\xi}) \frac{c_\reg}{(i |\vec{\xi}|)^L}
 \frac{\log^r (\varepsilon |\vec{\xi}|)}{\varepsilon^{L-1}} , 
\end{gather*}
where~$L$ is the degree of the simple fraction and~$c_\reg$, the so-called {\em{regularization parameter}},
is a real-valued function of the spatial direction~$\vec{\xi}/|\vec{\xi}|$ which also depends on
the simple fraction and on the regularization details
(the error of the approximation will be specified below). The integer~$r$ describes
a possible logarithmic divergence. Apart from this logarithmic divergence, the
scalings in both~$\xi$ and~$\varepsilon$ are described by the degree.

When analyzing a sum of expressions of the form~\eqref{sfr}, one must
know if the corresponding regularization parameters are related to each other.
In this respect, the {\em{integration-by-parts rules}}
are important, which are described
symbolically as follows. On the factors~$T^{(n)}_\circ$ we introduce a~derivation~$\nabla$ by
\[ \nabla T^{(n)}_\circ = T^{(n-1)}_\circ . \]
Extending this derivation with the Leibniz and quotient rules to simple fractions, the
integration-by-parts rules state that
\begin{gather*} 
\nabla \left( \frac{ T^{(a_1)}_\circ \cdots T^{(a_\alpha)}_\circ
\overline{T^{(b_1)}_\circ \cdots T^{(b_\beta)}_\circ} }
{ T^{(c_1)}_\circ \cdots T^{(c_\gamma)}_\circ
\overline{T^{(d_1)}_\circ \cdots T^{(d_\delta)}_\circ} }
\right) = 0 .
\end{gather*}
These rules give relations between simple fractions.

Using the above computation rules, one can evaluate and analyze the EL equations~\eqref{Qrel}
in detail, always working modulo error terms of the form
\begin{gather} \label{ap1}
\text{higher orders in~$\varepsilon/\ell_\text{macro}$} ,
\end{gather}
where~$\ell_\text{macro}$ denotes the length scale of macroscopic physics
(like the Compton scale~$m^{-1}$ or larger).
For more detailed explanations of these constructions we again refer to~\cite{cfs}.

\begin{figure}[t]\centering
\includegraphics{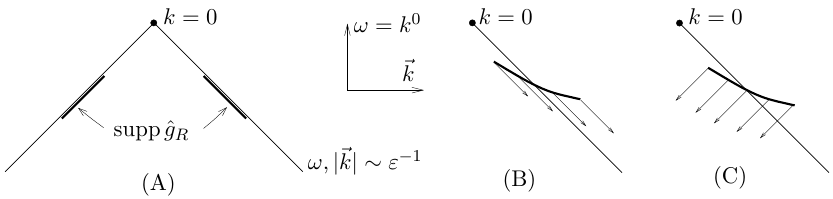}
\caption{Plots of~$\hat{g}_R$ exemplifying different regularization mechanisms in the neutrino sector.}
\label{l:fig1}
\end{figure}

Finally, we shall also encounter specific regularization effects of the {\em{neutrinos}},
which we now briefly explain (for details see~\cite[Section~4.2]{cfs}).
We first decompose the kernel of the fermionic projector describing neutrinos
into its left- and right-handed components as well as the scalar component,
\begin{gather*} 
P^\varepsilon(x,y) = \chi_L \slashed{g}_L(x,y) + \chi_R \slashed{g}_R(x,y) + h(x,y) .
\end{gather*}
If the neutrinos are massless and left-handed, then~$g_R$ and~$h$ are zero.
In order to describe massive neutrinos, both~$g_R$ and~$h$ are non-zero
(thus massive neutrinos have a right-handed components, but it does not couple to
any gauge fields, which are zero or left-handed).
Furthermore, we introduce a non-trivial regularization of the right-handed component~$g_R$.
Here ``non-trivial'' simply means that the regularization is designed with a specific purpose in mind.
More precisely, in the example of a spherically symmetric regularization,
these regularization effects are described by specific contributions to~$g_R$.
They are supported on an energy scale which is typically much smaller than the
Planck energy (see Fig.~\ref{l:fig1}(A)).
These contributions to~$g_R$ have the form
\begin{gather}
\hat{\slashed{g}}_R(k) \asymp -4 \pi^2 \big(\gamma^0 + \gamma^k\big) \hat{h}_1(\omega) \delta \big(
|\vec{k}|-K(\omega) \big) \qquad \text{general surface states}, \label{gss} \\
\hat{\slashed{g}}_R(k) \asymp -4 \pi^2 \big(\gamma^0 - \gamma^k\big) \hat{h}_2(\omega) \delta \big(
|\vec{k}|-K(\omega) \big) \qquad \text{shear of surface states} \label{shear}
\end{gather}
with weight functions~$\hat{h}_{1/2}(\omega)$.
Here the function~$K(\omega)$ parametrizes a surface in momentum space
on which the surface states are supported (see Fig.~\ref{l:fig1}(B)).
For the shear contributions, on the other hand, the vector field~$g_R$ is no longer tangential
to the mass cone, but the sign of the spatial component is flipped (see Fig.~\ref{l:fig1}(C)).
These contributions to~$g_R$ serve two purposes:
First, they make it possible to arrange that the EL equations are satisfied in the vacuum
(compensating for the fact that the masses of the neutrinos are smaller than that of charged leptons
and quarks).
Second, in curved space-time they give rise to curvature contributions which
yield the Einstein equations in the continuum limit,
with the gravitational constant~$\kappa$ given by~$\kappa \sim \delta^2$
with the length scale~$\delta$ (which can be identified with the Planck length)
as determined by the above regularization effects.

In the formalism of the continuum limit, the regularization effects in~\eqref{gss} and~\eqref{shear}
are taken into account by factors~$T^{(n)}_{[R,p]}$ and~$T^{(n)}_{\{ R,p \} }$
which come with corresponding computation rules. Here we do not need to
enter the details but refer the interested reader instead to~\cite[Section~4.2.5]{cfs}.

\section{Mathematical setup} \label{secsetup}
In this section we introduce the mathematical setup and recall a few constructions from~\cite{cfs}.

\subsection{The fermionic projector} \label{secPgen}
We denote the points of Minkowski space by~$x,y \in \scrM$.
In order to describe the vacuum, we consider the Dirac sea configuration of the standard model
as introduced in~\cite[Chapter~5]{cfs}. Thus we consider the kernel of the fermionic projector
\begin{gather} \label{q:Pvac}
P(x,y) = P^N(x,y) \oplus P^C(x,y) ,
\end{gather}
where~$P^C$ is composed of the Dirac seas of the charged leptons and quarks
\[
P^C(x,y) = \bigoplus_{a=1}^7 \sum_{\beta=1}^3 P^\text{vac}_{m_\beta}(x,y) , \]
where~$m_\beta$ are the masses of the fermions and~$P^\text{vac}_m$ is the distribution
\[ P^\text{vac}_m(x,y) = \int \frac{{\rm d}^4k}{(2 \pi)^4} (\slashed{k}+m) \delta\big(k^2-m^2\big) \Theta(-k^0) {\rm e}^{-{\rm i}k(x-y)} \]
(where~$\Theta$ denotes the Heaviside function).
The direct summand~$P^N$ in~\eqref{q:Pvac} describes the neutrinos
\[
P^N(x,y) = \sum_{\beta=1}^3 P^\text{vac}_{\tilde{m}_\beta}(x,y) , \]
where the neutrino masses~$\tilde{m}_\beta \geq 0$ will in general be different from
the masses~$m_\beta$ in the charged sector.

Before going on, we point out that~\eqref{q:Pvac} involves eight direct summands
(seven for the charged leptons and
quarks, and one for the neutrinos). Thus~$P(x,y)$ is a $(32 \times 32)$-matrix.
Consequently, constructing the corresponding causal fermion system as explained
for one Dirac sea in Sections~\ref{secquantumcfs} and~\ref{secsea}, one obtains a~causal fermion system of spin dimension~$16$.
The reason why eight direct summands give rise to the interactions of the standard
model is that, when gauge phases are taken into account,
the eight direct summands (also called sectors) form pairs (referred to
as blocks). One of the resulting four blocks describes the leptons,
whereas the remaining three blocks describe the three quark colors.
Here we cannot describe the underlying mechanisms, which are worked out
in~\cite[Sections~5.3 and 5.4]{cfs}.

In order to describe the interaction, we first introduce the {\em{auxiliary fermionic projector}} by
\[ P^\text{aux} = P^N_\text{aux} \oplus P^C_\text{aux} , \]
where
\[
P^N_\text{aux} = \left( \bigoplus_{\beta=1}^3 P^\text{vac}_{\tilde{m}_\beta} \right) \oplus 0
\qquad \text{and} \qquad
P^C_\text{aux} = \bigoplus_{a=1}^7 \bigoplus_{\beta=1}^3 P^\text{vac}_{m_\beta} \]
(the last direct summand of~$P^N_\text{aux}$ has the purpose
of describing the right-handed high-energy states; for details see~\cite[Sections~4.2.6 and~5.2.1]{cfs}).
Next, we introduce the {\em{chiral asymmetry matrix}}~$X$ and the {\em{mass matrix}}~$Y$ by
\begin{gather*}
X = \left( \1_{\C^3} \oplus \tau_\reg \chi_R \right) \oplus \bigoplus_{a=1}^7 \1_{\C^3}, \\
m Y = \operatorname{diag} ( \tilde{m}_1, \tilde{m}_2, \tilde{m}_3, 0 )
\oplus \bigoplus_{a=1}^7 \operatorname{diag} ( m_1, m_2, m_3 ) ,
\end{gather*}
where~$\chi_R := \big(1 + \gamma^5\big)/2$ is the projection on the right-handed component,
$m$ is an arbitrary mass parameter, and~$\tau_\reg \in (0,1]$ is a dimensionless parameter
which is used in order to keep track of the regularization effects in the neutrino sector
(for a detailed explanation see again~\cite[Section~4.2.6]{cfs}).
This allows us to rewrite the vacuum fermionic projector as
\begin{gather} \label{q:Pauxdef}
P^\text{aux} = X t = t X^* \qquad \text{with} \qquad t := \bigoplus_{\beta=1}^{25} P^\text{vac}_{m Y^\beta_\beta} .
\end{gather}
Now~$t$ is a solution of the Dirac equation
\[ ({\rm i} \Pdd - m Y) t = 0 . \]
We note for clarity that~$t(x,y)$ is an $(N \times N)$-matrix with~$N = 4\times 25=100$.
The matrices~$X$ and~$Y$, on the other hand, are $25 \times 25$-matrices which act
on the direct summands in~\eqref{q:Pauxdef} (thus these matrices act trivially on the spinor index).

In order to introduce the interaction, we insert an operator~$\B$ into the Dirac equation,
\begin{gather} \label{q:Dinteract}
({\rm i} \Pdd + \B - m Y) \tilde{t} = 0
\end{gather}
(for clarity, we always denote the objects in the presence of a bosonic potential
by an additional tilde). Now~$\tilde{t}$ can be computed in a perturbation expansion
in~$\B$. For example, to first order we obtain
\[ \tilde{t} = t - s \B t - t \B s , \]
where~$s$ is the direct sum of the Green's operators of each Dirac sea, i.e.,
\begin{gather} \label{sdef}
s = \bigoplus_{\beta=1}^{25} s_{m Y^\beta_\beta} ,
\end{gather}
and~$s_m$ is the integral operator whose kernel~$s_m(x,y)$ is the
mean of the advanced and retarded Green's function for Dirac particles of mass~$m$.
A systematic treatment of all orders in~$\B$, taking into account the
correct normalization of the Dirac states, leads to the so-called {\em{causal
perturbation expansion}} (for details see~\cite[Section~2.2]{pfp}, \cite[Sections~2.1.6 and~5.2.1,]{cfs}
or~\cite{norm}). Analyzing the singular behavior of the resulting distributions on the light cone
gives rise to a~representation of~$\tilde{t}$ of the form
\begin{gather*}
\tilde{t}(x,y) = \sum_{n=-1}^\infty\sum_{k} m^{p_k}
{\text{(nested bounded line integrals)}} \times T^{(n)}(x,y) \\
\hphantom{\tilde{t}(x,y) =}{} + \text{(smooth contributions)} ,
\end{gather*}
where the line integrals involve~$\B$ and its partial derivatives.
For details on this so-called light-cone expansion we refer to~\cite[Section~2.2]{cfs}.

Finally, the fermionic projector~$P$ in the presence of the potential~$\B$
is obtained by forming the {\em{sectorial projection}}
\[
( P )^i_j = \sum_{\alpha, \beta} \big(\tilde{P}^\text{aux}\big)^{(i,\alpha)}_{(j, \beta)} , \]
where~$i,j \in \{1, \ldots, 8\}$ is the sector index, and the indices~$\alpha$ and~$\beta$ run over
the corresponding generations (i.e., $\alpha \in \{1, \ldots, 4\}$ if~$i=1$ and~$\alpha \in \{1, 2, 3 \}$
if~$i=2, \ldots, 8$). For more details on the sectorial projection and its significance
we refer to~\cite[Section~3.4.1]{cfs}.

\subsection{Electromagnetic and gravitational interactions}
For simplicity, we here restrict attention to electromagnetic and gravitational interactions.
An electromagnetic field is described by the potential
\begin{gather} \B[A^{\rm em}] =
\slashed{A}^{\rm em} \operatorname{diag} \left(0,-1, \frac{2}{3}, -\frac{1}{3}, \frac{2}{3}, -\frac{1}{3},
\frac{2}{3}, -\frac{1}{3} \right) . \label{q:em}
\end{gather}
(thus the potential acts trivially on the generation index; see~\cite[Section~5.3.1]{cfs}).
For technical simplicity, we only consider linearized gravity. Thus, as in~\cite[Section~2.3]{cfs}
we consider a first order perturbation~$h_{jk}$ of the Minkowski
metric~$\eta_{jk}=\operatorname{diag}(1,-1,-1,-1)$,
\[ g_{jk}(x) = \eta_{jk} + h_{jk}(x) . \]
Working in the symmetric gauge, the resulting perturbation of the Dirac operator is
\[ \B[h_{jk}] = - \frac{{\rm i}}{2} \gamma^j h_{jk} \eta^{kl}
 \frac{\partial}{\partial x^l} +\frac{{\rm i}}{8} (\Pdd h) . \]

After including these potentials into the operator~$\B$ in the Dirac equation~\eqref{q:Dinteract},
the distribution~$\tilde{t}$ is computed with the help of the causal perturbation expansion
(for an introduction see~\cite[Section~2.1]{cfs}).
Before forming composite expressions in the kernel of the fermionic projector,
one must introduce an ultraviolet regularization.
We denote the length scale of this regularization by~$\varepsilon$.
Then the Lagrangian can be computed asymptotically as~$\varepsilon \searrow 0$
in the formalism of the continuum limit (as introduced in~\cite[Sections~2.4 and~2.6]{cfs}).
For completeness, we remark that we again assume the following regularization conditions
(see~\cite[equations~(4.6.36), (4.6.38), (4.9.2) and~(5.2.9), (5.2.10)]{cfs})
\begin{alignat}{3}
& T^{(0)}_{[0]} T^{(-1)}_{[0]} \overline{T^{(0)}_{[0]}} = 0 \qquad &&
\text{in a weak evaluation on the light cone},& \nonumber\\ 
& \big| L^{(n)}_{[0]} \big| = \big| T^{(n)}_{[0]} \big| \left( 1 + \O \big( (m \varepsilon)^{2 p_\reg} \big) \right)
\qquad && \text{for~$n=0, -1$ pointwise} ,& \label{q:Tpoint}
\end{alignat}
where the parameter~$p_\reg$ is in the range~$0 < p_\reg < 2$,
and the factors~$L^{(n)}_{[0]}$ are defined by
\begin{gather} \label{l:Lndef}
L^{(n)}_{[p]} = T^{(n)}_{[p]} + \frac{1}{3} \tau_\reg T^{(n)}_{[R,p]} .
\end{gather}

\section{The Lagrangian of the causal action in Minkowski space}
As in~\cite[Section~5.2.4]{cfs} we count the eigenvalues of the
closed chain~$A_{xy} := P(x,y) P(y,x)$ with algebraic multiplicities and denote them by~$\lambda^{xy}_{ncs}$,
where~$n \in \{1,\ldots, 8\}$, $c \in \{L,R\}$ and~$s \in \{+,-\}$.
Then the causal action reads
\[ \Sact[P] = \iint_{\scrM \times \scrM} \L[A_{xy}] \, {\rm d}^4 x {\rm d}^4y \]
with the Lagrangian (see~\cite[Section~5.2.4 and equation~(1.1.9)]{cfs})
\[ \L[A_{xy}] = \big|A_{xy}^2\big| - \frac{1}{32} |A_{xy}|^2
= \frac{1}{64} \sum_{n,n'=1}^8 \sum_{c,c' =L,R} \sum_{s,s'=\pm}
 \big( \big|\lambda^{xy}_{ncs} \big| - \big|\lambda^{xy}_{n'c's'} \big| \big)^2
 . \]
Using the relation (see~\cite[equation~(4.4.3)]{cfs})
\begin{gather} \label{l:ccp}
\lambda^{xy}_{nR\pm} = \overline{\lambda^{xy}_{nL\mp}} ,
\end{gather}
it suffices to consider the summands with~$s=s'=+$,
\begin{gather} \label{Lag}
\L[A_{xy}] = \frac{1}{16} \sum_{n,n'=1}^8 \sum_{c,c' =L,R}
\big( \big|\lambda^{xy}_{nc+} \big| - \big|\lambda^{xy}_{n'c'+} \big| \big)^2 .
\end{gather}

In the vacuum and to leading degree five on the light cone, the eigenvalues of the
close chain all have the same absolute value (see~\cite[Section~2.6.1 or Section~3.6.1 and Section~4.3.1]{cfs}. Therefore, when
perturbing the Lagrangian, we need to perturb each factor~$\big( \big|\lambda^{xy}_{nc+} \big| - \big|\lambda^{xy}_{n'c'+} \big|\big)$
in~\eqref{Lag}, i.e.,
\begin{gather} \label{DeltaLag}
\Delta \L[A_{xy}] = \frac{1}{8} \sum_{n,n'=1}^8 \sum_{c,c' =L,R}
\Delta \big( \big|\lambda^{xy}_{nc+} \big| - \big|\lambda^{xy}_{n'c'+} \big| \big)
 \Delta \big( \big|\lambda^{xy}_{nc+} \big| - \big|\lambda^{xy}_{n'c'+} \big| \big) .
\end{gather}
For a convenient notation, we write the contributions to~$\Delta \L$ in the form
\begin{gather} \label{notation}
\sim (\cdots) \cdot (\cdots) ,
\end{gather}
where the two brackets refer to the two factors in~\eqref{DeltaLag}.

Many contributions to the perturbation~$\Delta \big|\lambda^{xy}_{nc+}\big|$ of the absolute values of the
eigenvalues were already computed in~\cite{cfs}. In order to make these results applicable,
it is most convenient to use the following lemma.
\begin{Lemma} The perturbation of the absolute value of the eigenvalues
is related to the perturbation~${\mathscr{K}}_{nc}(x,y)$
computed in~{\rm \cite[Chapter~4]{cfs}} by
\[ \Delta |\lambda^{xy}_{nc+}| = \frac{1}{3} {\mathscr{K}}_{nc}(x,y)
\frac{\big| T^{(0)}_{[0]} \big|}{T^{(0)}_{[0]} \big|T^{(-1)}_{[0]} \big|} . \]
\end{Lemma}
\begin{proof} According to~\cite[equation~(4.4.6)]{cfs},
\[
{\mathscr{K}}_{nc}(x,y) := \frac{\Delta \big|\lambda^{xy}_{nc-}\big|}{\big|\lambda^{xy}_-\big|} 3^3
T^{(0)}_{[0]} T^{(-1)}_{[0]} \overline{T^{(-1)}_{[0]}} . \]
We now use~\eqref{l:ccp} together with the fact that the eigenvalues of the vacuum
have the absolute value~$\big|\lambda^{xy}_-\big| = 3^2 \big| T^{(-1)}_{[0]} T^{(0)}_{[0]} \big|$
(see~\cite[equation~(3.6.5)]{cfs} or~\eqref{l:lamp} below).
\end{proof}

\subsection{The eigenvalues of the closed chain in the vacuum}
In the vacuum, the eigenvalues of the closed chain can be computed separately in each sector.
They are given by (see~\cite[Section~4.4.2]{cfs})
\begin{gather}
\lambda^{xy}_{nL+} = \lambda^{xy}_{nR+} = 9 T^{(0)}_{[0]} \overline{T^{(-1)}_{[0]}} + m^2 (\cdots) + (\deg < 2)
 ,\qquad n=2,\ldots,8, \label{l:lamp} \\
\lambda^{xy}_{1L+} = 9 T^{(0)}_{[0]} \overline{L^{(-1)}_{[0]}}
+ 3 \tau_\reg \delta^{-2} T^{(0)}_{[0]} \overline{T^{(0)}_{[R, 2]}} + m^2 (\cdots) + (\deg < 2), \label{l:lamL} \\
\lambda^{xy}_{1R+} = 9 L^{(0)}_{[0]} \overline{T^{(-1)}_{[0]}}
- 3 \tau_\reg \delta^{-2} T^{(1)}_{\{R, 0 \}} \overline{T^{(-1)}_{[0]}} + m^2 (\cdots) + (\deg < 2) , \label{l:lamR}
\end{gather}
where~$\deg$ is the degree on the light cone (see~\eqref{degdef} and the explanation thereafter),
and the factors~$L^{(n)}_{[0]}$ are again given by~\eqref{l:Lndef}.
Here the terms~$(\cdots)$ stand for additional contributions whose explicit form will not be needed here
(for details see~\cite[equation~(5.3.24)]{pfp}).
However, it is important to keep in mind that in~\eqref{l:lamp}, these
terms depend on the masses $m_\beta$ of the charged leptons.
They coincide with the corresponding terms in~\eqref{l:lamL} and~\eqref{l:lamR}, except
that in the latter terms the masses~$m_\beta$ are to be replaced by the neutrino masses~$\tilde{m}_\beta$.
The parameter~$\delta$ in~\eqref{l:lamL} and~\eqref{l:lamR} describes the length scale on which the
regularization effects in the neutrino sector
(due to the shear and the general surface states; for details see below) come into play.
It scales like (for details see~\cite[Section~4.4.2]{cfs})
\[ \varepsilon \ll \delta \lesssim \frac{1}{m} (m \varepsilon)^{\frac{p_\reg}{2}} , 
\]
where~$p_\reg \in (0,2)$ is again the power in~\eqref{q:Tpoint},
and the notation~$\lesssim$ means ``smaller than a constant times the right side''.
The length scale~$\delta$ can be thought of as the Planck scale, because
the gravitational coupling constant~$\kappa$ obtained in the continuum limit scales like
(see~\cite[Sections~4.9 and~5.4.3]{cfs}){\samepage
\[ \kappa \simeq \frac{\delta^2}{\tau_\reg} . \]
(where $\simeq$ means that we omit irrelevant real prefactors).}

A straightforward computation using~\eqref{l:lamp}--\eqref{l:lamR}
as well as~\eqref{q:Tpoint} yields
\begin{gather}
\big|\lambda^{xy}_{nL+}\big| = \big|\lambda^{xy}_{nR+}
\big| = 9 \big| T^{(0)}_{[0]} T^{(-1)}_{[0]} \big| + m^2 (\deg<3) + (\deg < 2), \notag \\
\big| \lambda^{xy}_{1L+} \big| = 9 \big| T^{(0)}_{[0]} L^{(-1)}_{[0]} \big|
+ \re \left(3 \tau_\reg \delta^{-2} T^{(0)}_{[0]} \overline{T^{(0)}_{[R, 2]}}
\frac{9 \overline{T^{(0)}_{[0]}} L^{(-1)}_{[0]}}{9 \big| T^{(0)}_{[0]} L^{(-1)}_{[0]} \big|} \right) \notag \\
\hphantom{\big| \lambda^{xy}_{1L+} \big| =}{} + m^2 (\deg<3) + (\deg < 2) \notag \\
\hphantom{\big| \lambda^{xy}_{1L+} \big|}{} = 9 \big| T^{(0)}_{[0]} T^{(-1)}_{[0]} \big|
+\frac{3 \tau_\reg}{2 \delta^2} \frac{\big|T^{(0)}_{[0]} \big|}{\big| T^{(-1)}_{[0]} \big|}
\Big\{ T^{(0)}_{[R, 2]} \overline{L^{(-1)}_{[0]}} + \text{c.c.} \Big\} \notag\\
\hphantom{\big| \lambda^{xy}_{1L+} \big| =}{} + m^2 (\deg<3) + (\deg < 2), \label{needed} \\
\big| \lambda^{xy}_{1R+} \big| = 9 \big| L^{(0)}_{[0]} T^{(-1)}_{[0]} \big|
+ \re \left( - 3 \tau_\reg \delta^{-2} T^{(1)}_{\{R, 0 \}} \overline{T^{(-1)}_{[0]}}
\frac{9 \overline{L^{(0)}_{[0]}} T^{(-1)}_{[0]}}{9 \big| L^{(0)}_{[0]} T^{(-1)}_{[0]} \big|} \right) \notag \\
\hphantom{\big| \lambda^{xy}_{1R+} \big|=}{} + m^2 (\deg<3) + (\deg < 2) \notag \\
\hphantom{\big| \lambda^{xy}_{1R+} \big|}{} = 9 \big| T^{(0)}_{[0]} T^{(-1)}_{[0]} \big|
-\frac{3 \tau_\reg}{2 \delta^2} \frac{ \big|T^{(-1)}_{[0]} \big|}{\big| T^{(0)}_{[0]} \big|}
\Big\{ T^{(1)}_{\{R, 0 \}} \overline{L^{(0)}_{[0]}} + \text{c.c.} \Big\}\notag \\
\hphantom{\big| \lambda^{xy}_{1R+} \big|-}{} + m^2 (\deg<3) + (\deg < 2) . \label{noneed}
\end{gather}
We recall that~\eqref{needed} and~\eqref{noneed} involve the
effects of the general surface states and the shear
(see~\eqref{gss} and~\eqref{shear}). More precisely, the
 the factor~$\delta^{-2} T^{(0)}_{[R, 2]}$ in~\eqref{needed}
describes the effect of {\em{general surface states}}. This factor is essential for
getting the Einstein equations in the continuum limit (see~\cite[Sections~4.9 and~5.4.3]{cfs}).
The factor~$\delta^{-2} T^{(1)}_{\{R, 0 \}}$ in~\eqref{noneed}, on the other hand,
describes the {\em{shear}} of the sea states. At the moment, there is no compelling reason why
the regularization should involve a shear on the scale~${\sim} \delta^{-2}$. The shear is needed
merely in order to satisfy the EL equations to the order~${\sim} m^2$ in the vacuum.
Therefore, the factor in~\eqref{noneed} could be as small as
\[ \delta^{-2} T^{(1)}_{\{R, 0 \}} \simeq m^2 (\deg=0) . \]
We also point out that from the size of the factors~$\delta^{-2} T^{(0)}_{[R, 2]}$ or~$\delta^{-2} T^{(1)}_{\{R, 0 \}}$
we cannot infer how large the contributions~\eqref{needed} and~\eqref{noneed} are,
because there may be cancellations between the two summands inside the curly brackets
(see also~\cite[Section~4.4.2]{cfs}).

\subsection[The contributions ${\sim} \delta^{-2} \cdot \delta^{-2}$ and ${\sim} \delta^{-4} \cdot \delta^{-4}$]{The contributions $\boldsymbol{{\sim} \delta^{-2} \cdot \delta^{-2}}$ and $\boldsymbol{{\sim} \delta^{-4} \cdot \delta^{-4}}$}\label{secdd}

We now begin with the computation of different contributions to the Lagrangian,
using the notation~\eqref{notation}. In view of the contributions~\eqref{needed} and~\eqref{noneed},
the leading contribution to the Lagrangian~\eqref{Lag} is of the form
\begin{gather}
 \left( \frac{\tau_\reg}{\delta^2} \frac{\big|T^{(0)}_{[0]} \big|}{\big| T^{(-1)}_{[0]} \big|}
\Big\{ T^{(0)}_{[R, 2]} \overline{L^{(-1)}_{[0]}} + \text{c.c.} \Big\} \right)^2
+ \left( \frac{3 \tau_\reg}{2 \delta^2} \frac{ \big|T^{(-1)}_{[0]} \big|}{\big| T^{(0)}_{[0]} \big|}
\Big\{ T^{(1)}_{\{R, 0 \}} \overline{L^{(0)}_{[0]}} + \text{c.c.} \Big\} \right)^2 \notag \\
\qquad\quad{} + \left( \frac{\tau_\reg}{\delta^2} \frac{\big|T^{(0)}_{[0]} \big|}{\big| T^{(-1)}_{[0]} \big|}
\Big\{ T^{(0)}_{[R, 2]} \overline{L^{(-1)}_{[0]}} + \text{c.c.} \Big\}
- \frac{3 \tau_\reg}{2 \delta^2} \frac{ \big|T^{(-1)}_{[0]} \big|}{\big| T^{(0)}_{[0]} \big|}
\Big\{ T^{(1)}_{\{R, 0 \}} \overline{L^{(0)}_{[0]}} + \text{c.c.} \Big\} \right)^2 \label{d4} \\
\qquad{} \lesssim \frac{\tau_\reg}{\delta^4} (\deg = 4) . \label{scaleupper}
\end{gather}
This contribution is in general non-zero but, as explained after~\eqref{noneed}, it could vanish
for specific regularizations due to cancellations in the curly brackets in~\eqref{needed}
and~\eqref{noneed}.

For a better understanding of the contributions in~\eqref{d4}, it is instructive to compare them to
the contributions to the Lagrangian away from the light cone as computed in~\cite{reg, vacstab}.
The latter contributions arise independent of the regularization and reflect the fact that, due to the
different masses, the macroscopic behavior of the fermionic projector is necessarily different in the charged and
neutrino sectors. The resulting contribution to the Lagrangian scales like
(see~\cite[Section~2]{vacstab} or~\cite[Section~3]{reg})
\begin{gather}
\L[A_{xy}] \lesssim m^6 (\deg=3) \label{Lm6}
\end{gather}
(the precise scaling depends on the size of the bilinear-dominated region
as discussed in~\cite[Section~6]{reg}).
Clearly, this contribution is much smaller than the upper bound in~\eqref{scaleupper}.
More precisely, since decreasing the degree by one gives two factors of the dimension of length,
which when integrating give rise to a factor~$\varepsilon^2$, the
causal action per four-dimensional volume would become smaller by a scaling factor
\[ \lesssim m^6 \delta^4 \varepsilon^2 . \]
We remark that, as is worked out in detail in \cite[Appendix~A]{jacobson}, there is also a contribution to the
causal Lagrangian if~$x$ is close to~$y$.
This analysis shows that this contribution is also much smaller than~\eqref{d4}.

The fact that~\eqref{d4} is much larger than~\eqref{Lm6} suggests that
when minimizing the causal action, one should try to arrange that the contribution~\eqref{d4}
vanishes. The analysis in this paper gives strong indications that
it is indeed physically sensible to assume that the contribution~\eqref{d4} is zero
(for a detailed discussion of this point see Remark~\ref{remd4}).
More technically, this can be arranged as follows:
First, one should keep in mind that, being a sum of squares, the expression~\eqref{d4} is non-negative.
Therefore, it vanishes in a weak evaluation on the light cone only if it vanishes pointwise.
Note that the contributions in~\eqref{d4} are a consequence of
the shear and general surface states (see the contributions to the eigenvalues
in~\eqref{needed} and~\eqref{noneed}), which are needed in order to satisfy the
EL equations in the vacuum to degree four on the light cone (for details see~\cite[Section~4.4.2]{cfs}).
More precisely, the effect of the shear and general surface states on the absolute values
is to arrange that the EL equations hold in the vacuum by compensating
for the fact that the masses of the neutrinos are smaller than that of the charged leptons and quarks.
Then the contributions in~\eqref{needed} and~\eqref{noneed} scale like~${\sim} m^2$ $(\deg=2)$,
and exactly as explained in~\cite[Section~4.4.2]{cfs}, they can be compensated by the contributions by the rest mass
in the charged sectors.

We remark for clarity that contributions~${\sim} \delta^{-2}$ to the closed chain
are also needed in order to obtain the gravitational interaction with the coupling
constant~$\kappa \sim \delta^2$ (see~\cite[Sections~4.9 and~5.4.3]{cfs}).
However, going through the detailed computations, it turns out that
the resulting contributions to the Lagrangian have a different form than
the terms in~\eqref{d4}. Therefore, the assumption that~\eqref{d4} vanishes is not in conflict
with a gravitational constant~$\kappa \sim \delta^2$.

To the next lower degree on the light cone, we obtain contributions where
\begin{gather}
\Delta \big|\lambda^{xy}_{ncs}\big| \simeq \frac{1}{\delta^4} (\deg=1) . \label{Dl4}
\end{gather}
(where $\simeq$ again means that we omit irrelevant real prefactors).
Since~$\delta$ is to be chosen of the order of the Planck length,
these contributions need to be taken into account. The resulting contribution to the
Lagrangian scales like
\[ \L[A_{xy}] \simeq \frac{1}{\delta^8} (\deg=2) .
\]

\subsection[The contributions ${\sim} \delta^{-4} \cdot (J+j)$]{The contributions $\boldsymbol{{\sim} \delta^{-4} \cdot (J+j)}$} \label{secdJ}
We next consider the contributions where one of the brackets in~\eqref{notation}
contains the contribution~\mbox{${\sim} \delta^{-4}$} given in~\eqref{Dl4}, whereas the other
bracket involves contributions by the Maxwell or Dirac currents.
The perturbation by the Dirac current~$J$ is (see~\cite[equation~(B.2.21)]{cfs})
\[ \Delta \lambda^{xy}_{c+} = \frac{{\rm i}g}{8 \pi} J^j \xi_j \overline{T^{(-1)}_{[0]}}
= \frac{{\rm i}g}{16 \pi} \Sl \psi(y) | \slashed{\xi} \psi(x) \Sr \overline{T^{(-1)}_{[0]}} , \]
where~$g$ denotes the number of generations (thus here we may always set~$g=3$), and
in the last step we used the formula for~$J$ given in~\cite[beginning of Section~3.7.2]{cfs}).
Here~$\xi := y-x$, and~$\Sl \cdot |\cdot \Sr$ again denotes the spin scalar product.
Likewise, the perturbation by the Maxwell current~$j$ is (see~\cite[Proof of Lemma~3.7.3 in Appendix~B]{cfs})
\begin{gather} \label{lamj}
\Delta \lambda^{xy}_{c+} = \frac{{\rm i} g^2}{3} j^i \xi_i T^{(1)}_{[0]} \overline{T^{(-1)}_{[0]}}
-\frac{{\rm i} g^2}{6} j^i \xi_i T^{(0)}_{[0]} \overline{T^{(0)}_{[0]}} .
\end{gather}
where~$j_k = \partial_{kl} A^l - \Box A_k$.
The resulting contributions to the Lagrangian are of the general form
\begin{gather} \label{Ld2Jj}
\L(x,y) \simeq \frac{1}{\delta^4} \big( (J_i + c j_i ) \xi^i \big) (\deg = 3)
\end{gather}
with a real constant~$c$.

The contributions~\eqref{Ld2Jj} to the Lagrangian drop out of the causal action,
as we now explain. We consider the contributions by the Maxwell and Dirac currents
separately. The contribution by the Dirac current vanishes when integrating over~$y$
as a consequence of support properties of the Lagrangian in momentum space.
This will be worked out in detail in Section~\ref{seccurrent} below (see Theorem~\ref{thmIconserve}),
and we do not want to anticipate these arguments here.
Instead, we merely point out that these methods do {\em{not}} apply to the contributions by the Maxwell current
in~\eqref{Ld2Jj}.
But there is another simple reason why the contributions by the Maxwell current vanish in~\eqref{Ld2Jj}:
Namely, the corresponding gauge potential in~\eqref{q:em} is trace-free on the sectors
and vanishes in the neutrino sector. Therefore, it vanishes in~\eqref{Ld2Jj}
when the sums over the eigenvalues of the closed chain are carried out.
Turning this argument around, one can take the fact that the bosonic currents must vanish in the Lagrangian
to the order~${\sim} \delta^{-4} \cdot j$ as the reason why the
bosonic currents in the standard model are all trace-free.
More technically, this argument can be used as an alternative to the
trace condition derived in the analysis of the field tensor terms in the $\iota$-formalism
(see~\cite[Sections~4.2.7, 4.6.2 and~5.3.4]{cfs}).

\subsection[The contributions~${\sim} \delta^{-4} \cdot F$]{The contributions~$\boldsymbol{{\sim} \delta^{-4} \cdot F}$} \label{secdF}
We next consider the contributions where one of the brackets in~\eqref{notation}
contains the contribution~\mbox{${\sim} \delta^{-4}$} given in~\eqref{Dl4}, whereas the other
bracket contains a contribution by the Maxwell field tensor (as computed in~\cite[Section~4.6.2]{cfs}).
Note that these contributions vanish in the formalism of the continuum limit due
to the anti-symmetry of the field tensor (because both tensor indices of
the field tensor are contracted with outer factors~$\xi$).
But the contributions are in general non-zero in the $\iota$-formalism
which gives refined information on the singular behavior on the light cone
(see~\cite[Section~4.2.7]{cfs}). More precisely, the perturbation by the field tensor terms is (see~\cite[Lemma~4.6.6]{cfs})
\begin{gather}
\Delta \lambda^{xy}_{L-} = \frac{{\rm i}g^2}{2} \int_x^y (2 \alpha-1) F^{ij}
\check{\xi}_i \big( \iota^{(-1)}_{[0]} \big)_j T^{(0)}_{[0]} \overline{T^{(0)}_{[0]}} \nonumber\\
\hphantom{\Delta \lambda^{xy}_{L-}}{} \overset{(*)}{=} -\frac{{\rm i}g^2}{4} \int_x^y \big(\alpha^2-\alpha\big) \xi^k \partial_k F^{ij}
\check{\xi}_i \big( \iota^{(-1)}_{[0]} \big)_j T^{(0)}_{[0]} \overline{T^{(0)}_{[0]}} , \label{stareq}
\end{gather}
where we used a short notation for the integral along the line segment joining the points~$x$ and~$y$; i.e.,
for example
\[ \int_x^y (2 \alpha-1) F^{ij} := \int_0^1 (2 \alpha-1) F^{ij}|_{\alpha y + (1-\alpha)x} \,{\rm d}\alpha . \]
In order to verify the equality~$(*)$ in~\eqref{stareq}, one rewrites the directional derivative
as a derivative with respect to~$\alpha$,
\[ \xi^k \partial_k F^{ij}|_{\alpha y + (1-\alpha)x} = \frac{{\rm d}}{{\rm d}\alpha} F^{ij}|_{\alpha y + (1-\alpha)x} \]
and integrates by parts. A direct computation gives
\begin{gather*}
\Delta \big| \lambda^{xy}_{L+} \big|
 =\frac{1}{\big|\lambda^{xy}_{c+}\big|} \re \big( \big( \Delta \lambda^{xy}_{c-} \big) \lambda^{xy}_{c+} \big) \\
\hphantom{\Delta \big| \lambda^{xy}_{L+} \big|}{} =-\frac{{\rm i}g^2}{8} \int_x^y \big(\alpha^2-\alpha\big) \xi^k \partial_k F^{ij}
\check{\xi}_i \big( \iota^{(-1)}_{[0]} \big)_j \frac{\big|T^{(0)}_{[0]} \big|}{\big|T^{(-1)}_{[0]} \big|}
\Big( \overline{T^{(0)}_{[0]}} T^{(-1)}_{[0]} - \text{c.c.} \Big).
\end{gather*}

The resulting contributions to the Lagrangian again vanish
because the Maxwell field is trace-free on the sectors
and vanishes in the neutrinos sector.

\subsection[The contributions ${\sim} F \cdot F$]{The contributions $\boldsymbol{{\sim} F \cdot F}$}
We next consider the contributions where each of the brackets in~\eqref{notation}
contains a contribution by the Maxwell field tensor (as computed in~\cite[Section~4.6.2]{cfs};
see the formulas in Section~\ref{secdF} above).

The resulting contributions to the Lagrangian can be arranged to vanish in two different ways:
One method is to impose conditions on the regularization which imply that the
contributions~\mbox{${\sim} F \cdot F$} vanish in the $\iota$-formalism.
Alternatively, one can take the point of view that all the contributions obtained in the
$\iota$-formalism should be disregarded, because they depend on details of the regularization
which at present are unknown and seem out of reach.
However, discarding the $\iota$-formalism makes it necessary to
rely on the argument described at the end of Section~\ref{secdJ} to obtain the condition
that bosonic potentials must be trace-free.

It is an open question which of the above methods is physically more sensible.
Fortunately, this open question does not have any consequences on the results of the present paper.

\subsection[The contributions ${\sim} (J+j) \cdot (J+j)$]{The contributions $\boldsymbol{{\sim} (J+j) \cdot (J+j)}$} \label{secJJ}
We next consider the contributions where each of the brackets in~\eqref{notation}
contains a contribution by the Maxwell or Dirac current (as computed in~\cite[Section~3.7]{cfs};
see the formulas in Section~\ref{secdJ}).
The resulting contributions to the Lagrangian are of the general form
\begin{gather} \label{Jjcontrib}
\L(x,y) \simeq \big( (J_i + c j_i ) \xi^i \big) \cdot \big( (J_k + c j_k ) \xi^k \big) (\deg = 4) ,
\end{gather}
where again~$J$ is the Dirac current and~$j$ is the Maxwell current.
These contributions and their first variations vanish if and only if
\begin{gather} \label{maxwell1}
J_i + c j_i = 0
\end{gather}
(where, as explained in~\cite[Section~4.4.4]{cfs},
the logarithmic poles again vanish due to the microlocal chiral transformation).
These are the Maxwell equations, where the coupling constant~$c$ is a regularization parameter which
depends on the details of the regularization. We point out that in the continuum limit, the
classical field equations were obtained in a completely different way
(see~\cite[Sections~3.7, 4.8 and~5.4]{cfs}). The main difference is that in the continuum limit, one
analyzes the EL equations of the causal action, whereas here we compute the Lagrangian and
vary the classical potentials. It is remarkable that the results of both procedures give the same
structural results. However, in order to get complete agreement, the coupling constant~$c$ in~\eqref{maxwell1}
must coincide with the coupling constant as computed in~\cite{cfs}. This poses a condition on the
regularization.

We remark that contributions away from the diagonal~$x=y$
can be compensated by nonlocal potentials (i.e., potentials of the form of
an integral operator) as described in~\cite[Section~3.10]{cfs}.

\subsection[The contributions ${\sim} \delta^{-4} \cdot \big(F^2+T\big)$]{The contributions $\boldsymbol{{\sim} \delta^{-4} \cdot \big(F^2+T\big)}$} \label{secdFFT}
We next consider the contributions where one of the brackets in~\eqref{notation}
contains the contribution~${\sim} \delta^{-4}$ given in~\eqref{d4}, whereas the other
bracket involves contributions by the energy-momentum tensor as computed
in~\cite[Section~4.5]{cfs}.
Keeping in mind that there are also corresponding
curvature terms (also computed in~\cite[Section~4.5]{cfs}),
the contribution can be written as
\[ \L(x,y) \simeq \frac{1}{\delta^4} \cdot
\big( \delta^{-2} R_{ij} + c T_{ij}[\psi] + c' T_{ij}[A] \big) \xi^i \xi^j (\deg = 3) , \]
where~$T_{ij}[\psi]$ and~$T_{ij}[A]$ are the energy-momentum tensors of the
Dirac and Maxwell fields, respectively.
This contribution vanishes if the Einstein equations hold.
Exactly as explained in Section~\ref{secJJ} for the Maxwell equations,
here the Einstein equations appear in a quite different way than in the
analysis of the continuum limit. The fact that the coupling constants must coincide
poses constraints on the regularization.

\subsection[The contributions ${\sim} \big(F^2+T\big) \cdot \big(F^2+T\big)$]{The contributions $\boldsymbol{{\sim} \big(F^2+T\big) \cdot \big(F^2+T\big)}$} \label{secFFT}

There are also contributions where both brackets in~\eqref{notation}
involve the energy-momentum tensor and curvature as computed in~\cite[Section~4.5]{cfs},
\[ \L(x,y) \simeq \big( \big( \delta^{-2} R_{ij} + c T_{ij}[\psi] + c' T_{ij}[A] \big) \xi^i \xi^j \big)^2 (\deg = 4) . \]
These contributions as well as their first variation vanish again as a consequence of the
Einstein equation, provided that the regularization satisfies all consistency conditions
for the coupling constants.

\section{Computation of the conserved one-form} \label{seccurrent}
In~\cite[Section~5]{noether} a conserved surface integral was computed which generalized
the Dirac current. It has the form
\begin{gather} \label{Irho}
\gamma(\u) = \int_{-\infty}^{t_0} {\rm d}t \int_{\R^3} {\rm d}^3x \int_{t_0}^\infty {\rm d}t' \int_{\R^3} {\rm d}^3y\,
 ( D_{1,u} - D_{2,u} ) \L(x,y) ,
\end{gather}
where~$\u=(0,u)$ is a jet with vanishing scalar component, whose vector
component is described in terms of a vector~$u \in \Hil$ by
\[ \nabla_{1,\u} P(x,y) = -\nabla_{2,\u} P(x,y) = - {\rm i} |\psi^\u(x)\Sr \Sl \psi^\u(y)| . 
\]
Jets of this form have been introduced more generally in~\cite[Section~2.2]{fockfermionic};
see also~\cite[Section~7.5]{intro}.
Compared to the situation in~\cite{noether} where only one
sector was considered, now we must take into account that, as a consequence of the neutrino
sector, there is an additional contribution to the Lagrangian with the scaling
\begin{gather} \label{osicL}
\L(x,y) \simeq \frac{1}{\delta^4} J_i \xi^i (\deg = 3) ,
\end{gather}
where~$J$ is the current corresponding to the Dirac wave function~$\psi^u$,
\begin{gather} \label{JIdef}
J_i := \Sl \psi^\u(y) | \gamma_i \psi^\u(x) \Sr \pm \text{c.c.}
\end{gather}
(the sign of~$\pm$ depends on whether~\eqref{osicL} involves factors of~$i$;
we do not specify this here).
It was argued in Section~\ref{secdJ} that this contribution vanishes
for~$x \approx y$ as a consequence of the Maxwell equations (see
the explanation after~\eqref{Ld2Jj}). However, there are also contributions
if~$x$ and~$y$ are far apart, which might have an influence on the surface layer integral.
For this reason, we now compute the surface layer integral~\eqref{Irho} for the
contribution~\eqref{osicL}. We shall conclude that the resulting contribution to the surface layer integral vanishes
(see Theorem~\ref{thmIvanish} below). This justifies that the methods and results of~\cite[Section~5]{noether}
remain valid for systems involving neutrinos.

\begin{figure}[t]\centering
\includegraphics{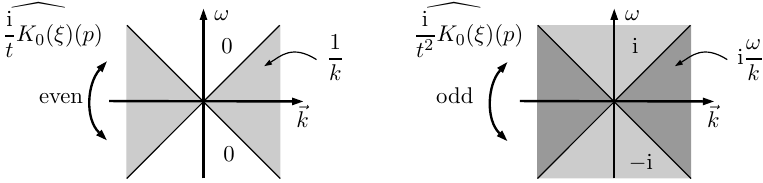}
\caption{The kernels involving~$K_0(\xi)$ in momentum space.}\label{figK12}
\end{figure}

We first give the scaling behavior:
\begin{gather}
( D_{1, \u} - D_{2, \u} ) \L(x,y) \simeq \frac{1}{\delta^4} (\deg=1)
 ( D_{1, \u} - D_{2, \u} ) \big|\lambda^{xy}_{ncs}\big| \simeq J^k \xi_k \frac{1}{\delta^4} (\deg=3)\nonumber\\
\hphantom{( D_{1, \u} - D_{2, \u} ) \L(x,y)}{} \simeq J^k \xi_k \frac{1}{\delta^4} \frac{1}{\varepsilon^2 t^3} \delta(|t|-r)
\simeq J^k \xi_k \frac{1}{\delta^4 \varepsilon^2} \frac{1}{t^2} {\rm i} K_0(\xi) , \label{ddJscaleI}
\end{gather}
where~$K_0$ is the causal fundamental solution of the scalar wave equation defined by
\begin{gather} \label{K0def}
K_0(\xi) := \frac{1}{2 \pi {\rm i}} \big( S^\vee_0- S^\wedge_0 \big)
= \frac{{\rm i}}{4 \pi^2} \epsilon(t) \delta\big(t^2-r^2\big) = \frac{{\rm i}}{4 \pi^2} \frac{1}{2 t} \delta(|t|-r)
\end{gather}
(and~$S^\vee_0$ and~$S^\wedge_0$ denote the advanced and retarded Green's operators,
respectively).
In momentum space, the distribution~$\hat{K}_0$ is supported on the mass shell.
More precisely, setting~$p=(\omega, \vec{k})$ and~$k=|\vec{p} |$, we have
\[ \hat{K}_0(p) \simeq \frac{1}{\omega} ( \delta(\omega-k) - \delta(\omega+k) ) . \]
The computation for the one-dimensional Fourier transform of a function~$f$
\begin{gather}
\widehat{\left( \frac{f(t)}{t} \right)}(\omega) = \int_{-\infty}^\infty \frac{f(t)}{t} {\rm e}^{{\rm i} \omega t} \,{\rm d}t
= \int_{-\infty}^\infty f(t) \left(\frac{1}{t} + {\rm i} \int_0^\omega {\rm e}^{{\rm i} \kappa t} \,{\rm d}\kappa \right) dt \nonumber\\
\hphantom{\widehat{\left( \frac{f(t)}{t} \right)}(\omega)}{} = \int_{-\infty}^\infty \frac{f(t)}{t} \,{\rm d}t + {\rm i} \int^\omega \hat{f}(\kappa) {\rm d}\kappa ,\label{tomega}
\end{gather}
shows that each factor~$1/t$ corresponds in momentum space to ${\rm i}$ times an integration over~$\omega$.
The integration constant can be determined by using the symmetries.
The resulting kernels are shown in Fig.~\ref{figK12}.

We now proceed in two steps: We first show that the resulting contribution
to the surface layer integral~\eqref{Irho} is conserved in time (i.e., that it is
independent of~$t_0$). The second step will be to show that it is even zero.
For the first step, we differentiate~\eqref{Irho} with respect to~$t_0$.
Using that the integrand is antisymmetric in the arguments~$x$ and~$y$, we obtain
\begin{gather} \label{pointwiseI}
\frac{{\rm d}I}{{\rm d}t_0} = \int_{\R^3} {\rm d}^3x \int_\scrM ( D_{1, \u} - D_{2, \u} ) \L (t_0, \vec{x}; y ) \, {\rm d}^4y .
\end{gather}
Therefore, the surface layer integral~\eqref{Irho} is conserved if and only if this
expression vanishes. Clearly, it is a sufficient condition to show that
\begin{gather} \label{Inecessary}
\int_\scrM ( D_{1, \u} - D_{2, \u} ) \L(x,y) \, {\rm d}^4y = 0 \qquad \text{for all~$x \in \scrM$} .
\end{gather}
This is the scalar component of the linearized field equations. Indeed, this equation was
used in~\cite{noether, osi} to derive the conservation law (for details see~\cite[Proof of Theorem~3.1]{osi}).
\begin{Thm} \label{thmIconserve}
The contribution~\eqref{ddJscaleI} with~$J$ according to~\eqref{JIdef} vanishes in~\eqref{Inecessary}.
\end{Thm}
\begin{proof} Using~\eqref{ddJscaleI} in~\eqref{Inecessary}, we obtain terms of the form
\[ \int_\scrM J^k \xi_k \frac{1}{\delta^4 \varepsilon^2} \frac{1}{t^2} {\rm i} K_0(\xi) \,{\rm d}^4y . \]
According to~\eqref{JIdef}, the current gives a factor~$\psi(y)$ or~$\overline{\psi}(y)$.
Therefore, applying Plancherel gives terms of the form
\begin{gather} \label{multizero}
\int \hat{\mathfrak{K}}(p) \hat{g}(k) \, {\rm d}^4p ,
\end{gather}
where~$\hat{g}$ is supported on the upper or lower mass shell.
Noting that the factor~$\xi_k$ in~\eqref{ddJscaleI} corresponds to a partial derivative in momentum space,
the kernel~${\mathfrak{K}}(p)$ is obtained from the distribution shown on the right of
Fig.~\ref{figK12} by differentiation. Hence it vanishes inside both the upper and lower mass cone.
We conclude that the two factors in the integrand in~\eqref{multizero} have disjoint supports.
Therefore, the integral~\eqref{multizero} vanishes, giving the result.
\end{proof}

Knowing that the surface layer integral is conserved, we can simplify its form
with the help of the following lemma, which is inspired by a similar computation
in~\cite[proof of Lemma~5.5]{noether}.

\begin{Lemma} \label{lemmaIconserve}
Using the conservation of the surface layer integral as proved in Theorems~{\rm \ref{thmIconserve}}, the surface layer integral can be written as
\begin{gather}
 \int_{-\infty}^{t_0} {\rm d}t \int_{\R^3} {\rm d}^3x \int_{t_0}^\infty {\rm d} t' \int_{\R^3} {\rm d}^3y
 ( D_{1, \u} - D_{2, \u} )\L(t,\vec{x}; t', \vec{y}) \notag \\
\qquad{} = \frac{1}{2} \lim_{T \rightarrow \infty} \frac{1}{T} \int_0^T {\rm d}t \int_{\R^3} {\rm d}^3x \int_\scrM {\rm d}^4y \big(y^0 - t\big) ( D_{1, \u} - D_{2, \u} ) \L(t,\vec{x}; y) . \label{ositI}
\end{gather}
\end{Lemma}
\begin{proof} In view of~\eqref{pointwiseI} and Theorem~\ref{thmIconserve},
we know that the above surface layer integral is time independent.
As a consequence, denoting the spatial integrals by
\[ A(t,t') := \int_{\R^3} {\rm d}^3x \int_{\R^3} {\rm d}^3y\,
 ( D_{1, \u} - D_{2, \u} ) \L(t,\vec{x}; t', \vec{y}) , \]
the surface layer integral can be written as
\begin{align*}
\int_{-\infty}^{t_0} {\rm d}t \int_{t_0}^\infty {\rm d}t'\, A(t,t') &= \frac{1}{T} \int_0^T {\rm d}\tau
\int_{-\infty}^\tau {\rm d}t \int_\tau^\infty {\rm d}t' \, A(t,t') \\
&= \frac{1}{T} \int_0^T {\rm d}\tau
\int_{-\infty}^\infty {\rm d}t \int_{-\infty}^\infty {\rm d}t'\, \Theta(\tau-t) \Theta(t'-\tau) A(t,t') \\
&= \frac{1}{T}
\int_{-\infty}^\infty {\rm d}t \int_{-\infty}^\infty {\rm d}t' \int_0^T {\rm d}\tau\,
 \Theta(\tau-t) \Theta(t'-\tau) A(t,t') .
\end{align*}
Now the $\tau$-integration can be carried out.
Going through the different cases where~$t$,~$t'$ is positive or negative
and~$t$,~$t'$ is larger or smaller than~$T$, a straightforward computation yields
\begin{gather}
 \int_{-\infty}^{t_0} {\rm d}t \int_{t_0}^\infty {\rm d}t'\, A(t,t')
= \frac{1}{T} \int_{-\infty}^0 {\rm d}t \int_0^T {\rm d}t'\, t' A(t,t')
 \frac{1}{T} \int_{-\infty}^0 {\rm d}t \int_T^\infty dt' T A(t,t') \notag \\
\hphantom{\int_{-\infty}^{t_0} {\rm d}t \int_{t_0}^\infty {\rm d}t'\, A(t,t')=}{} + \frac{1}{T} \int_{0}^T {\rm d}t \int_t^T {\rm d}t' (t'-t) A(t,t')
+ \frac{1}{T} \int_{0}^T {\rm d}t \int_T^\infty {\rm d}t' (T-t) A(t,t') \notag \\
\hphantom{\int_{-\infty}^{t_0} {\rm d}t \int_{t_0}^\infty {\rm d}t'\, A(t,t')}{} = \frac{1}{T} \int_{-\infty}^0 {\rm d}t \int_0^T {\rm d}t' \, t' A(t,t')
+ \frac{1}{T} \int_{-\infty}^0 {\rm d}t \int_T^\infty {\rm d}t' T A(t,t') \label{osi1} \\
\hphantom{\int_{-\infty}^{t_0} {\rm d}t \int_{t_0}^\infty {\rm d}t'\, A(t,t')=}{} + \frac{1}{T} \int_{0}^T {\rm d}t \int_t^\infty {\rm d}t'\, (t'-t) A(t,t') \label{vol} \\
\hphantom{\int_{-\infty}^{t_0} {\rm d}t \int_{t_0}^\infty {\rm d}t'\, A(t,t')=}{} + \frac{1}{T} \int_{0}^T {\rm d}t \int_T^\infty {\rm d}t' \, ((T-t)-(t'-t) ) A(t,t') . \label{osi2}
\end{gather}
The integrals in~\eqref{osi1} and~\eqref{osi2} are surface layer integrals.
Since~$A(t,t')$ has suitable decay properties in~$|t-t'|$,
these integrals are bounded uniformly in~$T$
(more precisely, these integrals exist in the Lebesgue sense provided that
the electromagnetic potentials decay~${\sim} 1/|t|$).
Therefore, in the limit~$T \rightarrow \infty$ only the summand~\eqref{vol} remains,
\begin{gather} \label{Apos}
\int_{-\infty}^{t_0} {\rm d}t \int_{t_0}^\infty {\rm d}t'\, A(t,t')
= \lim_{T \rightarrow \infty} \frac{1}{T} \int_{0}^T {\rm d}t\, \int_t^\infty {\rm d}t' \, (t'-t) A(t,t') .
\end{gather}

Using that~$A(t,t')$ is anti-symmetric in its arguments, we can use the same argument with the time direction reversed to obtain
\begin{align}
\int_{-\infty}^{t_0} {\rm d}t \int_{t_0}^\infty {\rm d}t'\, A(t,t')&
= -\int_{t_0}^\infty {\rm d}t \int_{-\infty}^{t_0} {\rm d}t'\, A(t,t') \notag \\
&= -\lim_{T \rightarrow \infty} \frac{1}{T} \int_{0}^T {\rm d}t \int_{-\infty}^t {\rm d}t'\, (t-t') A(t,t') . \label{Aneg}
\end{align}
Taking the arithmetic mean of~\eqref{Apos} and~\eqref{Aneg} gives the result.
\end{proof}

We finally show that the surface layer integral vanishes:
\begin{Thm} \label{thmIvanish}
The contribution~\eqref{ddJscaleI} with~$J$ according to~\eqref{JIdef} vanishes in the surface layer integral~\eqref{Irho}.
\end{Thm}
\begin{proof} The integrand in~\eqref{ositI} differs from that in~\eqref{Inecessary}
by an additional factor~$\big(y^0 - t\big)$. In momentum space, this factor corresponds to an
additional $\omega$-derivative. As a consequence, the resulting kernel in momentum
space again vanishes inside the upper and lower mass cone.
Therefore, the method of proof of Theorem~\ref{thmIconserve} again applies, giving the result.
\end{proof}

\section{Computation of bosonic conserved surface layer integrals} \label{secosib}
We now proceed with the analysis of contributions to the Lagrangian
of degree three on the light cone.
As we shall see, these contributions are in general non-zero.
Their significance is that they give rise to physically sensible expressions
for conserved surface layer integrals. More precisely, we shall compute the
surface layer integral~\eqref{osicombined} which is composed
of both the {\em{symplectic form}}~\eqref{siggen} and the
{\em{surface layer inner product}}~\eqref{osispgen}.
For clarity, we first consider the bosonic contributions; the fermionic contributions
will be computed in Section~\ref{secosif} below.

\subsection{Computation of unbounded line integrals} \label{secunbound}
In the computation of surface layer integrals, there is the complication that the
two arguments of the Lagrangian are varied differently. In particular, our task is to
compute the variation
\[ (\nabla_{1,\v} - \nabla_{2,\v} ) P(x,y) \]
for bosonic jets. Writing out the jets as variations of the wave functions, we obtain
the expression
\begin{gather} \label{nonlocal}
 ( -s \slashed{A} P + P \slashed{A} s )(x,y)
\end{gather}
(where~$s$ is the symmetric Dirac Green's operator~\eqref{sdef}).
The light-cone expansion of this bi-distribution involves unbounded line integrals,
as is made precise in the following lemma.

\begin{Lemma} \label{lemmaunbounded}
The light-cone expansions of the distributions
\[ (-s \slashed{A} P - P \slashed{A} s )(x,y) \qquad \text{and} \qquad
 (-s \slashed{A} P )(x,y) \]
are obtained from each other by replacing the line integrals according to
\[ \int_0^1 (\cdots)|_{z=\alpha y + (1-\alpha)x} \,{\rm d}\alpha
 \longrightarrow \frac{1}{2} \int_{-\infty}^\infty \epsilon(\alpha) (\cdots)|_{z=\alpha y + (1-\alpha)x}\, {\rm d}\alpha . \]
\end{Lemma}
\begin{proof} Light-cone expansions involving unbounded line integrals
were carried out in explicit detail in the unpublished preprint~\cite{lightint} using so-called light-cone integrals.
A more systematic and more compact method was developed in~\cite[Appendix~F]{pfp},
where a light-cone expansion involving unbounded line integrals was derived
for an operator product involving the causal fundamental solution~$K_a$
defined as a multiple of the difference of the advanced and retarded Klein--Gordon Green's distribution,
\begin{gather} \label{KSdecomp}
K_a = \frac{1}{2 \pi {\rm i}} \big( S^\vee_a - S^\wedge_a \big) ,
\end{gather}
where~$a$ is the mass squared (thus~$K$ satisfies the equation~$(\Box_x + a) K_a(x,y) = 0$).
Differentiating with respect to the parameter~$a$ gives the distributions
(for details on this method see~\cite{light})
\[ K^{(\ell)} := \frac{{\rm d}^\ell}{{\rm d}a^\ell} K_a \Big|_{a=0} . \]
In~\cite[Lemma~F.3]{pfp} it was shown that for any~$l, r \geq 0$ and any scalar potential~$V$
(here and in what follows, we assume for simplicity that the potential
is a Schwartz function or a smooth function with compact support),
\begin{gather}
\big(K^{(l)} V K^{(r)}\big)(x,y) := \int K^{(l)}(x,z) V(z) K^{(r)}(z,y) \, {\rm d}^4z \label{KTformel} \\
\qquad{} = -\frac{1}{2 \pi^2} \sum_{n=0}^\infty
\frac{1}{n!} \int_{-\infty}^\infty {\rm d}\alpha\, \alpha^l (1-\alpha)^r
 \big(\alpha - \alpha^2\big)^n (\Box^n
V)_{| \alpha y + (1-\alpha) x} S^{(l+r+n+1)}(x,y) \nonumber \\
\qquad\quad{} + ({\mbox{contributions smooth for $x \neq y$}}) \qquad \label{eq:bbv}
\end{gather}
(where we rearranged the smooth contribution and the bounded line integrals
in~\cite[Lemma~F.3]{pfp} in a way most convenient for our purposes).
Inserting the decomposition~\eqref{KSdecomp} into the first factor in~\eqref{KTformel}, multiplying out and
using the support properties of the Green's functions, we obtain corresponding
light-cone expansions for the operator product~$S^{\vee,(l)} V K^{(r)}$,
and~$S^{\wedge,(l)} V K^{(r)}$,
\begin{gather*}
\big(S^{\vee,(l)} V K^{(r)}\big)(x,y) = ({\mbox{contributions smooth for $x \neq y$}}) \\
 \hphantom{\big(S^{\vee,(l)} V K^{(r)}\big)(x,y) =}{} +\sum_{n=0}^\infty \frac{1}{n!} \int_{-\infty}^\infty
\big( \Theta\big(y^0-x^0\big) \Theta(\alpha) - \Theta\big({-}\big(y^0-x^0\big) \big) \Theta(-\alpha) \big) \\
\qquad\quad\hphantom{\big(S^{\vee,(l)} V K^{(r)}\big)(x,y) =}{} \times \alpha^l (1-\alpha)^r \big(\alpha - \alpha^2\big)^n
 (\Box^n V)_{| \alpha y + (1-\alpha) x} \, {\rm d}\alpha\, K^{(l+r+n+1)}(x,y), \\
\big(S^{\wedge,(l)} V K^{(r)}\big)(x,y) = ({\mbox{contributions smooth for $x \neq y$}}) \\
\hphantom{\big(S^{\wedge,(l)} V K^{(r)}\big)(x,y) =}{} +\sum_{n=0}^\infty \frac{1}{n!} \int_{-\infty}^\infty
\big( {-}\Theta\big(y^0-x^0\big) \Theta(-\alpha) + \Theta\big({-}\big(y^0-x^0\big) \big) \Theta(\alpha) \big) \\
\quad\qquad\hphantom{\big(S^{\wedge,(l)} V K^{(r)}\big)(x,y) =}{} \times \alpha^l (1-\alpha)^r \big(\alpha - \alpha^2\big)^n
 (\Box^n V)_{| \alpha y + (1-\alpha) x} \,{\rm d}\alpha\, K^{(l+r+n+1)}(x,y) .
\end{gather*}
Taking the difference of these formulas and using~\eqref{KSdecomp}, we recover~\eqref{eq:bbv}.
However, taking the mean of these formulas, we conclude that
\begin{gather}
\big(S^{(l)} V K^{(r)}\big)(x,y)\nonumber\\
\qquad{} = \frac{1}{2} \sum_{n=0}^\infty
\frac{1}{n!} \int_{-\infty}^\infty {\rm d}\alpha\, \epsilon(\alpha) \alpha^l (1-\alpha)^r
 \big(\alpha - \alpha^2\big)^n (\Box^n
V)_{| \alpha y + (1-\alpha) x} K^{(l+r+n+1)}(x,y) \notag \\
 \qquad\quad {} + ({\mbox{contributions smooth for $x \neq y$}}) , \qquad \label{eq:bbv2}
\end{gather}
where~$S:=\big(S^\vee + S^\wedge\big)/2$.

It remains to replace the factors~$K^{(p)}$ in~\eqref{eq:bbv2} by corresponding factors~$T^{(p)}$.
To this end, one can proceed exactly as in~\cite[Proof of Lemma~F.4]{pfp} and multiply from the right
by the projection operator on the negative frequencies. This gives the result.
\end{proof}

Applying this lemma to both summands in~\eqref{nonlocal}, one obtains the line integrals
\[ \int_{-\infty}^\infty \epsilon(\alpha) (\cdots)|_{z=\alpha y + (1-\alpha)x} \,{\rm d}\alpha
- \int_{-\infty}^\infty \epsilon(\beta) (\cdots)|_{z=\beta x + (1-\beta)y} \,{\rm d}\alpha . \]
Changing variables in the second integral according to~$\beta \rightarrow \alpha=1-\beta$,
the integrals can be combined to
\[ 2 \left( \int_{-\infty}^0 - \int_1^\infty \right) (\cdots)|_{z=\alpha y + (1-\alpha)x} \, {\rm d}\alpha . \]

\subsection[The contributions ${\sim} \delta^{-4} \cdot F^2 \varepsilon^2/t^2$]{The contributions $\boldsymbol{{\sim} \delta^{-4} \cdot F^2 \varepsilon^2/t^2}$} \label{secdF22}

Before entering the detailed computations, it might be instructive to
consider the scalings, starting from the contributions to the fermionic projector
as given in~\cite[equations~(B.5.1) and~(B.5.2)]{cfs}:
\begin{gather*}
\Delta P(x,y) \sim F^2 \slashed{\xi} \xi \xi T^{(0)}, \qquad
\Delta \lambda^{xy}_{ncs} \sim F^2 \xi \xi (\deg=2)
\end{gather*}
(here the tensor indices of the factors~$\xi$ are contracted to
the field tensor; in other words, they are outer factors; see the explanation after~\eqref{eq54}
in Section~\ref{seccl}).
These contributions were already taken into account in Sections~\ref{secdFFT} and~\ref{secFFT},
where they were compensated by corresponding curvature terms.
With this in mind, it remains to consider the contributions obtained by a regularization expansion.
More precisely, the terms with the correct scaling behavior are those of second order
in~$\varepsilon/t$, i.e.,
\begin{gather*}
\L(x,y) \sim F^2 \xi \xi (\deg=5) \frac{\varepsilon^2 t^2}{\delta^4} \frac{\varepsilon^2}{t^2}
 \sim F^2 \xi \xi \frac{\varepsilon^2 t^2}{\delta^4} \frac{\varepsilon^2}{t^2} \frac{1}{\varepsilon^4 t^5} \delta(|t|-r) \\
\hphantom{\L(x,y)}{} \sim F^2 \frac{1}{\delta^4} \frac{1}{t^3} \delta(|t|-r)
 \sim F^2 \frac{1}{\delta^4} \frac{1}{t^2} {\rm i} K_0(\xi) ,
\end{gather*}
where~$K_0$ is again the causal fundamental solution of the wave equation~\eqref{K0def}
(the terms first order in~$\varepsilon/t$ will be analyzed in Section~\ref{secdF21}).
One should keep in mind that in the regularization expansion, the outer factors~$\xi$
(i.e., those factors contracted with the field tensor; see again the explanation after~\eqref{eq54}
in Section~\ref{seccl}) need not be taken into account,
because they appear in the same way in the Ricci tensor, and thus they drop
out exactly as explained in Sections~\ref{secdFFT} and~\ref{secFFT} above.
With this in mind, it suffices to compute the contraction of the fermionic projector
with an outer factor~$\xi$.
We now compute these contributions and study their effect
on the Lagrangian. For clarity, we treat the contributions which involve and do not
involve logarithmic poles after each other.

We point out that there are also contributions which are linear in~$\varepsilon/t$.
For clarity of presentation, we will analyze these contributions in Section~\ref{secdF21} below.

\subsubsection{Contributions with logarithmic poles}
We begin by computing the contributions to the fermionic projector with a logarithmic pole:
\begin{figure}[t]\centering
\includegraphics{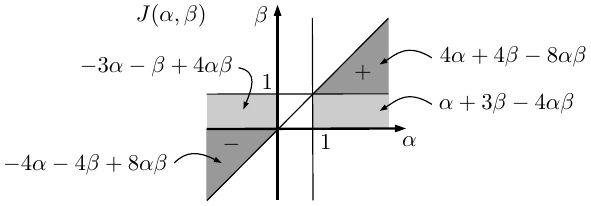}
\caption{The function~$J(\alpha, \beta)$.}\label{figJ}
\end{figure}

\begin{Lemma} \label{lemmalog}
The contributions to the fermionic projector which involve the electromagnetic field strength
and have a logarithmic pole on the light cone have the form
\begin{gather*}
( \nabla_{1, \u} - \nabla_{2, \u} )
( \nabla_{1, \v} + \nabla_{2, \v} ) \frac{1}{4} \Tr ( \slashed{\xi} P(x,y) ) \\
\qquad{} = -2{\rm i} \int_{-\infty}^\infty {\rm d}\alpha \int_{-\infty}^\infty {\rm d}\beta\, J(\alpha, \beta)
(F^\u)_{ij}|_{\alpha y + (1-\alpha) x} (F^\v)^j_{ k}|_{\beta y + (1-\beta) x} \xi^i \xi^k
T^{(1)}_{[0]} , 
\end{gather*}
where $($see Fig.~{\rm \ref{figJ})}
\begin{gather}
J(\alpha, \beta) = (\alpha +3 \beta -4 \alpha \beta) \chi_{(1,\infty)}(\alpha) \chi_{(0,1)}(\beta)\nonumber\\
\hphantom{J(\alpha, \beta) =}{}
 + (-3 \alpha -\beta +4 \alpha \beta) \chi_{(-\infty,0)}(\alpha) \chi_{(0,1)}(\beta)\nonumber \\
\hphantom{J(\alpha, \beta) =}{} + (4 \alpha +4 \beta -8 \alpha \beta) \chi_{(1,\infty)}(\alpha) \chi_{(1,\alpha)}(\beta)\nonumber \\
\hphantom{J(\alpha, \beta) =}{}
 + (-4 \alpha -4 \beta +8 \alpha \beta) \chi_{(-\infty,0)}(\alpha) \chi_{(\alpha,0)}(\beta) .\label{Jform}
\end{gather}
\end{Lemma}
\begin{proof} Our starting point is the contribution to the fermionic projector
of second order in the electromagnetic field strength as given in~\cite[equation~(B.5.2)]{cfs},
\begin{gather*}
\frac{1}{4} \Tr ( \slashed{\xi} P(x,y) ) \asymp
 -2 {\rm i} \int_x^y [0,2 | 0]\, {\rm d}z \int_{z}^y [1,0 | 0] \,{\rm d}\tilde{z}\, F_{ki}(z) F^{kj}(\tilde{z}) \xi^i \xi_j T^{(1)}_{[0]} \\
\hphantom{\frac{1}{4} \Tr ( \slashed{\xi} P(x,y) )\asymp}{} -8{\rm i} \int_x^y [0,1 | 1] \, {\rm d}z \int_{z}^y [0,1 | 0] \, {\rm d} \tilde{z} \, F_{ki}(z) F^{kj}(\tilde{z}) \xi^i \xi_j
 T^{(1)}_{[0]} .
\end{gather*}
Here we use the notation for the nested line integrals (for details see~\cite[Section~2.2]{cfs})
\begin{gather*}
\int_x^y [p,q | r] \,{\rm d}z \int_{z}^y [\tilde{p},\tilde{q} | \tilde{r}] \,{\rm d}\tilde{z} F_{ki}(z) F^{kj}(\tilde{z}) \\
\qquad{} := \int_0^1 {\rm d}\tau \,\tau^p (1-\tau)^q \big(\tau-\tau^2\big)^r
\int_0^1 {\rm d}\tilde{\tau} \tilde{\tau}^{\tilde{p}} (1-\tilde{\tau})^{\tilde{q}} \big(\tilde{\tau}-\tilde{\tau}^2\big)^{\tilde{r}} \\
 \qquad \quad {}\times F_{ki}(z) F^{kj}(\tilde{z}) \Big|_{z=\tau y + (1-\tau)x, \tilde{z} = \tilde{\tau} y + (1-\tilde{\tau}) z} .
\end{gather*}
Transforming to the new integration variables
\begin{gather} \label{alphabetadef}
\alpha = \tau \qquad \text{and} \qquad
\beta = \tau+ (1-\tau) \tilde{\tau} ,
\end{gather}
the above contribution to~$P(x,y)$ can be written as
\[ \frac{1}{4} \Tr ( \slashed{\xi} P(x,y) ) \asymp
-2{\rm i} \int_0^1 {\rm d}\alpha \int_\alpha^1 {\rm d}\beta \, ( 3 \alpha +\beta -4 \alpha \beta )
F_{ki}(z) F^{kj}(\tilde{z}) \xi^i \xi_j T^{(1)}_{[0]} , \]
to be evaluated at
\begin{gather} \label{zeval}
z = \alpha y + (1-\alpha)x \qquad \text{and} \qquad \tilde{z} = \beta y + (1-\beta)x .
\end{gather}

The contribution symmetric in~$A_\u$ and~$A_\v$ is obtained immediately by
applying the polarization identity (i.e., the identity~$B(u,v) = (B(u+v, u+v) - B(u-v, u-v))/4$ for a bilinear form~$B$),
\begin{gather*}
 ( \nabla_{1, \u} + \nabla_{2, \u} )
 ( \nabla_{1, \v} + \nabla_{2, \v} ) \frac{1}{4} \Tr ( \slashed{\xi} P(x,y) )
 = -2{\rm i} \int_0^1 {\rm d}\alpha \int_\alpha^1 {\rm d}\beta \, ( 3 \alpha +\beta -4 \alpha \beta ) \xi^i \xi_j T^{(1)}_{[0]} \\
 \hphantom{( \nabla_{1, \u} + \nabla_{2, \u} )
 ( \nabla_{1, \v} + \nabla_{2, \v} ) \frac{1}{4} \Tr ( \slashed{\xi} P(x,y) ) =} {}\times \big( (F_\u)_{ki}(z) (F_\v)^{kj} (\tilde{z})
+ (F_\v)_{ki}(z) (F_\u)^{kj} (\tilde{z}) \big) ,
\end{gather*}
again to be evaluated at~\eqref{zeval}.
Changing the sign before~$\nabla_{2,\u}$ amounts to replacing the integrals by unbounded
integrals, similar as derived in Lemma~\ref{lemmaunbounded} in first order perturbation theory.
In order to apply this result to second order perturbation theory, we arrange the
contributions as follows,
\begin{gather*}
 ( \nabla_{1, \u} - \nabla_{2, \u} )
 ( \nabla_{1, \v} + \nabla_{2, \v} ) P(x,y) \\
 \qquad{} = s \slashed{A}_\u s \slashed{A}_\v P + s \slashed{A}_\v s \slashed{A}_\u P
+ s \slashed{A}_\u P \slashed{A}_\v s - s \slashed{A}_\v P \slashed{A}_\u s
- P \slashed{A}_\u s \slashed{A}_\v s - P \slashed{A}_\v s \slashed{A}_\u s \\
\qquad{} = ( s \slashed{A}_\v s ) \slashed{A}_\u P
+ s \slashed{A}_\u \big( P \slashed{A}_\v s + s \slashed{A}_\v P \big)
- ( s \slashed{A}_\v + P \slashed{A}_\v s ) \slashed{A}_\u s - P \slashed{A}_\u ( s \slashed{A}_\v s )
\end{gather*}
(we remark that, being smooth, we can omit the so-called high-energy contributions involving three factors~$k$ or~$p$;
for details and our notation see~\cite[Section~2.2]{cfs}).
Now the terms inside the brackets involve convex line integrals, whereas the
operator products outside the brackets can be handled with the help of Lemma~\ref{lemmaunbounded}.
Using that
\[ \int_0^1 {\rm d}\alpha \int_\alpha^1 {\rm d}\beta \cdots =
\int_0^1 {\rm d}\beta \int_0^\beta {\rm d}\alpha \cdots , \]
we bring the integral corresponding to the product outside the brackets to the left
and apply the replacement rules
\begin{alignat*}{3}
 &( s \slashed{A}_\v s ) \slashed{A}_\u P \colon \quad &&
\int_0^1 {\rm d}\beta \int_0^\beta {\rm d}\alpha \longrightarrow \frac{1}{2}
\int_{-\infty}^\infty \epsilon(\beta) \, {\rm d}\beta \int_{(0,\beta) \cup (\beta,0)} {\rm d}\alpha,& \\
&s \slashed{A}_\u ( P \slashed{A}_\v s + s \slashed{A}_\v P ) \colon \quad &&
\int_0^1 {\rm d}\alpha \int_\alpha^1 {\rm d}\beta
 \longrightarrow \frac{1}{2} \int_{-\infty}^\infty \epsilon(\alpha)\, {\rm d}\alpha \int_{(\alpha,1) \cup (1,\alpha)} {\rm d}\beta,& \\
& ( s \slashed{A}_\v + P \slashed{A}_\v s ) \slashed{A}_\u s \colon \quad &&
\int_0^1 {\rm d}\beta \int_0^\beta {\rm d}\alpha \longrightarrow \frac{1}{2}
\int_{-\infty}^\infty \epsilon(1-\beta) \, {\rm d}\beta \int_{(0,\beta) \cup (\beta,0)} {\rm d}\alpha, & \\
& P \slashed{A}_\u ( s \slashed{A}_\v s ) \colon \quad &&
\int_0^1 {\rm d}\alpha \int_\alpha^1 {\rm d}\beta \longrightarrow \frac{1}{2}
\int_{-\infty}^\infty \epsilon(1-\alpha) \, {\rm d}\alpha \int_{(\alpha,1) \cup (1,\alpha)} {\rm d}\beta .&
\end{alignat*}
Moreover, by suitably renaming the integration variables~$\alpha$
and~$\beta$ we arrange that~$A_\u$ is always evaluated at~$\alpha y + (1-\alpha)x$,
whereas~$A_\v$ is always evaluated at~$\beta y + (1-\beta)x$.
A straightforward computation gives the result.
\end{proof}

We now compute the resulting effect on the Lagrangian:
\begin{Prp} \label{prplog}
The contribution to the fermionic projector of Lemma~{\rm \ref{lemmalog}}
affects the second variation of the Lagrangian
to the order~${\sim} \delta^{-4} \cdot F^2 \varepsilon^2/t^2$ by a term of the form
\begin{gather*}
 ( \nabla_{1, \u} - \nabla_{2, \u} )
 ( \nabla_{1, \v} + \nabla_{2, \v} ) \L(x,y) \\
 \qquad{} \asymp \int_{-\infty}^\infty {\rm d}\alpha \int_{-\infty}^\infty {\rm d}\beta \, J(\alpha, \beta)
 (F^\u)_{ij}|_{\alpha y + (1-\alpha) x} (F^\v)^j_{ k}|_{\beta y + (1-\beta) x} \xi^i \xi^k
 \frac{c}{\delta^4} \frac{1}{t^4} \delta \big( \xi^2 \big) , ,
\end{gather*}
where~$c$ is a real constant.
\end{Prp}
\begin{proof} Computing the perturbation of the eigenvalues of the closed chain
as in~\cite[Appendix~B.5]{cfs}, to the considered order on the light cone we obtain
for the perturbation of the Lagrangian an expression of the form
\begin{gather}
 ( \nabla_{1, \u} - \nabla_{2, \u} )
 ( \nabla_{1, \v} + \nabla_{2, \v} ) \L(x,y) \notag \\
\qquad{} \asymp \int_{-\infty}^\infty {\rm d}\alpha \int_{-\infty}^\infty {\rm d}\beta \, J(\alpha, \beta)
(F^\u)_{ij}|_{\alpha y + (1-\alpha) x} (F^\v)^j_{ k}|_{\beta y + (1-\beta) x} \xi^i \xi^k \nonumber\\
 \qquad \quad{} \times \frac{1}{\delta^4} \frac{1}{t^4} {\rm i} K_0(\xi)
\re \big( C T^{(1)}_{[0]} \big) \label{extraterm}
\end{gather}
with a complex constant~$C$.
Here the factor~$i$ can be understood from the fact that the left side is real-valued.

Let us analyze the symmetry when exchanging~$x$ and~$y$.
Obviously, the left side is anti-symmetric. In the line integrals, exchanging~$x$ and~$y$
corresponds to the replacements
\[ \alpha \rightarrow 1-\alpha \qquad \text{and} \qquad \beta \rightarrow 1-\beta . \]
From~\eqref{Jform} one sees that the line integrals are anti-symmetric.
This shows that the term~\eqref{extraterm} must be symmetric when exchanging~$x$ and~$y$.
Since~$K_0$ is anti-symmetric, we conclude that from the factor
\begin{gather} \label{T1form}
T^{(1)}_{[0]} = \frac{1}{32 \pi^3} \big( \log \big|\xi^2 \big| + c + {\rm i} \pi \Theta\big(\xi^2\big) \epsilon(\xi^0) \big)
\end{gather}
(with a real constant~$c$; for details for how this formula comes about see~\cite[equation~(2.2.3)]{cfs}) only the last summand contributes.
This gives the result.
\end{proof}

\subsubsection{Contributions without logarithmic poles}
\begin{figure}[t]\centering
\includegraphics{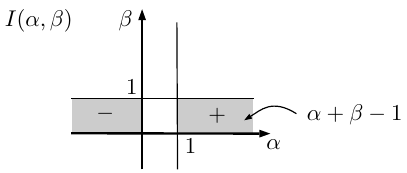}
\caption{Unbounded line integrals for the $FF$-terms without logarithmic pole.}\label{figI}
\end{figure}

\begin{Prp} \label{prpnolog}
The contributions to the fermionic projector involving no logarithmic poles on the light cone
$($i.e., all contributions except for those in Lemma~{\rm \ref{lemmalog})}
affect the second variation of the Lagrangian
to the order~${\sim} \delta^{-4} \cdot F^2 \varepsilon^2/t^2$ by a term of the form
\begin{gather*}
 ( \nabla_{1, \u} - \nabla_{2, \u} )
 ( \nabla_{1, \v} + \nabla_{2, \v} ) \L(x,y) \\
\qquad{} \asymp \int_{-\infty}^\infty {\rm d}\alpha \int_{-\infty}^\infty {\rm d}\beta \, I(\alpha, \beta)
(F^\u)_{ij}|_{\alpha y + (1-\alpha) x} (F^\v)^j_{ k}|_{\beta y + (1-\beta) x} \xi^i \xi^k
 \frac{c}{\delta^4} \frac{1}{t^4} {\rm i} K_0(\xi) ,
\end{gather*}
where~$c$ is a real constant and $($see Fig.~{\rm \ref{figI})}
\begin{gather} \label{Idef}
I(\alpha, \beta) = (\alpha +\beta-1) ( \chi_{(1,\infty)}(\alpha) - \chi_{(-\infty,0)}(\alpha) )\chi_{(0,1)}(\beta) .
\end{gather}
\end{Prp}

\begin{proof} Computing the perturbation of the eigenvalues of the closed chain
as in~\cite[Appendix~B.5]{cfs}, to the considered degree on the light cone we obtain
for the perturbation of the Lagrangian an expression of the form
\begin{gather}
( \nabla_{1, \u} - \nabla_{2, \u})
( \nabla_{1, \v} + \nabla_{2, \v} ) \L(x,y)\nonumber\\
\qquad{} \asymp \int_{-\infty}^\infty {\rm d}\alpha \int_{-\infty}^\infty {\rm d}\beta \, I(\alpha, \beta)
(F^\u)_{ij}|_{\alpha y + (1-\alpha) x} (F^\v)^j_{ k}|_{\beta y + (1-\beta) x} \xi^i \xi^k
 \frac{1}{\delta^4} \frac{c}{t^4} {\rm i} K_0(\xi)\label{uvL}
\end{gather}
with a real-valued function~$I(\alpha, \beta)$ (to be determined below)
and a real constant~$c$. Note that the factor~$i$ can again be understood from the
fact that the whole expression is real-valued.

Obviously, the left side of the above equation is anti-symmetric in~$x$ and~$y$.
Moreover, the factor~$K_0(\xi)$ is anti-symmetric. Therefore, the line integrals
must be symmetric when exchanging~$x$ and~$y$.
Let us consider how this can come about.
To second order in perturbation theory, we need to take into account
the contribution given in~\cite[equation~(B.5.1)]{cfs},
\begin{gather} \label{P2cont}
P(x,y) \asymp -i \slashed{\xi} \int_x^y [0,1 | 1] \, {\rm d}z \int_{z}^y [0,1 | 0] \, {\rm d}\tilde{z}\,
F_{ki}(z) F^{kj}(\tilde{z}) \xi^i \xi_j T^{(0)}_{[0]} .
\end{gather}
Transforming to the integration variables~$\alpha$ and~$\beta$
as in~\eqref{alphabetadef}, one sees that the line integrals are symmetric in~$x$ and~$y$
(as can be verified alternatively by taking the conjugate and using that~$P(x,y)^* = P(y,x)$).
Repeating the method in the proof of Lemma~\ref{lemmalog},
after applying the replacement rules the resulting unbounded line integrals
are anti-symmetric in~$x$ and~$y$. Therefore, the second order
contributions to~$P(x,y)$ do not enter~\eqref{uvL}.

It remains to consider the contributions to~$P(x,y)$ to first order in the field strength.
These are given in~\cite[equation~(B.2.4) and~(B.2.5)]{cfs},
\[ P(x,y) = \frac{1}{4} \slashed{\xi} \int_x^y F^{ij} \gamma_i \gamma_j T^{(0)}
-\xi_i \int_x^y [0,1 | 0] F^{ij} \gamma_j T^{(0)} . \]
All we need here is that the integrands are linear polynomials. Therefore, the resulting
line integrals in the symmetric derivatives of the Lagrangian are of the general form
\begin{gather*}
 ( \nabla_{1, \u} + \nabla_{2, \u} )
 ( \nabla_{1, \v} + \nabla_{2, \v} ) \L(x,y) \\
 \asymp \int_0^1 (a\alpha+b) {\rm d}\alpha \int_0^1 (c \beta + d) {\rm d}\beta\,
\big( (F_\u)_{ki}(z) (F_\v)^{kj} (\tilde{z})
+ (F_\v)_{ki}(z) (F_\u)^{kj} (\tilde{z}) \big) \xi^i \xi^k
 \frac{1}{\delta^4} \frac{c}{t^4} {\rm i} K_0(\xi)
\end{gather*}
with four parameters~$a,\ldots,d$.
Since only first order perturbations of~$P(x,y)$ are involved,
the sign of~$\nabla_{2, \u}$ can be changed with the help of
Lemma~\ref{lemmaunbounded}.
Using that the resulting line integrals must be symmetric in~$x$ and~$y$,
we conclude that~$I(\alpha, \beta)$ must be of the form~\eqref{Idef}.
\end{proof}

\subsection{Analysis in momentum space}
In order to gain more insight into how to compute the surface layer integrals,
it is useful to transform the formulas of the previous section to momentum space.
To this end, we write the contributions to the second variation of the Lagrangian
as computed in Propositions~\ref{prplog} and~\ref{prpnolog} (as well other
expressions to be introduced later) in the general form
\begin{gather} \label{Axydef}
A(x,y) :=
\frac{1}{\delta^4} \mathfrak{K}(\xi) \int_{-\infty}^\infty {\rm d}\alpha\, \int_{-\infty}^\infty {\rm d}\beta\, \P(\alpha, \beta)
(F^\u)_{ij}|_{\alpha y + (1-\alpha) x} (F^\v)^j_{ k}|_{\beta y + (1-\beta) x} ,
\end{gather}
where~$\mathfrak{K}(\xi)$ is a distribution (which may involve tensor indices), and the integrand~$\P$ is either~$I(\alpha, \beta)$ or~$J(\alpha, \beta)$
(or other similar functions to be introduced later).

\begin{figure}[t]\centering
\includegraphics{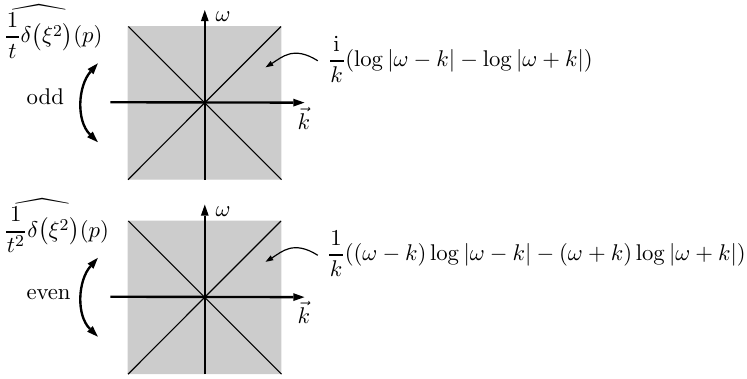}
\caption{The kernels involving~$\delta\big(\xi^2\big)$ in momentum space.}\label{figS12}
\end{figure}

For the distribution~$\mathfrak{K}(\xi)$ we need to consider two essentially different cases:
In Proposition~\ref{prpnolog}, it has the form
\begin{gather} \label{Kgen}
\mathfrak{K}(\xi) \simeq \frac{1}{t^p} {\rm i} K_0(\xi) \qquad \text{with} \quad K_0(\xi) \simeq {\rm i} \epsilon\big(\xi^0 \big) \delta\big(\xi^2\big) .
\end{gather}
The distribution appearing on the right was already computed in
momentum space in Section~\ref{seccurrent}; see Fig.~\ref{figK12}.
In Proposition~\ref{prplog}, on the other hand, the kernels are of the form
\[
\mathfrak{K}(\xi) \simeq \frac{1}{t^p} \delta \big(\xi^2 \big) . \]
Noting that~$\delta \big(\xi^2 \big)$ is the Green's operator of the scalar wave equation,
one immediately sees that its Fourier transform is given by the principal value~$\simeq \PP/p^2$
(defined for example by~$\PP/p^2 := \lim\limits_{\kappa \searrow 0} \big(1/\big(p^2+{\rm i} \kappa\big) + 1/\big(p^2-{\rm i} \kappa\big)\big)/2$).
Translating the factors~$1/t$ again into $\omega$-integrations
(as explained after~\eqref{tomega}), one obtains the kernels shown in Fig.~\ref{figS12}.

Before going on, we clarify how to handle the integration constants when
rewriting factors~$1/t$ as $\omega$-integrals: Clearly, integrating over~$\omega$
while preserving the spherical symmetry of the kernel gives us the freedom to add an
arbitrary function of~$k$. In position space, this corresponds to a distributional
contribution supported at~$t=0$, which vanishes when multiplying by~$t$.
Since in position space, the distribution is supported on the light cone,
this distributional contribution at~$t=0$ must vanish away from the origin~$\xi=0$.
This condition can be satisfied in momentum space by demanding that the
distribution be harmonic, i.e.,
\[ \left( \frac{\partial^2}{\partial \omega^2} - \frac{\partial^2}{\partial k^2}
- \frac{2}{k} \frac{\partial}{\partial k} \right) \hat{\mathfrak{K}}(\omega, k) = 0 . \]
When integrating over~$\omega$, we always choose the integration constant
such that this equation holds.

In the following sections, our task is to evaluate integrals of kernels~$A(x,y)$
of the form~\eqref{Axydef}. More precisely, the expressions of interest are
\[ \int_\scrM A(x,y)\, {\rm d}^4y \qquad \text{and} \qquad \int_{\R^3} {\rm d}^3x \int_\scrM {\rm d}^4y \, A(t,\vec{x};y) .\]
In momentum space, these expressions become
\begin{gather}
 \int_\scrM A(x,y) \, {\rm d}^4y = \frac{1}{\delta^4} \int \frac{{\rm d}^4p_\u}{(2 \pi)^4}
\int \frac{{\rm d}^4p_\v}{(2 \pi)^4} {\rm e}^{-{\rm i} (p_\u + p_\v) x} \notag \\
 \qquad{}\times \int_{-\infty}^\infty {\rm d}\alpha \int_{-\infty}^\infty {\rm d}\beta\, \P(\alpha, \beta)
\big(\hat{F}^\u(p_\u) \big)_{ij} \big(\hat{F}^\v(p_\v) \big)^j_{ k} \hat{\mathfrak{K}} (-\alpha p_\u -\beta p_\v ),\nonumber\\
 \int_{\R^3} {\rm d}^3x \int_\scrM {\rm d}^4y\, A(t,\vec{x};y)
= \frac{1}{\delta^4} \int \frac{{\rm d}^4p_\u}{(2 \pi)^4}
\int \frac{{\rm d}^4p_\v}{(2 \pi)^4} {\rm e}^{-{\rm i} (p^0_\u + p^0_\v) x^0} \delta^3 ( \vec{p}_\u + \vec{p}_\v ) \notag \\
 \qquad {}\times \int_{-\infty}^\infty {\rm d}\alpha \int_{-\infty}^\infty {\rm d}\beta\, \P(\alpha, \beta)
\big(\hat{F}^\u(p_\u) \big)_{ij} \big(\hat{F}^\v(p_\v) \big)^j_{ k} \hat{\mathfrak{K}} (-\alpha p_\u -\beta p_\v) .
\label{i7}
\end{gather}

\subsection{Conservation of surface layer integrals} \label{secosiconserve}
The conservation law for the surface layer integral~\eqref{osicombined}
as established in~\cite{osi} is based on the identity
\begin{gather} \label{pointwise}
\int_\scrM ( \nabla_{1, \u} - \nabla_{2, \u} )
 ( \nabla_{1, \v} + \nabla_{2, \v} ) \L(x,y) \, {\rm d}\rho(y) = 0 .
\end{gather}
Namely, it is obtained by integrating over~$\Omega$ and using the anti-symmetry
of the integrand (for details see~\cite[Proof of Theorem~3.1]{osi}).
In our setting where~$\Omega$ is the past of a surface~$\{t=\text{const}\}$
in Minkowski space, the conservation law follows already if we know that the spatial
integral of~\eqref{pointwise} vanishes,
\begin{gather} \label{integrated}
\int_{\R^3} {\rm d}^3x \int_\scrM ( \nabla_{1, \u} - \nabla_{2, \u} )
 ( \nabla_{1, \v} + \nabla_{2, \v} ) \L(t,\vec{x};y) \, {\rm d}^4y = 0 .
\end{gather}
Our strategy is to first show that, under suitable assumptions on the potentials
and on the regularization, the surface layer integrals are conserved in the sense that
the integral~\eqref{integrated} vanishes (see Theorems~\ref{thmconserved1}
and~\ref{thmconserved2} below).
This makes it possible to express the surface layer integral in a convenient
way (see Lemma~\ref{lemmaconservedform} in Section~\ref{secconservedform}).
Sections~\ref{secreduce} and~\ref{secbounded} are then devoted to the
detailed computations leading to Theorem~\ref{thmbose}.

\subsubsection{Contributions without logarithmic poles}
In order to explain our method and the involved assumptions, we proceed step by step, beginning
with the contributions involving no logarithmic poles as computed in Proposition~\ref{prpnolog}.
Rewriting~\eqref{integrated} according to~\eqref{i7} in momentum space,
our task is to show that
\begin{gather}
0 =\int \frac{{\rm d}^4p_\v}{(2 \pi)^4} {\rm e}^{-{\rm i} (p^0_\u + p^0_\v) x^0} \delta^3 ( \vec{p}_\u + \vec{p}_\v ) \nonumber\\
\hphantom{0=}{} \times \int_{-\infty}^\infty {\rm d}\alpha \int_{-\infty}^\infty {\rm d}\beta \, I(\alpha, \beta)
\big(\hat{F}^\u(p_\u) \big)_{ij} \big(\hat{F}^\v(p_\v) \big)^j_{ k} \hat{\mathfrak{K}} (-\alpha p_\u -\beta p_\v ) ,\label{nologconserve}
\end{gather}
where~$I(\alpha, \beta)$ is the function~\eqref{Idef} and~$\hat{\mathfrak{K}}$ is the Fourier transform of the kernel
\begin{gather} \label{Knolog}
\mathfrak{K}(\xi) \simeq \xi^i \xi^k \frac{1}{\delta^4} \frac{1}{t^4} {\rm i} K_0(\xi) .
\end{gather}
Recall that we consider a non-interacting region of spacetime, where the potentials~$A_\u$
and~$A_\v$ describe electromagnetic waves. Therefore, choosing the Lorenz gauge,
the momenta are on the mass cone. Moreover, for technical reasons we assume that the
momenta are non-zero:
\begin{itemize}\itemsep=0pt
\item[(a)] \label{(a)}
The momenta lie on the double mass cone away from the origin,
\[ p_\u^2 = 0 = p_\v^2 \qquad \text{and} \qquad p_\u \neq 0, \qquad p_\v \neq 0 . \]
\end{itemize}
As a consequence, the Lorentz inner product of the argument of~$\hat{\mathfrak{K}}$
in~\eqref{nologconserve} simplifies to
\begin{gather} \label{inprod}
( -\alpha p_\u -\beta p_\v )^2 = 2 \alpha \beta \la p_\u, p_\v \ra .
\end{gather}

\begin{figure}[t]\centering
\includegraphics{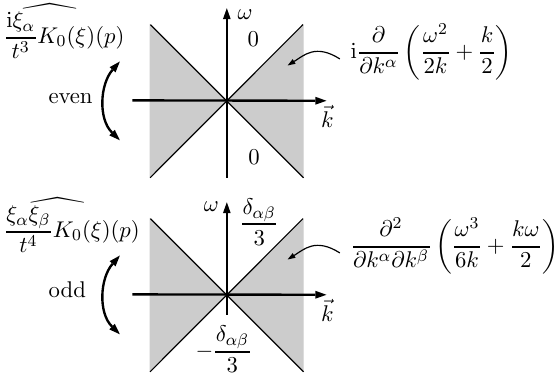}
\caption{Handling of the spatial tensor indices.}\label{figK34}
\end{figure}

The main difficulty in evaluating~\eqref{nologconserve} is that it involves unbounded
line integrals. Our first step is to show that the unbounded part of the line integrals
vanishes, leaving us with an expression involving convex line integrals of the form
\[ \int_0^1 {\rm d}\alpha \int_0^1 {\rm d}\beta\, (\alpha+\beta-1) \cdots . \]
To this end, we first consider an the integral of a polynomial in~$\alpha$,
\begin{gather} \label{poly}
\int_{-\infty}^\infty \text{(polynomial in $\alpha$)} \times {\rm e}^{-{\rm i} p_\u (
\alpha y + (1-\alpha) x )} \, {\rm d}\alpha .
\end{gather}
Carrying out the integral, we obtain a distribution supported on the
hypersurface $\{p_\u \xi = 0\}$. Since~$p_\u$ is on the mass cone,
this hypersurface is null. Since the distribution~${\mathfrak{K}}(\xi)$ is supported
on the light cone, we conclude that the hypersurface~$\{p_\u \xi = 0\}$
intersects the support of~${\mathfrak{K}}(\xi)$ only on the straight line~$\R p_\u$.
Therefore, when regularizing, by making the support of~${\mathfrak{K}}(\xi)$ slightly
smaller one can arrange that the supports no longer intersect
(for a discussion of this point see Remark~\ref{remreg} in Section~\ref{secdiscuss}).
With this in mind, in what follows we shall make use of the following assumption:
\begin{itemize}\itemsep=0pt
\item[(b)] Polynomial integrals over the whole real line~\eqref{poly} vanish in~\eqref{nologconserve}.
\end{itemize}
We next consider integrals over the half line,
\begin{gather} \label{ray}
\int_0^\infty \text{(polynomial in $\alpha$)} \times {\rm e}^{-{\rm i} p_\u (
\alpha y + (1-\alpha) x )} \, {\rm d}\alpha .
\end{gather}
Knowing that the integral over the whole real line vanishes, our task is to handle
Heaviside functions~$\Theta(\alpha)$ or~$\Theta(-\alpha)$ in the integrand.
Using~\eqref{inprod}, we can rewrite these Heaviside functions as
the characteristic function of the inner mass shell in the argument of~$\hat{\mathfrak{K}}$,
\[ \Theta\big( ( -\alpha p_\u -\beta p_\v )^2 \big) = \Theta(\alpha) \Theta ( \beta \la p_\u, p_\v \ra ) . \]
We analyze the effect of this characteristic function for different choices of the tensor indices.
If~$i=0$ and~$j=0$, the distribution~\eqref{Knolog}
simplifies to~\eqref{Kgen} for~$p=2$ as depicted on the right of Fig.~\ref{figK12}.
Multiplying by the characteristic function of the inner mass shell gives a distribution
which is constant inside the upper and lower mass cones and vanishes otherwise.
In position space, this distribution is causal and is again supported on the light cone.
If one or both tensor indices are spatial, the resulting distributions~$\hat{\mathscr{K}}(p)$
are shown in Fig.~\ref{figK34}.
Multiplying by the characteristic function of the inner mass shell gives zero.
We conclude that for any choice of the tensor indices~$i$ and~$j$,
the distribution~$\Theta\big(p^2\big) \hat{\mathfrak{K}}(p)$ is supported on the light cone.
Therefore, we can argue just as after~\eqref{poly}
(this will again be discussed in Remark~\ref{remreg} in Section~\ref{secdiscuss}), leading us to the
following assumption:
\begin{itemize}\itemsep=0pt
\item[(c)] Polynomial integrals over the half line~\eqref{ray} vanish in~\eqref{nologconserve}.
\end{itemize}
Using the above assumptions~(a)--(c), we can simplify~\eqref{nologconserve}
to the condition
\begin{gather*}
0 =\int \frac{{\rm d}^4p_\v}{(2 \pi)^4} {\rm e}^{-{\rm i} (p^0_\u + p^0_\v) x^0} \delta^3 ( \vec{p}_\u + \vec{p}_\v ) \\
\hphantom{0 =}{} \times \int_0^1 {\rm d}\alpha \int_0^1 {\rm d}\beta\, (\alpha +\beta-1)
\big(\hat{F}^\u(p_\u) \big)_{ij} \big(\hat{F}^\v(p_\v) \big)^j_{ k} \hat{\mathfrak{K}} (-\alpha p_\u -\beta p_\v ) .
\end{gather*}

Evaluating the remaining compact line integrals gives the following result:
\begin{Thm} \label{thmconserved1}
Under the above assumptions~{\rm (a)--(c)},
the relation~\eqref{integrated} holds
for the contributions involving no logarithms as computed in Proposition~{\rm \ref{prpnolog}}.
\end{Thm}
\begin{proof} We must consider the two cases that the momenta~$p_\u$ and~$p_\v$
lie on the same mass cone (i.e., both on the upper or both on the lower mass cone)
and that they lie on different mass cones.
In the first case, we see from the right of Fig.~\ref{figK12} as well as from Fig.~\ref{figK34}
that the distribution~$\hat{\mathfrak{K}}$ vanishes or is constant for all momenta~$-\alpha p_\u - \beta p_\v$.
As a consequence, the line integrals can be carried out to obtain zero,
\[ \int_0^1 {\rm d}\alpha \int_0^1 {\rm d}\beta \, (\alpha +\beta-1) = 0 . \]

In the remaining case that~$p_\u$ and~$p_\v$ lie on different mass cones, we
must make use of the fact that, due to the $\delta$-distribution in~\eqref{nologconserve},
it suffices to consider the case~$p_\u = -p_\v$. Then~$\hat{\mathfrak{K}}$ depends
only on~$\alpha-\beta$. As a consequence, the resulting integrals vanish by symmetry,
\[ \int_0^1 {\rm d}\alpha \int_0^1 {\rm d}\beta\, (\alpha +\beta-1) \hat{\mathfrak{K}} (-(\alpha-\beta) p_\u ) = 0 , \]
because the integrand is odd under the
transformation~$\alpha \rightarrow 1-\beta$ and~$\beta \rightarrow 1-\alpha$.
\end{proof}

\subsubsection{Contributions with logarithmic poles}
We now turn our attention to the contributions involving logarithms as computed in Proposition~\ref{prplog}.
Rewriting~\eqref{integrated} according to~\eqref{i7} in momentum space,
our task is to show that
\begin{gather}
0 =\int \frac{{\rm d}^4p_\v}{(2 \pi)^4} {\rm e}^{-{\rm i} (p^0_\u + p^0_\v) x^0} \delta^3 ( \vec{p}_\u + \vec{p}_\v ) \nonumber\\
\hphantom{0=}{} \times \int_{-\infty}^\infty {\rm d}\alpha \int_{-\infty}^\infty {\rm d}\beta\, J(\alpha, \beta)
\big(\hat{F}^\u(p_\u) \big)_{ij} \big(\hat{F}^\v(p_\v) \big)^j_{ k} \hat{\mathfrak{K}} (-\alpha p_\u -\beta p_\v ) ,
\label{logconserve}
\end{gather}
where~$J(\alpha, \beta)$ is the function in~\eqref{Jform}
(see Fig.~\ref{figJ}) and~$\hat{\mathfrak{K}}$ is the Fourier transform of the kernel
\begin{gather} \label{Klog}
\mathfrak{K}(\xi) \simeq \xi^i \xi^k
 \frac{1}{\delta^4} \frac{1}{t^4} \delta \big(\xi^2 \big) .
\end{gather}

Since the distribution~$\mathfrak{K}(\xi)$ is again supported on the light cone,
we can argue exactly as after~\eqref{poly} to justify the following assumption:
\label{bp}
\begin{itemize}\itemsep=0pt
\item[(b$'$)] Polynomial integrals over the whole real line~\eqref{poly} vanish in~\eqref{logconserve}.
\end{itemize}

\begin{figure}[t]\centering
\includegraphics{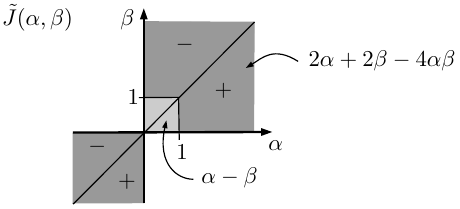}
\caption{The function~$\tilde{J}(\alpha, \beta)$.}\label{figJtilde}
\end{figure}

\begin{Lemma} \label{lemmaantisymm}
Using~{\rm (b$'$)}, the function~$J$ in~\eqref{logconserve} can be replaced by
the function~$\tilde{J}$ given by $($see Fig.~{\rm \ref{figJtilde})}
\begin{gather*}
\tilde{J}(\alpha, \beta) = (\alpha-\beta) \chi_{(0,1)}(\alpha) \chi_{(0,1)}(\beta) + ( 4 \alpha \beta - 2 \alpha - 2 \beta ) \epsilon(\beta-\alpha) \\
\hphantom{\tilde{J}(\alpha, \beta) =}{} \times
\big( \chi_{(0, \infty)}(\alpha) \chi_{(0, \infty)}(\beta) +
\chi_{(-\infty,0)}(\alpha) \chi_{(-\infty,0)}(\beta) - \chi_{(0,1)}(\alpha) \chi_{(0,1)}(\beta) \big) .
\end{gather*}
\end{Lemma}
\begin{proof} For clarity, we proceed in several steps.
First, we subtract from~$J(\alpha, \beta)$ (see Fig.~\ref{figJ}) the polynomial in~$\alpha$
\[ (-3 \alpha - \beta + 4 \alpha \beta) \chi_{(0,1)}(\beta) . \]
to obtain the following function:
\begin{center}
\includegraphics{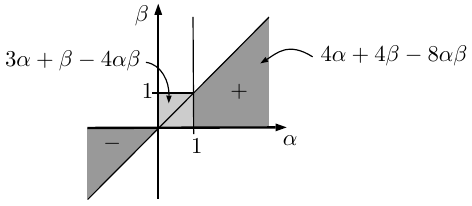}
\end{center}

\noindent
Next, we subtract the polynomial in~$\beta$
\[ \big(-2\alpha -2 \beta + 4 \alpha \beta \big) \chi_{(-\infty,0)}(\alpha) \]
to obtain
\begin{center}
\includegraphics{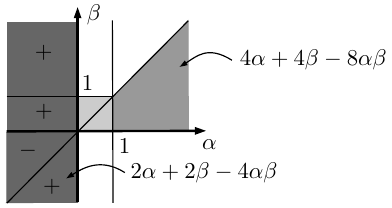}
\end{center}

\noindent
Finally, we subtract the polynomial in~$\alpha$
\[ \big(2\alpha +2 \beta - 4 \alpha \beta \big) \chi_{(0, \infty)}(\beta) \]
to obtain the result.
\end{proof}

\begin{figure}[t]\centering
\includegraphics{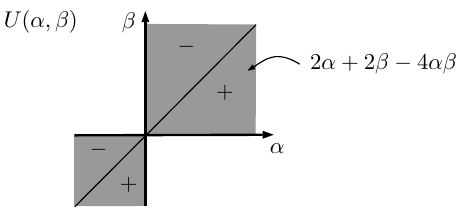}
\caption{The function~$U(\alpha, \beta)$.}\label{figU}
\end{figure}

The resulting line integrals are still unbounded. In order to analyze the effect of the
unbounded contribution of the line integrals to~\eqref{logconserve},
it is useful to introduce the function (see Fig.~\ref{figU})
\begin{gather} \label{Udef}
U(\alpha, \beta) = 2 (\alpha+\beta-2\alpha \beta)\Theta(\alpha \beta) \epsilon(\alpha-\beta) ,
\end{gather}
because the function~$J-U$ has compact support. Moreover, the Fourier integral of
the line integrals of~$U$ can be computed explicitly:

\begin{Lemma} \label{lemmaUterm}
 \begin{gather}
\int_{-\infty}^\infty {\rm d}\alpha \int_{-\infty}^\infty {\rm d}\beta \, U(\alpha, \beta)
{\rm e}^{-{\rm i} p_\u (\alpha y + (1-\alpha) x)} {\rm e}^{-{\rm i} p_\v (\beta y + (1-\beta) x)} \nonumber\\
\qquad{} = {\rm e}^{-{\rm i} p_\u x -{\rm i} p_\v x} A(p_\u \xi, p_\v \xi) ,
 \label{UA}
\end{gather}
where~$A$ is the bi-distribution
\begin{gather} \label{Auvdef}
A(u,v) = 4 \pi {\rm i} ({\rm i} \partial_u +{\rm i} \partial_v + \partial_u \partial_v )
\left( -\frac{\PP}{u} \delta(v) + \delta(u) \frac{\PP}{v} +8 \delta(u+v) \frac{\PP}{u-v} \right)
\end{gather}
in the new variables~$u:= p_\u \xi$ and~$v:=p_\v \xi$.
\end{Lemma}

\begin{proof} Multiplying~\eqref{UA} by~${\rm e}^{{\rm i} p_\u x +{\rm i} p_\v x}$, we obtain
\[ A(u,v) = \int_{-\infty}^\infty {\rm d}\alpha \int_{-\infty}^\infty {\rm d}\beta\, U(\alpha, \beta)
{\rm e}^{-{\rm i} \alpha u -{\rm i} \beta v} . \]
Our task is to show that this equation agrees with~\eqref{Auvdef}. To this end, we
first rewrite the polynomials as derivatives,
\begin{gather*}
A = 2 \big({\rm i} \partial_u +{\rm i} \partial_v + \partial_u \partial_v \big)
\int_{-\infty}^\infty {\rm d}\alpha \int_{-\infty}^\infty {\rm d}\beta \, \Theta(\alpha \beta) \epsilon(\alpha-\beta)
{\rm e}^{-{\rm i} \alpha u -{\rm i} \beta v} .
\end{gather*}
The remaining Fourier integrals can be computed with the help of the relations
\begin{gather*}
4 \Theta(\alpha \beta) \epsilon(\alpha-\beta)
 = \epsilon(\alpha) - \epsilon(\beta) - 2 \epsilon(\alpha-\beta), \\
\int_{-\infty}^\infty {\rm e}^{-{\rm i} \alpha u} {\rm d}\alpha = 2 \pi \delta(u), \\
\int_{-\infty}^\infty \epsilon(u) {\rm e}^{-{\rm i} \alpha u} {\rm d}\alpha = -{\rm i} \frac{\PP}{u} .
\end{gather*}
We thus obtain
\begin{gather*}
 \int_{-\infty}^\infty {\rm d}\alpha \int_{-\infty}^\infty {\rm d}\beta \, (\epsilon(\alpha) - \epsilon(\beta) )
{\rm e}^{-{\rm i} \alpha u -{\rm i} \beta v} = -2 \pi {\rm i} \frac{\PP}{u} \delta(v) + 2 \pi {\rm i} \delta(u) \frac{\PP}{v}, \\
 \int_{-\infty}^\infty {\rm d}\alpha \int_{-\infty}^\infty {\rm d}\beta\, \epsilon(\alpha-\beta)
{\rm e}^{-{\rm i} \alpha u -{\rm i} \beta v} = \left\{\begin{matrix} a = (\alpha+\beta)/2 \\ b = (\alpha-\beta)/2 \end{matrix} \right\} \\
\qquad{} = 2 \int_{-\infty}^\infty {\rm d}a \int_{-\infty}^\infty {\rm d}b \, \epsilon(b) {\rm e}^{-{\rm i} a (u+v) -{\rm i} b(u-v)}
= -4 \pi {\rm i} \delta(u+v) \frac{\PP}{u-v} .
\end{gather*}
Collecting all the terms gives the result.
\end{proof}

We now explain what this result means. The contributions containing factors~$\delta(u)$
and~$\delta(v)$ (and distributional derivatives of these factors)
are supported on the null hypersurfaces~$\{p_\u \xi=0\}$ and~\mbox{$\{p_\v \xi=0\}$}, respectively.
Therefore, we can argue again just as for the polynomial contributions in~$\alpha$
or~$\beta$ (assumption~{(b$'$)} above) to conclude that these contribution vanish
in~\eqref{logconserve}. The factors~$\delta(u+v)$ (and distributional derivatives thereof),
on the other hand, are supported on the hypersurface
\[ \Hil := \{ (p_\u+p_\v) \xi =0\} . \]
If~$p_\u$ and~$p_\v$ are on the same mass cone, then the
momentum~$p_\u + p_\v$ is timelike, so that the hypersurface~$\Hil$ is spacelike.
As a consequence, this hypersurface intersects the support of~$\hat{\mathfrak{K}}(p)$
only in the origin. Therefore, choosing the regularization such that~$\hat{\mathfrak{K}}(p)$
vanishes at the origin, the resulting contribution
to~\eqref{logconserve} is zero
(alternatively, the vanishing of this contribution
can be derived by a scaling argument when the regularization is removed).
The remaining case that~$p_\u$ and~$p_\v$ are
supported on different mass cones is more subtle:
We first note that the space spanned by~$p_\u$ and~$p_\v$ is a two-dimensional
timelike hypersurface. As a consequence, its orthogonal complement
\[ \big\{ \xi \big| \la \xi, p_\u \ra = 0 = \la \xi, p_\v \ra \big\} \]
is a two-dimensional spacelike hypersurface of Minkowski space.
On the left of~\eqref{i7}, we integrate~${\mathfrak{K}}(\xi)$ over this subspace.
Similar as in Hadamard's method of descent (see for example~\cite[Section~5.1(b)]{john}),
the kernel obtained after carrying out this integration
is again causal. With this in mind, it suffices to consider the case
\[ \xi \in \operatorname{span}(p_\u, p_\v) . \]
Then, due to the $\delta$-distribution in~\eqref{logconserve}, we only get a contribution if
\[ \xi \sim p_\u - p_\v \approx 2 p_\u \]
(note that~$(p_\u - p_\v)(p_\u + p_\v) = p_\u^2 - p_\v^2 = 0$ because~$p_\u$ and~$p_\v$ are lightlike).
Hence in the limit~$p_\u \rightarrow -p_\v$ we only get a contribution if~$\xi$ is tangential to the
light cone. Arguing again as explained after~\eqref{inprod},
we are led to the following assumption: \label{dp}
\begin{itemize}\itemsep=0pt
\item[(d)] Replacing the function~$J(\alpha, \beta)$ in~\eqref{logconserve}
by the function~$U(\alpha, \beta)$ introduced in~\eqref{Udef},
the resulting line integrals vanish in~\eqref{logconserve}.
\end{itemize}

Using this assumption, we may replace the function~$J(\alpha, \beta)$ in~\eqref{logconserve}
by~$J-U$. Hence it remains to show that
\begin{gather}
0 =\int \frac{{\rm d}^4p_\v}{(2 \pi)^4} {\rm e}^{-{\rm i} (p^0_\u + p^0_\v) x^0} \delta^3 ( \vec{p}_\u + \vec{p}_\v ) \nonumber\\
\hphantom{0 =}{} \times \int_0^1 {\rm d}\alpha \int_0^1 {\rm d}\beta\, V(\alpha, \beta)
\big(\hat{F}^\u(p_\u) \big)_{ij} \big(\hat{F}^\v(p_\v) \big)^j_{ k} \hat{\mathfrak{K}} (-\alpha p_\u -\beta p_\v ) ,
\label{logconserve2}
\end{gather}
where
\begin{gather} \label{Vdef}
V(\alpha, \beta) = \begin{cases} -\alpha - 3 \beta + 4 \alpha \beta & \text{if $\alpha>\beta$}, \\
3\alpha + \beta - 4 \alpha \beta & \text{if $\beta>\alpha$} .
\end{cases}
\end{gather}

Our final task is to evaluate the resulting compact line integrals. Here we proceed similar as in the
proof of Theorem~\ref{thmconserved1}. However, the symmetry argument is a bit more
subtle and makes it necessary to use the Maxwell equations:
\begin{itemize}\itemsep=0pt
\item[(e)] \label{(e)}%
The field tensors~$F^\u$ and~$F^\v$ satisfy the source-free Maxwell equations
\[ \big( \hat{F}^\u(p_\u) \big)_{ij} p_\u^j = 0 = \big( \hat{F}^\v(p_\v) \big)_{ij} p_\v^j . \]
\end{itemize}
As mentioned in the introduction,
the EL equations corresponding to the causal action principle imply that the Maxwell equations
hold in the continuum limit (as is worked out in~\cite[Chapter~5]{cfs}). Therefore, we may use
the Maxwell equations in the computation
of the surface layer integrals. Consequently, the main assumption in~(e) is that
there are no sources inside the surface layer. In other words, we disregard the influence
of electromagnetic currents on the bosonic surface layer integral. This is a simplifying assumption,
which is sensible at least in situations when the surface layer integral
is computed for a scattering process for the non-interacting incoming or outgoing fields.

\begin{Thm} \label{thmconserved2}
Under the above assumptions~{\rm (a)}, {\rm (b$'$)}, {\rm (d)} and~{\em (e)},
the relation~\eqref{integrated} holds for the contributions involving logarithms as computed
in Proposition~{\rm \ref{prplog}}.
\end{Thm}
\begin{proof} Due to the $\delta$-distribution in~\eqref{logconserve},
it again suffices to consider the case~$\vec{p}_\u = -\vec{p}_\v$.
We distinguish the two cases that the momenta~$p_\u$ and~$p_\v$
lie on the same mass cone and that they lie on different mass cones.
In the latter case, we have~$p_\u = -p_\v$. Then~$\hat{\mathfrak{K}}$ depends
only on~$\alpha-\beta$, giving rise to the line integrals
\begin{gather} \label{Vint}
\int_0^1 {\rm d}\alpha \int_0^1 {\rm d}\beta \, V(\alpha,\beta) \hat{\mathfrak{K}} (-(\alpha-\beta) p_\u ) .
\end{gather}
As is obvious from~\eqref{Vdef}, the factor~$V(\alpha, \beta)$ is odd under the replacements~$\alpha
\leftrightarrow \beta$. Thus in order to prove that~\eqref{Vint} vanishes, it suffices to
show that the factor~$\hat{\mathfrak{K}}(-(\alpha-\beta) p_\u)$ is even under the transformations~$\alpha
\leftrightarrow \beta$. If both tensor indices in~\eqref{Klog} are zero, this is obvious from
the lower plot in Fig.~\ref{figS12}.
Exactly as shown in Fig.~\ref{figK34} for the contributions without logarithms,
the spatial indices can be handled by integrating in~$\omega$ and differentiating in the spatial momenta.
Since this does not change the symmetry of~$\hat{\mathfrak{K}}$ about the origin,
the distribution~$\hat{\mathfrak{K}}(-(\alpha-\beta) p_\u)$ is again even under the transformations~$\alpha
\leftrightarrow \beta$. This concludes the proof in the case that the momenta~$p_\u$ and~$p_\v$
lie on different mass cones.

In the remaining case that~$p_\u$ and~$p_\v$ lie on the same mass cone,
the transformation~$\alpha \leftrightarrow \beta$ corresponds to an inversion of the spatial component
of the argument of~$\hat{\mathfrak{K}}$. If both tensor indices in~\eqref{Klog} are zero,
it follows from spherical symmetry of the integral kernels that~$\hat{\mathfrak{K}}$ is even under this transformation.
Using again that~$V(\alpha,\beta)$ is odd under this transformation, we obtain zero.
The same argument applies if both tensor indices in~\eqref{Klog} are spatial.
If exactly one of the indices is spatial, integrating in~$\omega$ and differentiating in the spatial momenta
gives a factor~$k_\alpha$ (similar as shown in Fig.~\ref{figK34}).
This factor is contracted with a field tensor. Therefore, using the Maxwell equations~(e)
\[ \big( \hat{F}^\u(p_\u) \big)_{i \alpha} (p_\u)^\alpha = - \big( \hat{F}^\u(p_\u) \big)_{i0} (p_\u)^0 \]
(and similarly for~$\hat{F}^\v$), we get back to the case where both tensor indices are zero.
This concludes the proof.
\end{proof}

\subsection{An integral formula for conserved surface layer integrals} \label{secconservedform}
Having proved that the surface layer integrals are time-independent,
we may simplify the formula for the surface layer integrals
with the help of the following lemma, which is a straightforward generalization
of Lemma~\ref{lemmaIconserve}.
\begin{Lemma} \label{lemmaconservedform}
Using the conservation of the surface layer integral
as proved in Theorems~{\rm \ref{thmconserved1}} and~{\rm \ref{thmconserved2}},
the surface layer integral can be written as
\begin{gather}
 \int_{-\infty}^{t_0} {\rm d}t \int_{\R^3} {\rm d}^3x \int_{t_0}^\infty {\rm d}t' \int_{\R^3} {\rm d}^3y\,
 ( \nabla_{1, \u} - \nabla_{2, \u} )
 ( \nabla_{1, \v} + \nabla_{2, \v} ) \L(t,\vec{x}; t', \vec{y}) \notag \\
 \qquad{} = \frac{1}{2} \lim_{T \rightarrow \infty} \frac{1}{T} \int_0^T {\rm d}t
 \int_{\R^3} {\rm d}^3x \int_\scrM {\rm d}^4y \, \big(y^0 - t\big)
 ( \nabla_{1, \u} - \nabla_{2, \u} )
 ( \nabla_{1, \v} + \nabla_{2, \v} ) \L(t,\vec{x}; y) . \!\!\label{osit}
\end{gather}
\end{Lemma}
\begin{proof} Differentiating the left side of~\eqref{osit} with respect to~$t_0$ and using that the integrand is
antisymmetric in the arguments~$x$ and~$y$, we obtain precisely
the expression in~\eqref{integrated} evaluated at~$t=t_0$.
Therefore, the results of Theorems~\ref{thmconserved1} and~\ref{thmconserved2}
show that the above surface layer integral is indeed time-independent.
As a consequence, denoting the spatial integrals by
\[ A(t,t') := \int_{\R^3} {\rm d}^3x \int_{\R^3} {\rm d}^3y\,
 ( \nabla_{1, \u} - \nabla_{2, \u} )
 ( \nabla_{1, \v} + \nabla_{2, \v} ) \L(t,\vec{x}; t', \vec{y}) , \]
we can proceed exactly as in the proof of Lemma~\ref{lemmaIconserve}.
\end{proof}

\subsection{Reduction to bounded line integrals} \label{secreduce}
Our strategy is to compute the surface layer integral using the formula of Lemma~\ref{lemmaconservedform}.
The remaining task is to compute the expression in the last line in~\eqref{osit}
for any given~$t$,
\begin{gather} \label{osival}
\int_{\R^3} {\rm d}^3x \int_\scrM \big(y^0 - t\big) ( \nabla_{1, \u} - \nabla_{2, \u} )
 ( \nabla_{1, \v} + \nabla_{2, \v} ) \L(t,\vec{x};y) \, {\rm d}^4y .
\end{gather}
This expression coincides with the integral expression~\eqref{integrated}
up to the factor~$\big(y^0 - t\big)$ in the integrand.
This is very convenient, because we can apply many methods and results
of Section~\ref{secosiconserve}. Indeed, the expression~\eqref{osival}
can again be written in momentum space in the form~\eqref{nologconserve}
and~\eqref{logconserve}, respectively, where the kernels~${\mathfrak{K}}$
in~\eqref{Knolog} and~\eqref{Klog} are multiplied by~$t$, i.e.,
\begin{alignat}{3}
& \mathfrak{K}(\xi)\simeq \xi^i \xi^k \frac{c}{\delta^4} \frac{1}{t^3} {\rm i} K_0(\xi)
\qquad && \text{contributions without logarithmic poles}, & \label{Cnolog} \\
& \mathfrak{K}(\xi)\simeq \xi^i \xi^k
 \frac{c}{\delta^4} \frac{1}{t^3} \delta \big(\xi^2 \big)\qquad
&& \text{contributions with logarithmic poles} .& \label{Clog}
\end{alignat}
In the case~$i=k=0$, the Fourier transforms of these kernels are shown
in Figs.~\ref{figK12} and~\ref{figS12}. The tensor indices can
be handled again by integrating in~$\omega$ and
taking $k$-derivatives (as illustrated in Fig.~\ref{figK34}).

The similarity between~\eqref{osival} and~\eqref{integrated} implies that
the method for handling the unbounded line integrals in Section~\ref{secosiconserve}
can be applied without changes, giving the following result:

\begin{Thm} \label{thmosicompact}
Again under the above assumptions~{\rm (a)}--{\rm (e)} and~{\rm (b$'$)}
$($see pp.~{\rm \pageref{(a)}--\pageref{(e)})}, the surface layer integrals can be written as
\begin{gather}
 \int_{-\infty}^{t_0} {\rm d}t \int_{\R^3} {\rm d}^3x \int_{t_0}^\infty {\rm d}t' \int_{\R^3} {\rm d}^3y
 ( \nabla_{1, \u} - \nabla_{2, \u} )
 ( \nabla_{1, \v} + \nabla_{2, \v} ) \L(t,\vec{x}; t', \vec{y}) \notag \\
 \simeq \frac{1}{\delta^4} \int_{\R^3} {\rm d}^3x \int_\scrM {\rm d}^4y\,
\mathfrak{K} (y-x ) \int_0^1 {\rm d}\alpha \int_0^1 {\rm d}\beta \, V(\alpha, \beta)
(F^\u)_{ij}|_{\alpha y + (1-\alpha) x} (F^\v)^j_{ k}|_{\beta y + (1-\beta) x} .\!\!\!
\label{osiform}
\end{gather}
For the contributions without logarithmic poles $($as computed in Proposition~{\rm \ref{prpnolog})},
the kernel is given by~\eqref{Cnolog} and
\[ V(\alpha, \beta) = \alpha+\beta-1 . \]
Likewise, for the contributions with logarithmic poles $($as computed in Proposition~{\rm \ref{prplog})},
the kernel is given by~\eqref{Clog}, and the function~$V(\alpha, \beta)$ is given by~\eqref{Vdef}.
\end{Thm}

\subsection{Computation of bounded line integrals} \label{secbounded}
The remaining task is to compute the bounded line integrals in~\eqref{osiform}.
Here we face a difficulty which is in some sense complementary to the
difficulties in the previous Sections~\ref{secunbound}--\ref{secreduce}, as we now explain.
When analyzing the unbounded line integrals,
the contributions to the line integrals for large~$\alpha$
and/or~$\beta$ led to poles if~$y$ approaches~$x$. Such difficulties
for small distances can be regarded as ultraviolet problems.
Accordingly, in momentum space the difficulty was to make sense of the
convolution integrals for large momenta.
When analyzing the bounded integrals in~\eqref{osiform}, however, the
difficulties arise if combinations of~$\alpha$ and~$\beta$ are {\em{small}}.
More precisely, let us again assume that~$F^\u$ and~$F^\v$ have compact support.
Then both arguments~$\alpha y + (1-\alpha)x$ and~$\beta y + (1-\beta)y$
must be in a fixed compact set. But if~$\alpha \approx \beta$,
we nevertheless get contributions for arbitrarily large~$y$ and~$x$,
provided that only the convex combinations above lie inside the compact set.
This consideration shows that there are {\em{infrared problems}}
if~$\alpha \approx \beta$. Likewise, in momentum space these problems
will become apparent as poles at zero momentum.

\subsubsection{An infrared regularization} \label{secinfrared}
In order to treat the infrared problems in a clean way, we introduce an
infrared regularization by considering the system in finite spatial volume
(as we shall see, the contribution to the surface layer integral will
remain finite if the infrared regularization is removed).
To this end, we insert a cutoff function into the spatial integrals.
Thus let~$\eta \in C^\infty_0\big(\R^3\big)$ be a non-negative test function with
\[ \int_{\R^3} \eta(\vec{x}) \, {\rm d}^3x = 1 . \]
For technical simplicity, we assume that~$\eta$ is {\em{spherically symmetric}}
(i.e., depends only on~$|\vec{x}|$). For a parameter~$R>0$ we set
\[ \eta_R(\vec{x}) = \eta\left( \frac{\vec{x}}{R} \right) . \]
Then its Fourier transform is
\begin{gather} \label{hateta}
\hat{\eta}_R(\vec{k}) = R^3 \hat{\eta}(R \vec{k}) .
\end{gather}
Clearly, this family of functions converges to a multiple of the $\delta$ distribution,
\[ \lim_{R \rightarrow \infty} \hat{\eta}_R(\vec{k}) = (2 \pi)^3 \delta^3(\vec{k}) . \]
Inserting~$\eta_R$ into the spatial integrals in~\eqref{osiform}, we obtain the integrals
 \begin{gather}
A := \int_{\R^3} {\rm d}^3x \eta_R(\vec{x}) \int_\scrM {\rm d}^4y \, \eta_R(\vec{y}) \mathfrak{K}(y-x)\nonumber\\
 \hphantom{A :=}{} \times
\int_0^1 {\rm d}\alpha \int_0^1 {\rm d}\beta\, V(\alpha, \beta)
 (F_\u)_{ij} (\alpha y + (1-\alpha)x ) (F_\v)^j_{ k} (\beta y + (1-\beta)x ) ,
\label{infrared}
\end{gather}
where~$x=(0,\vec{x})$. Using that multiplication
in position space corresponds to convolution in momentum space, we get
\begin{gather*}
A = \int_0^1 {\rm d}\alpha \int_0^1 {\rm d}\beta\, V(\alpha, \beta)
\int \frac{{\rm d}^4p_\u}{(2 \pi)^4} \big(\hat{F}_\u\big)_{ij}(p_\u) \int \frac{{\rm d}^4p_\v}{(2 \pi)^3}
\big(\hat{F}_\v\big)^j_{ k}(p_\v) \\
\hphantom{A =}{} \times \int_{\R^3} \frac{{\rm d}^3q}{(2 \pi)^3}
\hat{\eta}_R(\vec{q}) \hat{\eta}_R (-\vec{p}_\u - \vec{p}_\v - \vec{q} )
\hat{\mathfrak{K}} (-\alpha p_\u - \beta p_\v - q ) ,
\end{gather*}
where~$q=(0, \vec{q})$. We again assume that the momenta of the field tensors
lie on the mass cone (see assumption~(a) on p.~\pageref{(a)}).
Moreover, knowing that the surface layer integrals are time independent
(see Theorems~\ref{thmconserved1} and~\ref{thmconserved2}),
we only get a contribution if the frequencies of the momenta~$p_\u$ and~$p_\v$
have opposite signs. Hence, writing~$p_\u=(\omega_\u, \vec{p}_\u)$ and~$p_\v=(\omega_\v, \vec{p}_\v)$,
it suffices to consider the cases
\begin{gather} \label{ouvdef}
\omega_\u = \pm | \vec{p}_\u | ,\qquad \omega_\v = \mp | \vec{p}_\v | .
\end{gather}

Next, in order to clarify the $R$-dependence, it is useful to introduce new integration
variables~$\vec{\tilde{q}}$,~$\vec{p}$ and $\Delta p$ by
\begin{gather} \label{pppdef}
\vec{q} = \frac{\vec{\tilde{q}}}{R} \qquad \text{and} \qquad
\vec{p}_\u = \vec{p} + \frac{\Delta \vec{p}}{2R} ,\qquad
\vec{p}_\v = -\vec{p} + \frac{\Delta \vec{p}}{2R} .
\end{gather}
We thus obtain
\begin{gather*}
A = \int_0^1 {\rm d}\alpha \int_0^1 {\rm d}\beta \, V(\alpha, \beta)
\int_{-\infty}^\infty \frac{{\rm d}\omega_\u}{2 \pi} \int_{-\infty}^\infty \frac{{\rm d}\omega_\v}{2 \pi}
\int_{\R^3} \frac{{\rm d}^3p}{(2 \pi)^3} \int_{\R^3} \frac{{\rm d}^3 \Delta p}{R^3 (2 \pi)^3}
 (\hat{F}_\u)_{ij}(p_\u) (\hat{F}_\v)^j_{ k}(p_\v) \\
\hphantom{A=}{}\times
\int_{\R^3} \frac{{\rm d}^3 \tilde{q}}{R^3 (2 \pi)^3} \hat{\eta}_R \left(\frac{\vec{\tilde{q}}}{R} \right) \hat{\eta}_R \left(
\frac{-\Delta \vec{p} - \vec{\tilde{q}}}{R} \right)
\hat{\mathfrak{K}}\left(-\alpha p_\u - \beta p_\v - \frac{\tilde{q}}{R} \right) \\
\hphantom{A}{} = \int_0^1 {\rm d}\alpha \int_0^1 {\rm d}\beta \, V(\alpha, \beta)
\int_{-\infty}^\infty \frac{{\rm d}\omega_\u}{2 \pi} \int_{-\infty}^\infty \frac{{\rm d}\omega_\v}{2 \pi}
\int_{\R^3} \frac{{\rm d}^3p}{(2 \pi)^3} \int_{\R^3} \frac{{\rm d}^3 \Delta p}{(2 \pi)^3}
 \big(\hat{F}_\u\big)_{ij}(p_\u) \big(\hat{F}_\v\big)^j_{ k}(p_\v) \\
\hphantom{A=}{} \times \int_{\R^3} \frac{{\rm d}^3q}{(2 \pi)^3}
\hat{\eta}(\vec{q}) \hat{\eta} ( -\Delta \vec{p} - \vec{q} )
\hat{\mathfrak{K}} \left(-\alpha p_\u - \beta p_\v - \frac{q}{R} \right) ,
\end{gather*}
where~$\tilde{q}=(0, \vec{\tilde{q}})$, and where
in the last step we used~\eqref{hateta} and omitted the tildes.
In order to concentrate on the line integrals in this equation, we set
\begin{gather} \label{Adef}
A = \int_{\R^3} \frac{{\rm d}^3p}{(2 \pi)^3} \int_{-\infty}^\infty \frac{{\rm d}\omega_\u}{2 \pi} \int_{-\infty}^\infty \frac{{\rm d}\omega_\v}{2 \pi}
\int_{\R^3} \frac{{\rm d}^3 \Delta p}{(2 \pi)^3} \int_{\R^3} \frac{{\rm d}^3q}{(2 \pi)^3}
\hat{\eta}(\vec{q}) \hat{\eta} ( -\Delta \vec{p} - \vec{q} ) B
\end{gather}
with
\[ B = \int_0^1 {\rm d}\alpha \int_0^1 {\rm d}\beta \, V(\alpha, \beta)
 \big(\hat{F}_\u\big)_{ij}(p_\u) \big(\hat{F}_\v\big)^j_{ k}(p_\v) \hat{\mathfrak{K}}\left(-\alpha p_\u - \beta p_\v - \frac{q}{R} \right) , \]
where the momenta~$p_\u$ and~$p_\v$ are given by~\eqref{pppdef} and~\eqref{ouvdef}.

In order to analyze the asymptotics for large~$R$, we first
denote the arguments of~$\hat{\mathfrak{K}}$ by~$\omega$ and~$\vec{k}$,
\begin{gather*}
\omega := \mp \alpha \left| \vec{p} + \frac{\Delta \vec{p}}{2R} \right|
\pm \beta \left| \vec{p} - \frac{\Delta \vec{p}}{2R} \right| ,\\
\vec{k} := -\alpha \vec{p}_\u - \beta \vec{p}_\v - \frac{\vec{q}}{R}
= -(\alpha-\beta) \vec{p} - (\alpha+\beta) \frac{\Delta \vec{p}}{2R} - \frac{\vec{q}}{R} .
\end{gather*}
Introducing the new integration variables~$u$ and~$v$ by
\begin{gather*}
u = \alpha+\beta - 1, \qquad
v = (\alpha-\beta) R |\vec{p} | ,
\end{gather*}
the integration measure transforms according to
\[ {\rm d}\alpha {\rm d}\beta = \frac{1}{2R |\vec{p} |} {\rm d}u {\rm d}v . \]
We thus obtain
\begin{gather} \label{Bprelim}
B = \frac{1}{2R |\vec{p} |} \int_{-1}^1 {\rm d}u \int_{-v_{\max}(u)}^{v_{\max}(u)} {\rm d}v \, V(\alpha, \beta)
 \big(\hat{F}_\u\big)_{ij}(p_\u) \big(\hat{F}_\v\big)^j_{ k}(p_\v) \hat{\mathfrak{K}} (\omega, \vec{k} ) ,
\end{gather}
where the boundaries of integration are given by
\begin{gather} \label{vmax}
v_{\max} = ( 1 - |u| ) R |\vec{p} | .
\end{gather}

Next, we make use of the fact that~$\hat{\mathfrak{K}}$ is homogeneous of degree minus one, i.e.,
\begin{gather} \label{hmo}
\hat{\mathfrak{K}} ( \omega, \vec{k} ) = R \hat{\mathfrak{K}} (R \omega, R \vec{k} ) .
\end{gather}
For the contributions without logarithms~\eqref{Cnolog}, this is obvious
by differentiating the formula in the lower plot in Fig.~\ref{figK34} with respect too~$\omega$.
Likewise, for the contributions with logarithms, this follows from~\eqref{Clog}
and the explicit formula
\[ 
\Big( \widehat{ \frac{\xi_j \xi_k}{t^3} \delta \big(\xi^2 \big) } \Big)(p) \simeq \frac{\partial^2}{\partial p^j \partial p^k}
\frac{i}{k} \big( (\omega-k)^2 \log |\omega-k| - (\omega+k)^2 \log |\omega+k| \big) \]
(which is obtained by integrating the formula in the lower plot in Fig.~\ref{figS12}
with respect to~$\omega$ and differentiating twice). Using~\eqref{hmo} in~\eqref{Bprelim}, we obtain
\begin{gather} \label{Bform}
B = \frac{1}{2 |\vec{p} |} \int_{-1}^1 {\rm d}u \int_{-v_{\max}(u)}^{v_{\max}(u)} {\rm d}v \, V(\alpha, \beta)
 \big(\hat{F}_\u\big)_{ij}(p_\u) \big(\hat{F}_\v\big)^j_{ k}(p_\v) \hat{\mathfrak{K}} (R \omega, R \vec{k} ) \big|_{(u,v)} .
\end{gather}

Now the arguments of~$\hat{\mathfrak{K}}$ can be expanded as follows,
\begin{gather}
R \omega = \mp \alpha \left| R \vec{p} + \frac{\Delta \vec{p}}{2} \right|
\pm \beta \left| R\vec{p} - \frac{\Delta \vec{p}}{2} \right| \notag \\
\hphantom{R \omega}{}= \mp v \big( 1 + \O\big(R^{-2} \big) \big) \mp \frac{1+u}{2}
 \la \hat{\vec{p}}, \Delta \vec{p} \ra \big( 1 + \O\big(R^{-2} \big) \big), \label{Rom} \\
R \vec{k} = -(\alpha-\beta) R \vec{p} - (\alpha+\beta) \frac{\Delta \vec{p}}{2} - \vec{q}
 = -v \hat{\vec{p}} - \frac{1+u}{2} \Delta \vec{p} - \vec{q}, \label{Rk} \\
\big\la \hat{\vec{p}}, R \vec{k} \big\ra = -v- \frac{1+u}{2} \big\la \hat{\vec{p}}, \Delta \vec{p} \big\ra -
\big\la \hat{\vec{p}}, \vec{q} \big\ra ,\nonumber
\end{gather}
where~$\hat{\vec{p}} := \vec{p}/|\vec{p} |$ is the unit vector pointing in the direction of~$\vec{p}$.

\subsubsection{Reduction to a scalar expression}
With the above transformations, we have arranged that the integrand in~\eqref{Bform}
converges pointwise in the limit~$R \rightarrow \infty$.
Therefore, the integrals converge in this limit on every compact set.
A~remaining difficulty is that the integration range also depends on~$R$
(see~\eqref{vmax}), making it necessary to analyze the behavior of the integrand in~\eqref{Bform}
for large~$v$. Since~$\hat{\mathfrak{K}}$ is homogeneous of degree minus one
and~$R \vec{k}$ and~$R \omega$ grow linearly in~$v$, the integrand decays at least like~$v^{-1} \log v$,
leading to an at most logarithmic divergence,
\begin{gather} \label{logpole}
|B| \lesssim \log^2 R .
\end{gather}
The question is whether this divergence really occurs. This question is related to the
contractions of the tensor indices, as we now explain:
Being a solution of the source-free Maxwell equations, the field tensor
has the property~(e) on p.~\pageref{(e)}. Moreover, from~\eqref{ouvdef} and~\eqref{pppdef}
we know that the momenta~$p_\u$ and~$p_\v$ coincide with the momentum~$(\pm |\vec{p}|, \vec{p})$
up to signs and corrections of order~$1/R$. Since~$B$ diverges at most logarithmically,
corrections of the order~$1/R$ tend to zero as~$R \rightarrow \infty$. Therefore, we may use the
computation rules
\begin{gather} \label{parules}
\pm \hat{F}_\u^{i0}(p_\u) |\vec{p}| - \hat{F}_\u^{i\alpha}(p_\u) p^\alpha = 0 \qquad \text{and} \qquad
\pm \hat{F}_\v^{i0}(p_\v) |\vec{p}| - \hat{F}_\v^{i\alpha}(p_\v) p^\alpha = 0
\end{gather}
(where we sum over~$\alpha=1,2,3$). Using these rules, one can get rid of
all factors~$\hat{p}^\alpha$. But factors~$\Delta p^\alpha$ and~$q^\alpha$ remain.
They can be treated with the following symmetry argument.
By assumption, the cutoff function~$\eta$ is spherically symmetric. Therefore,
carrying out the integrals over~$\Delta \vec{p}$ and~$\vec{q}$, the
spherical symmetry is broken only by the vector~$\vec{p}$.
Therefore, the spatial vector and tensor indices can be decomposed as
\begin{gather*}
\int_{\R^3} {\rm d}^3 \Delta p (\cdots) \Delta p^\alpha = a_1 p^\alpha, \\
\int_{\R^3} {\rm d}^3q (\cdots) q^\alpha = a_2 p^\alpha, \\
\int_{\R^3} {\rm d}^3\, \Delta p \int_{\R^3} {\rm d}^3q (\cdots) \Delta p^\alpha \Delta p^\beta
 = a_3 p^\alpha p^\beta + a_4 \delta^{\alpha \beta}, \\
\int_{\R^3} {\rm d}^3 \, \Delta p \int_{\R^3} {\rm d}^3q (\cdots) q^\alpha \Delta p^\beta = a_5 p^\alpha p^\beta + a_6 \delta^{\alpha \beta} ,\\
\int_{\R^3} {\rm d}^3 \, \Delta p \int_{\R^3} {\rm d}^3q (\cdots) q^\alpha q^\beta = a_7 p^\alpha p^\beta + a_8 \delta^{\alpha \beta}
\end{gather*}
with real constants~$a_1,\ldots, a_8$. The appearing indices~$p^\alpha$ and~$p^\beta$
can again be treated with the help of~\eqref{parules}.

In order to handle the remaining spatial tensor~$\delta_{\alpha \beta}$,
we make use of the fact that in the contributions to Lagrangian, the field
tensors are contracted to outer factors~$\xi$ (see Propositions~\ref{prplog} and~\ref{prpnolog}).
Since these factors are null vectors, it is obvious that a contribution to the
tensor~$(F_\u)_{ij} (F_\v)^j_{ k}$ which is proportional to the metric~$g_{ik}$ drops out.
In other words, only the trace-free part of the tensor comes up in our computations.
Therefore, we may set
\begin{gather}\label{tensor}
\big(\hat{F}_\u\big)_{\alpha j} \big(\hat{F}_\v\big)^j_{ \beta} \delta^{\alpha \beta} = \big(\hat{F}_\u\big)_{0j} \big(\hat{F}_\v\big)^j_{ 0} .
\end{gather}
After the above transformations, it remains to compute one scalar expressions.
On the other hand, knowing from the above arguments
that the corresponding tensor expression is spherically symmetric and only involves
the trace-free part of the field tensor squared, the form of this tensor expression is
determined uniquely from the scalar expression.

\subsubsection{Completing the computation}
For the further analysis of~\eqref{Bform}, it is important to take into account
the symmetries under the transformations~$u \rightarrow -u$ and~$v \rightarrow -v$.
Therefore, it is helpful to rewrite~$B$ as
\begin{gather} \label{Bformalt}
B = \frac{1}{2 |\vec{p} |} \int_0^1 {\rm d}u \int_0^{v_{\max}(u)} {\rm d}v
\sum_{s,s'=\pm1} V(\alpha, \beta)
 \big(\hat{F}_\u\big)_{ij}(p_\u) \big(\hat{F}_\v\big)^j_{ k}(p_\v) \hat{\mathfrak{K}} (R \omega, R \vec{k} ) \big|_{(su, s'v)} .
\end{gather}
Now a straightforward computation shows that the logarithmic terms in~\eqref{logpole} drop out
of the integrand. This is best verified right after applying the computation rules~\eqref{parules}.
Then the leading contributions as~$R \rightarrow \infty$ are of the form
\[ \sim \int_0^{v_{\max}} \frac{v}{v_{\max}^2} {\rm d}v . \]
Thus the limit~$R \rightarrow \infty$ exists.
Furthermore, due to spherical symmetry of~$\hat{\mathfrak{K}}$ and~$\hat{\eta}$,
the value of this limit is independent of~$\vec{p}$. Moreover, a straightforward computation shows that
this limit is in general non-zero. We thus obtain
\[ A \simeq \sum_{s=\pm}
\int_{s \R^+} \frac{{\rm d}\omega_\u}{2 \pi} \int_{s \R^-} \frac{{\rm d}\omega_\v}{2 \pi}
\int_{\R^3} \frac{{\rm d}^3p}{(2 \pi)^3} \big(\hat{F}_\u\big)_{ij} (\omega_\u, \vec{p} )
\big(\hat{F}_\v\big)^j_{ k} (\omega_\v, -\vec{p} ) \frac{g(s)}{|\vec{p}|} , \]
where our notation with~$-\R^- = \R^+$ implements that, according to~\eqref{ouvdef},
$\omega_\u$ and~$\omega_\v$ have opposite signs, and~$s$ is the sign of~$\omega_\u$.
Finally, the function~$g(s)$ is determined from the following symmetry consideration:
According to~\eqref{Cnolog}, for the contributions without logarithmic poles
the kernel~${\mathfrak{K}}(\xi)$ is even under the transformation~$\xi \rightarrow -\xi$.
Likewise, also its Fourier transform~$\hat{\mathfrak{K}}(p)$ is even under sign flips~$p \rightarrow -p$
(as is also obvious from the left of Fig.~\ref{figK12}).
Conversely, the contributions with logarithmic poles, the kernels~${\mathfrak{K}}$
and~${\hat{\mathfrak{K}}}$ are odd (see~\eqref{Clog} and the upper plot in Fig.~\ref{figS12}).
Keeping in mind that with the notation~$\simeq$ we disregard real prefactors, we thus obtain
\[ g(s) = \left\{ \begin{matrix} 1 \\ {\rm i} s \end{matrix} \right\} \qquad \text{for contributions} \
\left\{ \begin{matrix} \text{without} \\ \text{with} \end{matrix} \right\}
\ \text{logarithmic poles} . \]
Moreover, as explained after~\eqref{tensor}, the correct tensor expression is obtained by the
replacement
\begin{gather} \label{tensorrep}
\big(\hat{F}_\u\big)_{ij} \big(\hat{F}_\v\big)^j_{ k} \xi^i \xi^k \rightarrow
\left( \big(\hat{F}_\u\big)_{0i} \big(\hat{F}_\v\big)_0^{ i} -\frac{1}{4} \big(\hat{F}_\u\big)_{ij}
 \big(\hat{F}_\v\big)^{ij} \right) \xi^0 \xi^0 .
\end{gather}
We thus obtain the following result:
\begin{Thm} \label{thmbose}
The contributions with logarithmic poles $($as computed in Proposition~{\rm \ref{prplog})}
give rise to the symplectic form~\eqref{symplecticbose}.
On the other hand, the contributions without logarithmic poles $($as computed in Proposition~{\rm \ref{prpnolog})}
give rise to the inner product~\eqref{bose}.
\end{Thm}

\begin{proof} In the case~$g=1$, we obtain precisely the inner product~\eqref{bose}.
In the case~$g=is$, on the other hand, the second summand in~\eqref{tensorrep} drops out because the
integrand is anti-symmetric under the replacements~$\omega_\u \rightarrow-\omega_\u$,
$\omega_\v \rightarrow-\omega_\v$ and~$\vec{p} \rightarrow -\vec{p}$.
We thus obtain
\begin{gather*}
A \simeq \sum_{s=\pm}
\int_{s \R^+} \frac{{\rm d}\omega_\u}{2 \pi} \int_{s \R^-} \frac{{\rm d}\omega_\v}{2 \pi}
\int_{\R^3} \frac{{\rm d}^3p}{(2 \pi)^3} \big(\hat{F}_\u\big)_{0k} \big|_{(\omega_\u, \vec{p})}
\big(\hat{F}_\v\big)_0^{ k} \big|_{(\omega_\v, -\vec{p})} \frac{{\rm i}}{\omega_\u} \\
\hphantom{A}{} = \frac{1}{2} \int_{\R^+ \times \R^- \cup \R^- \times \R^+}
\frac{{\rm d}\omega_\u}{2 \pi} \frac{{\rm d}\omega_\v}{2 \pi}
\int_{\R^3} \frac{{\rm d}^3p}{(2 \pi)^3} \big(\hat{F}_\u\big)_{0k} \big|_{(\omega_\u, \vec{p})}
\big(\hat{F}_\v\big)_0^{ k} \big|_{(\omega_\v, -\vec{p})} \left( \frac{{\rm i}}{\omega_\u}
- \frac{{\rm i}}{\omega_\v} \right) \\
\hphantom{A}{} = \frac{1}{2} \int_{-\infty}^\infty \frac{{\rm d}\omega_\u}{2 \pi} \int_{-\infty}^\infty \frac{{\rm d}\omega_\v}{2 \pi}
\int_{\R^3} \frac{{\rm d}^3p}{(2 \pi)^3} \big(\hat{F}_\u\big)_{0\alpha} \big|_{(\omega_\u, \vec{p})}
\big(\hat{F}_\v\big)_0^{ \alpha} \big|_{(\omega_\v, -\vec{p})} \left( \frac{{\rm i}}{\omega_\u}
- \frac{{\rm i}}{\omega_\v} \right) ,
\end{gather*}
where in the last step we used that the integrand is anti-symmetric.
Next, using that~$F_{0\alpha} = \partial_0 A_\alpha - \partial_\alpha A_0$ and rewriting the
derivatives as factors of the momentum variables, we obtain
\begin{gather*}
A = \frac{1}{2} \int_{-\infty}^\infty \frac{{\rm d}\omega_\u}{2 \pi} \int_{-\infty}^\infty \frac{{\rm d}\omega_\v}{2 \pi}
\int_{\R^3} \frac{{\rm d}^3p}{(2 \pi)^3} \\
\hphantom{A =}{} \times
\bigg( (\hat{A}_\u)_\alpha \big|_{(\omega_\u, \vec{p})} \big(\hat{F}_\v\big)_0^{ \alpha} \big|_{(\omega_\v, -\vec{p})}
+ \frac{p_\alpha}{\omega_u} \big(\hat{A}_\u\big)_0 \big|_{(\omega_\u, \vec{p})} \big(\hat{F}_\v\big)_0^{ \alpha}
\big|_{(\omega_\v, -\vec{p})} \\
\hphantom{A =}{} -\big(\hat{F}_\u\big)_{0\alpha} \big|_{(\omega_\u, \vec{p})} \big(\hat{A}_\v\big)^\alpha \big|_{(\omega_\v, -\vec{p})}
-\big(\hat{F}_\u\big)_{0\alpha} \big|_{(\omega_\u, \vec{p})} \frac{p^\alpha}{\omega_v} \big(\hat{A}_\v\big)_0
\big|_{(\omega_\v, -\vec{p})} \bigg) .
\end{gather*}
Next, the homogeneous Maxwell equations imply that~$p_\alpha \big(\hat{F}_\v\big)_0^{ \alpha}
= 0 = p^\alpha \big(\hat{F}_\u\big)_{0\alpha}$. We conclude that
\begin{gather*}
A = \frac{1}{2} \int_{-\infty}^\infty \frac{{\rm d}\omega_\u}{2 \pi} \int_{-\infty}^\infty \frac{{\rm d}\omega_\v}{2 \pi}
\int_{\R^3} \frac{{\rm d}^3p}{(2 \pi)^3} \\
\hphantom{A =}{} \times
\Big( (\hat{A}_\u)_\alpha \big|_{(\omega_\u, \vec{p})} \big(\hat{F}_\v\big)_0^{ \alpha} \big|_{(\omega_\v, -\vec{p})}
-\big(\hat{F}_\u\big)_{0\alpha} \big|_{(\omega_\u, \vec{p})} \big(\hat{A}_\v\big)^\alpha \big|_{(\omega_\v, -\vec{p})} \Big) \\
\hphantom{A}{} = \frac{1}{2} \int_{\R^3}
\Big( (A_\u)_\alpha(0,\vec{x}) (F_\v)_0^{ \alpha}(0,\vec{x})
- (F_\u)_{0\alpha}(0,\vec{x}) (A_\v)^\alpha(0,\vec{x}) \Big) {\rm d}^3x .
\end{gather*}
This is precisely the symplectic form~\eqref{symplecticbose}.
\end{proof}

We point out that the prefactors in the resulting formulas may well depend on the choice of~$\eta$,
in agreement with our general concept that the regularization has a physical significance and may
determine coupling constants, masses and prefactors in conservation laws.

\subsection[The contributions ${\sim} \delta^{-4} \cdot F^2 \varepsilon/t$]{The contributions $\boldsymbol{{\sim} \delta^{-4} \cdot F^2 \varepsilon/t}$} \label{secdF21}

We now come to the analysis of the contributions which involve one factor~$\varepsilon/t$.
Compared to the contributions in Section~\ref{secdF22}, these contributions are larger by a scaling
factor~$t/\varepsilon$, i.e.,
\begin{gather*}
\L(x,y) \sim F^2 \xi \xi (\deg=5) \frac{\varepsilon^2 t^2}{\delta^4} \frac{\varepsilon}{t}
 \sim F^2 \frac{1}{\varepsilon \delta^4} \frac{1}{t^2} \delta(|t|-r)
 \sim F^2 \frac{1}{\varepsilon \delta^4} \frac{1}{t} {\rm i} K_0(\xi) .
\end{gather*}

We first consider the contributions with logarithmic poles.
\begin{Prp} \label{prpnolog2}
The contributions to the kernel of the fermionic projector as computed in Lemma~{\rm \ref{lemmalog}}
affect the second variation of the Lagrangian
to the order~${\sim} \delta^{-4} \cdot F^2 \varepsilon/t$ by a term of the form
\begin{gather*}
 ( \nabla_{1, \u} - \nabla_{2, \u} )
 ( \nabla_{1, \v} + \nabla_{2, \v} ) \L(x,y)\\
\qquad{} \simeq \int_{-\infty}^\infty {\rm d}\alpha \int_{-\infty}^\infty {\rm d}\beta\, J(\alpha,\beta)
 (F^\u)_{ij}|_{\alpha y + (1-\alpha) x} (F^\v)^j_{ k}|_{\beta y + (1-\beta) x} \xi^i \xi^k
 \frac{1}{\varepsilon \delta^4} \frac{1}{t^3} {\rm i} K_0(\xi) \log|\varepsilon t|
\end{gather*}
with~$J(\alpha, \beta)$ as in~\eqref{Jform}.
\end{Prp}
\begin{proof} Similar to~\eqref{extraterm}, we can write the contribution as
\begin{gather*}
 ( \nabla_{1, \u} - \nabla_{2, \u} ) ( \nabla_{1, \v} + \nabla_{2, \v} ) \L(x,y) \\
\qquad{} \asymp \int_{-\infty}^\infty {\rm d}\alpha \int_{-\infty}^\infty {\rm d}\beta \, J(\alpha, \beta)
(F^\u)_{ij}|_{\alpha y + (1-\alpha) x} (F^\v)^j_{ k}|_{\beta y + (1-\beta) x} \xi^i \xi^k \\
 \qquad \quad {} \times
 \frac{1}{\varepsilon \delta^4} \frac{1}{t^3} {\rm i} K_0(\xi)
\re \big( C T^{(1)}_{[0]} \big) .
\end{gather*}
Compared to the proof of Proposition~\ref{prplog},
the only difference is that the additional factor~$t/\varepsilon$ flips the symmetry
under the replacements~$x \leftrightarrow y$.
Consequently, from the factor~$T^{(1)}$ in~\eqref{T1form}
only the {\em{real part}} contributes. Evaluating the resulting logarithm on the light cone
with a~regularization on the scale~$\varepsilon$, the computation
\[ \big( (t-{\rm i} \varepsilon)^2 - |\vec{x}|^2 \big)\big|_{t=|\vec{x}|} = -2 {\rm i} \varepsilon t - \varepsilon^2 , \]
shows that we obtain a scaling factor~$\log \big|\xi^2\big| \simeq \log | \varepsilon t|$.
\end{proof}

\begin{Prp} \label{prpsympdiv}
Introducing an infrared regularization~\eqref{infrared},
the contributions to the conserved surface layer integral~\eqref{osicombined}
of the order~${\sim} \delta^{-4} \cdot F^2 \varepsilon/t$ give the symplectic form,
\begin{gather}
 \int_{-\infty}^{t_0} {\rm d}t \int_{\R^3} {\rm d}^3x \int_{t_0}^\infty {\rm d}t' \int_{\R^3} {\rm d}^3y\,
 ( \nabla_{1, \u} - \nabla_{2, \u} )
 ( \nabla_{1, \v} + \nabla_{2, \v} ) \L(x,y) \notag \\
\qquad{} \asymp
\frac{c}{\varepsilon \delta^4}
\int_{\R^3} \big( A_\u^i (F_\v)_{i0} - A_\v^i (F_\u)_{i0}\big) \, {\rm d}^3x , \label{sympdiv}
\end{gather}
where the constant~$c$ diverges if the infrared regularization is removed.
\end{Prp} \noindent
The significance of this conservation law will be explained and discussed
in Remark~\ref{remsympdiv}.
\begin{proof}[Proof of Proposition~\ref{prpsympdiv}]
The proof consists of two parts: proving that the surface layer integral is conserved~\eqref{integrated}
and showing that the resulting integral in~\eqref{osit} gives~\eqref{sympdiv}.

In order to show conservation of the surface layer integral~\eqref{integrated},
we first proceed exactly as in Section~\ref{secreduce} to reduce to bounded line integrals.
The remaining task is to prove that the equation~\eqref{logconserve2} holds, where~${\mathfrak{K}}$
is the kernel in Proposition~\ref{prpnolog2},
\[ {\mathfrak{K}}(\xi) = \xi^i \xi^k \frac{1}{\varepsilon \delta^4} \frac{1}{t^3} {\rm i} K_0(\xi) \log|\varepsilon t| . \]
As in Section~\ref{secbounded}, we introduce an infrared regularization and
analyze the expression~\eqref{Adef} with~$B$ according to~\eqref{Bform}.
After treating the spatial tensor indices as explained after~\eqref{parules}, it suffices
to consider the case that both indices are timelike, i.e.,
\begin{gather} \label{K00symm}
{\mathfrak{K}}(\xi) = \frac{1}{\varepsilon \delta^4} \frac{1}{t} {\rm i} K_0(\xi) \log|\varepsilon t| .
\end{gather}
Now we can use the following symmetry argument:
We consider the transformation
\begin{gather*}
v \rightarrow -v ,\qquad \Delta \vec{p} \rightarrow -\Delta \vec{p} ,\qquad
\vec{q} \rightarrow -\vec{q} ,
\end{gather*}
keeping~$u$ unchanged. This corresponds to the transformation
\[ \alpha \leftrightarrow \beta . \]
Obviously, the function~$V(\alpha, \beta)$ is anti-symmetric under these transformations.
According to~\eqref{Bformalt} and the formulas for~$R \omega$ and~$R \vec{k}$
in~\eqref{Rom} and~\eqref{Rk}, the argument of~${\mathfrak{K}}$ flips sign.
According to~\eqref{K00symm}, ${\mathfrak{K}}$ and thus also~$\hat{\mathfrak{K}}$ are even,
so that~${\mathfrak{K}}$ is left invariant under the above transformations.
Due to spherical symmetry, the same is true for the factors~$\hat{\eta}$ in~\eqref{Adef}.
It follows that~$A$ vanishes, proving that the condition for conservation~\eqref{integrated}
holds.

Knowing that the surface layer integral is conserved,
according to Lemma~\ref{lemmaconservedform} we can again compute it
by evaluating the integral~\eqref{osival}. Therefore, our task is to
analyze again the integral in~\eqref{osit}, but now with the kernel
\begin{gather} \label{Koncone}
{\mathfrak{K}}(\xi) = \frac{1}{\varepsilon \delta^4} {\rm i} K_0(\xi) \log |\varepsilon t| .
\end{gather}
We decompose the logarithm as~$\log |\varepsilon t|=\log \varepsilon + \log |t|$
and consider the resulting terms after each other.
For the contribution involving~$\log \varepsilon$, after
transforming to momentum space, the resulting kernel is supported on the mass shell.
Therefore, the factor~$\hat{\mathfrak{K}}$ in~\eqref{logconserve} vanishes
no matter if~$p_\u$ and~$p_\v$ are supported on the upper or lower mass shell,
respectively. Hence we do not get a contribution to~\eqref{osit}.

It remains to consider the contribution involving the factor~$\log t$.
Transforming to momentum space, the resulting kernel~$\hat{\mathfrak{K}}$
is supported also outside the mass shell. Therefore, we need to compute
the integrals as explained in Section~\ref{secbounded} by introducing
an infrared regularization~\eqref{infrared}. There seems no symmetry argument
showing that the resulting integrals vanish. Therefore, we get a contribution
which is in general non-zero. Since~$V(\alpha, \beta)$ is anti-symmetric in~$\alpha$ and~$\beta$,
this contribution is anti-symmetric in~$\u$ and~$\v$.
This shows that we obtain again the symplectic form.
However, since~$\hat{\mathfrak{K}}$ is homogeneous of degree minus two,
the contribution diverges in the limit~$R \rightarrow \infty$ when the infrared regularization
is removed. This gives the result.
\end{proof}

For the contributions without logarithms, we have the following result:
\begin{Prp} \label{prp414}
If the contributions without logarithms
in Proposition~{\rm \ref{prpnolog}} to the conserved surface layer integral~\eqref{osicombined}
of the order~${\sim} \delta^{-4} \cdot F^2 \varepsilon/t$
are conserved in time, then they vanish.
\end{Prp}
\begin{proof} Assuming that the surface layer integral is conserved, we can again
compute it by evaluating the integral~\eqref{osival}. Therefore, our task is to show
again that~\eqref{logconserve2} holds, but now for the kernel
\[ {\mathfrak{K}}(\xi) = \frac{1}{\varepsilon \delta^4} {\rm i} K_0(\xi) . \]
In momentum space, this kernel is again supported on the mass shell.
Therefore, we can argue exactly as after~\eqref{Koncone} to conclude that
the resulting contributions to the surface layer integral vanish.
\end{proof}

The remaining question is why the contributions to the surface layer integral
considered in the previous proposition are conserved in time.
This is a subtle and difficult question. With present knowledge, it seems necessary to
arrange conservation by imposing a regularization condition.
For brevity, we only sketch the construction and point out the reason for the regularization condition:
Due to the additional factor~$t/\varepsilon$, the kernel in
Proposition~\ref{prpnolog} must be modified to
\begin{gather} \label{kernelflip}
\xi^i \xi^k \frac{1}{\varepsilon \delta^4} \frac{1}{t^3} {\rm i} K_0(\xi) .
\end{gather}
Since this kernel is even in~$\xi$, the corresponding line integrals
must be odd when exchanging~$x$ with~$y$.
Therefore, instead of~\eqref{Idef}, the contributions to~$P(x,y)$ of first order in the field strength
gives rise to the integrand
\[ 
I(\alpha, \beta) = \big( c_1 ( 2 \alpha \beta - \alpha - \beta ) + c_2 \big)
 ( \chi_{(1,\infty)}(\alpha) - \chi_{(-\infty,0)}(\alpha) )
\chi_{(0,1)}(\beta) , \]
involving {\em{two}} free parameters~$c_1$ and~$c_2$.
These two parameters could be determined by a~lengthy computation,
which we do not want to enter here.
Arguing only with the symmetry properties does not determine~$I(\alpha, \beta)$
up to a prefactor. As a consequence, it is impossible to evaluate the
integrals obtained after the infrared regularization~\eqref{Adef} and~\eqref{Bformalt}
(there seems no symmetry argument which would imply that~\eqref{Bformalt} vanishes).
Moreover, when adapting Proposition~\ref{prpnolog}, due to the reversed symmetry
of the kernel~\eqref{kernelflip}, we do need to take into account the
contributions of second order in perturbation theory~\eqref{P2cont}.
This can be done similar as explained in Lemma~\ref{lemmalog}
for the logarithmic contributions. Moreover, proceeding similar as
in Lemmas~\ref{lemmaantisymm} and~\ref{lemmaUterm}, one could
again reduce to bounded line integrals. However, it is not obvious what
would be the symmetries of the resulting line integrals as well as the resulting
contributions to~\eqref{Adef} and~\eqref{Bformalt}.
Therefore, without considerably higher computational efforts,
it is impossible to compute~\eqref{Adef} for the contributions without logarithms.
But it can clearly be arranged by a~suitable regularization condition
that~$A$ vanishes.

\section{Computation of fermionic conserved surface layer integrals} \label{secosif}
In this section we shall compute the surface layer integral~\eqref{osicombined}
for fermionic jets.
We use the same strategy as in the computation of the bosonic surface layer integrals:
We first show that, under suitable assumptions on the fermionic wave functions
as well as on the regularization, the surface layer integrals are conserved, meaning
that the integral~\eqref{integrated} vanishes (see Section~\ref{secfermiconserve}).
This makes it possible to again express the surface layer integral using the
formula of Lemma~\ref{lemmaconservedform} (see Section~\ref{secfermicompute}).
Sections~\ref{secfd4}--\ref{secconv} are devoted to preparatory calculations and
the analysis of the scaling behavior.

\subsection[The contributions ${\sim} \delta^{-4} \cdot J^2$]{The contributions $\boldsymbol{{\sim} \delta^{-4} \cdot J^2}$} \label{secfd4}
In Section~\ref{secdJ} we already considered the contribution where
one of the brackets in~\eqref{notation}
contains the contribution~${\sim} \delta^{-4}$ given in~\eqref{d4}, whereas the other
bracket involves contributions by the Maxwell or Dirac currents.
We now consider the contributions~${\sim} \delta^{-4}$ which
are quadratic in the currents.

We first give the scaling behavior:
\begin{gather}
 ( \nabla_{1, \u} - \nabla_{2, \u} ) ( \nabla_{1, \v} + \nabla_{2, \v} ) \L(x,y)
 \simeq \frac{1}{\delta^4} (\deg=1)
 ( \nabla_{1, \u} - \nabla_{2, \u} ) ( \nabla_{1, \v} + \nabla_{2, \v} ) |\lambda^{xy}_{ncs}| \notag \\
\qquad {} \simeq J^2 \xi \xi \frac{1}{\delta^4} (\deg=2) \simeq J^2 \xi \xi \frac{1}{\delta^4} \frac{1}{\varepsilon t^2} \delta(|t|-r)
\simeq J^2 \xi \xi \frac{1}{\delta^4 \varepsilon} \frac{1}{t} {\rm i} K_0(\xi) . \label{ddJscale}
\end{gather}
In order to compute the contributions in more detail, we first note that
\begin{gather}
\nabla_{1,\u} P(x,y) = -\big|\delta \psi^\u(x)\Sr \Sl \psi^\u(y)\big|, \label{nabP1} \\
\nabla_{2,\u} P(x,y) = -\big|\psi^\u(x) \Sr \Sl \delta \psi^\u(y)\big|, \label{nabP2} \\
\nabla_{1,\u} \nabla_{2,\v} P(x,y) = -\big|\delta \psi^\u(x) \Sr \Sl \delta \psi^\v(y)\big| ,\nonumber
\end{gather}
where~$\psi^\u$ denotes the physical wave function which is perturbed by the jet~$\u$.
Exactly as explained after~\eqref{Ld2Jj} for the contribution~${\sim} \delta^{-4} J$,
the term~$\nabla_{1,\u} \nabla_{2,\v} P(x,y)$ does not contribute to the Lagrangian
in the formalism of the continuum limit
(we remark, however, that, going beyond the error terms of the form~\eqref{ap1},
it does give rise to a finite contribution to the Lagrangian,
which could be computed similar to the contribution by the Dirac current in~\cite[Section~5.2]{noether}).

It remains to analyze the effect of~\eqref{nabP1} and~\eqref{nabP2}
on the Lagrangian. Recall that one of the brackets in~\eqref{notation} involves a factor~$\delta^{-4}$,
whereas the other bracket involves the Dirac current squared.
Therefore, we need to compute the second order perturbation of the absolute values of the eigenvalues
of the closed chain:

\begin{Lemma} \label{lemmadel}
In the formalism of the continuum limit, for any~$i,j \in \{1,2\}$ and~$n \in \{2,\ldots, 8\}$,
\[ \nabla_{i,\u} \nabla_{j,\v} \big| \lambda^{xy}_{ncs} \big|^2 = 0 \]
$($where~$i,j=1$ denotes the $x$-derivative, and~$i,j=2$ denotes the $y$-derivative$)$.
\end{Lemma}
\begin{proof}
We consider the different contributions after each other:

(i) Second order perturbation of the eigenvalues:
In the second order perturbation calculation of the eigenvalues, we consider the
linear perturbation of the closed chain
\begin{gather} \label{Alin}
\nabla A_{xy} = \nabla P \cdot P^* + P \cdot \nabla P^* ,
\end{gather}
where for ease in notation we omitted the arguments~$(x,y)$ and wrote~$P^*=P(y,x)$.
Clearly, it suffices to consider the eigenvalues~$\lambda^{xy}_{nL+}$, because
the other eigenvalues are obtained by flipping the chirality and taking the complex conjugate.
The second order perturbation of~$\lambda^{xy}_{nL+}$ is given by
(see~\cite[Section~2.6.3]{cfs})
\begin{gather} \label{lambda2}
\Delta \lambda^{xy}_{nL+} = \sum_{c=L,R} \frac{1}{\lambda^{xy}_{nL+} - \lambda^{xy}_{nc-}}
 \Tr\big( F^{xy}_{nL+} \nabla A_{xy} F^{xy}_{nc-} \nabla A_{xy} \big) ,
\end{gather}
where~$F^{xy}_{ncs}$ are the spectral projectors of
the closed chain of the vacuum defined by the relations
(for details see~\cite[Sections~2.6.1 and~5.2.4]{cfs})
\[ A_{xy} F^{xy}_{ncs} = \lambda_{ncs}^{xy} F^{xy}_{ncs} ,\qquad
\big(F^{xy}_{nc+} \big)^* = F^{xy}_{n\bar{c}-} \qquad \text{and} \qquad
\sum_{n=1}^8 \sum_{s=\pm} \sum_{c=L,R} F^{xy}_{ncs} = \1 \]
(where~$\bar{L}=R$ and~$\bar{R}=L$).
Note that we only consider perturbations which are diagonal in the sector index~$n$.
For ease in notation, the constant index~$n$ will be omitted in what follows.
Then the spectral projectors can be written more explicitly as
(see~\cite[equation~(2.6.19)]{cfs})
\[ F^{xy}_{cs} = \chi_c \frac{1}{2} \left( \1 \pm \frac{[\slashed{\xi}, \overline{\slashed{\xi}}]}{z-\overline{z}} \right)
+ \slashed{\xi} (\deg \leq 0) + (\deg < 0) . \]
Thus our task is to compute traces of the form
\begin{gather} \label{2trace}
\Tr \big( F^{xy}_{L+} \nabla A_{xy} F^{xy}_{c-} \nabla A_{xy} \big) .
\end{gather}
Substituting~\eqref{Alin} into~\eqref{2trace} and expanding, we get
summands involving different combinations of~$\nabla P$ and $\nabla P^*$.
We consider these summands after each other.
The contribution involving two factors of~$\nabla P^*$ to the above trace is
proportional to
\begin{gather} \label{zerotrace}
\Tr \big( F^{xy}_{L+} \slashed{\xi} \nabla P^* F^{xy}_{c-}
\slashed{\xi} \nabla P^* \big) .
\end{gather}
This expression vanishes for the following reason:
To the considered degree on the light cone, every factor~$\nabla P^*$ must be contracted with a factor~$\xi$.
Therefore, we may treat the terms $F^{xy}_{L+} \slashed{\xi}$
and~$F^{xy}_{c-} \slashed{\xi}$
as scalar multiples of an outer factor~$\slashed{\xi}$
(as defined after~\eqref{eq54}).
Hence we may use the relation~$F_{L+} \slashed{\xi}=0$ (see~\cite[equation~(2.6.17)]{cfs})
to infer that~\eqref{zerotrace} vanishes.
The summands involving two factors of~$\nabla P$ can be treated similarly.

It remains to consider the summands
which involve one factor~$\nabla P$ and one factor~$\nabla P^*$,
like for example
\[ \Tr \big( F^{xy}_{L+} \slashed{\xi} \nabla P^* F^{xy}_{c-} \nabla P
\slashed{\xi} \big) . \]
Applying the contraction rules, one sees that either the
tensor indices of the currents are contracted with each other,
or else the totally anti-symmetric $\epsilon$-tensor appears.
In both cases, the resulting contributions are by a factor~$\varepsilon/t$ smaller
than the contributions to be taken into account in the formalism of the continuum limit.

(ii) First order perturbation of the eigenvalues:
One contribution is obtained by taking the absolute value
of the linearly perturbed eigenvalue, i.e.,
\begin{gather}
 \big| \lambda^{xy}_{L+} + \nabla \lambda^{xy}_{L+} \big|
= \big( \big|\lambda^{xy}_{L+}\big|^2 + 2 \re \big( \nabla \lambda^{xy}_{L+} \overline{\lambda^{xy}_{L+}} \big)
+ \big|\nabla \lambda^{xy}_{L+} \big|^2 \big)^{\frac{1}{2}} \label{firstorder} \\
 \hphantom{\big| \lambda^{xy}_{L+} + \nabla \lambda^{xy}_{L+} \big|}{}
 = \big|\lambda^{xy}_{L+}\big| + \frac{1}{2 \big|\lambda^{xy}_{L+}\big|}
\big( 2 \re \big( \nabla \lambda^{xy}_{L+} \overline{\lambda^{xy}_{L+}} \big)
+ \big|\nabla \lambda^{xy}_{L+} \big|^2 \big)
-\frac{\big(\re \big( \nabla \lambda^{xy}_{L+}
\overline{\lambda^{xy}_{L+}} \big) \big)^2}{2 \big|\lambda^{xy}_{L+}\big|^3} .\notag
\end{gather}
The linear perturbation of the eigenvalues is compensated by the corresponding
Maxwell term (see Section~\ref{secJJ}). As a consequence, the
contribution~\eqref{firstorder}, which involves the linear perturbations quadratically,
also vanishes.
Therefore, all the terms in~\eqref{firstorder} vanish.

It remains to consider the contribution
by a first order perturbation calculation in~$\Delta A_{xy}$,
where~$\Delta A_{xy}$ is quadratic in~$\nabla P$, i.e.,
\[
\Delta A_{xy} = \nabla P \cdot \nabla P^* . \]
In this case, computing the trace in the formula
\begin{gather} \label{foDelta}
\Delta \lambda^{xy}_{L+} = \Tr \big( F^{xy}_{L+} \Delta A_{xy} \big) ,
\end{gather}
either the vector components of~$\nabla P$ and~$\nabla P^*$ are contracted
with each other, or else the commutator~$[\slashed{\xi}, \overline{\slashed{\xi}}]$
in~$F^{xy}_{L+}$ comes into play (see~\cite[equation~(2.6.16)]{cfs}).
In both cases, the resulting contributions are by a factor~$\varepsilon/t$ smaller
than the contributions taken into account in the formalism of the continuum limit.

This concludes the proof.
\end{proof}

Applying this finding to the Lagrangian immediately gives the following result:
\begin{Corollary} \label{corfermid}
The contributions~${\sim} \delta^{-4} \cdot J^2$ to the Lagrangian vanish.
\end{Corollary}

\subsection[The contributions ${\sim} \delta^{-4} \cdot J^2 \varepsilon/t$]{The contributions $\boldsymbol{{\sim} \delta^{-4} \cdot J^2 \varepsilon/t}$} \label{secd4JJ}

Compared to~\eqref{ddJscale}, the contributions~${\sim} \delta^{-4} \cdot J^2 \varepsilon/t$
involve an additional factor~$\varepsilon/t$, i.e.
\[ ( \nabla_{1, \u} - \nabla_{2, \u} ) ( \nabla_{1, \v} + \nabla_{2, \v} ) \L(x,y)
\simeq J^2 \xi \xi \frac{1}{\delta^4} \frac{1}{t^2} {\rm i} K_0(\xi) . \]
Here the factors~$\xi$ no longer need to be treated as outer factors.
We now compute this contribution in detail.
Since the factor~$\delta^{-4}$ arises in the neutrino sector, whereas the
currents are in the charged sectors, \eqref{DeltaLag} simplifies to
\[ \Delta \L[A_{xy}] = \frac{1}{4} \sum_{n=1}^7 \sum_{c,c' =L,R}
\Delta \big|\lambda^{xy}_{8c'+} \big| \Delta \big|\lambda^{xy}_{nc+} \big| . \]
This shows in particular that we may simplify the following calculations
by summing over the chiral index~$c$.

\begin{Lemma} \label{lemmadel2}
The perturbation of the absolute values of the eigenvalues of the closed chain
in the charged sectors of the order~${\sim} J^2 \varepsilon/t$ are given by
\begin{gather}
\sum_{c=L,R} \Delta \big| \lambda^{xy}_{nc+} \big|^2
\asymp -\Tr\big( F^{xy}_{n+} (\nabla P) P^* F^{xy}_{n-} (\nabla P) P^* \big) + \text{\rm c.c.} \label{s1} \\[-2ex]
\hphantom{\sum_{c=L,R} \Delta \big| \lambda^{xy}_{nc+} \big|^2\asymp}{}
-\Tr\big( F^{xy}_{n+} (\nabla P) P^* F^{xy}_{n-} P (\nabla P^*) \big) \label{s3} \\
\hphantom{\sum_{c=L,R} \Delta \big| \lambda^{xy}_{nc+} \big|^2\asymp}{}
-\Tr\big( F^{xy}_{n+} P (\nabla P^*) F^{xy}_{n-} (\nabla P) P^* \big) \label{s4} \\
\hphantom{\sum_{c=L,R} \Delta \big| \lambda^{xy}_{nc+} \big|^2\asymp}{}
+ \Tr \big( I_n P^* P (\nabla P) (\nabla P^*) \big) \label{s5} ,
\end{gather}
valid for any~$n \in \{2,\ldots, 8\}$.
\end{Lemma}
\begin{proof} We again leave out the sector index~$n$.
Also, for ease in notation, we again omit the arguments~$(x,y)$ and write~$P^*=P(y,x)$.
We must be careful to apply only those computation rules which are valid
to higher order in~$\varepsilon/t$. In particular, we may still use the relations
\[
F^{xy}_{L+} P = P F^{xy}_{R-} ,\qquad
F^{xy}_{L+} P^* = P^* F^{xy}_{R-} . \]
Moreover, the products~$F^{xy}_{L-} P$ and~$F^{xy}_{L-} P^*$ are multiples of each other,
\begin{gather} \label{FPPs}
F^{xy}_{L-} P = c F^{xy}_{L-} P^* \qquad \text{with~$c \in \C$} .
\end{gather}
This follows from the general fact that the operator products on the left and right
have the same one-dimensional image.
Moreover, they are both vectorial. Hence this vector must be the same null vector.
Alternatively, the relation~\eqref{FPPs} can be verified by a direct computation. Indeed,
\begin{gather}
P = \frac{i}{2} \slashed{\xi} T^{(-1)}_{[0]} ,\qquad
P^* = -\frac{i}{2} \overline{\slashed{\xi}} \overline{T^{(-1)}_{[0]}}, \nonumber\\
F_\pm = \frac{\1}{2} \pm \frac{[\slashed{\xi}, \overline{\slashed{\xi}}]}{2d} \qquad \text{with} \qquad
d := 2 \sqrt{(\xi \overline{\xi})^2 - \xi^2 \overline{\xi}^2}, \label{ddef} \\
F_{-} \overline{\slashed{\xi}} =
\frac{\overline{\slashed{\xi}}}{2} - \frac{[\slashed{\xi}, \overline{\slashed{\xi}}] \overline{\slashed{\xi}}}{2d}
= \frac{\overline{\slashed{\xi}}}{2} - \frac{1}{2d}
 \big( 2 \overline{\xi}^2 \slashed{\xi} - 2 (\xi \overline{\xi}) \overline{\slashed{\xi}} \big) \notag \\
\hphantom{F_{-} \overline{\slashed{\xi}}}{} = \frac{\overline{\slashed{\xi}}}{2} -\frac{\overline{\xi}^2}{d}
 \slashed{\xi} +\frac{\xi \overline{\xi}}{d} \overline{\slashed{\xi}}
= \left( \frac{1}{2} +\frac{\xi \overline{\xi}}{d} \right) \overline{\slashed{\xi}}
-\frac{\overline{\xi}^2}{d} \slashed{\xi}, \nonumber\\ 
F_{-} \slashed{\xi} =
\frac{\slashed{\xi}}{2} - \frac{[\slashed{\xi}, \overline{\slashed{\xi}}] \slashed{\xi}}{2d}
= \frac{\slashed{\xi}}{2} - \frac{1}{2d}
 \big( {-}2 \xi^2 \overline{\slashed{\xi}} + 2 (\xi \overline{\xi}) \slashed{\xi} \big) \notag \\
\hphantom{F_{-} \slashed{\xi}}{} = \frac{\slashed{\xi}}{2} + \frac{\xi^2}{d}
 \overline{\slashed{\xi}} - \frac{\xi \overline{\xi}}{d} \slashed{\xi}
= \left( \frac{1}{2} - \frac{\xi \overline{\xi}}{d} \right) \slashed{\xi} + \frac{\xi^2}{d} \overline{\slashed{\xi}}, \nonumber\\
F_{+} \slashed{\xi} =
\frac{\slashed{\xi}}{2} + \frac{[\slashed{\xi}, \overline{\slashed{\xi}}] \slashed{\xi}}{2d}
= \frac{\slashed{\xi}}{2} + \frac{1}{2d}
 \big( {-}2 \xi^2 \overline{\slashed{\xi}} + 2 (\xi \overline{\xi}) \slashed{\xi} \big) \notag \\
\hphantom{F_{+} \slashed{\xi}}{} = \frac{\slashed{\xi}}{2} - \frac{\xi^2}{d}
 \overline{\slashed{\xi}} + \frac{\xi \overline{\xi}}{d} \slashed{\xi}
= \left( \frac{1}{2} + \frac{\xi \overline{\xi}}{d} \right) \slashed{\xi} - \frac{\xi^2}{d} \overline{\slashed{\xi}} ,\nonumber
\end{gather}
showing that~\eqref{FPPs} holds with
\[ c = \frac{2 \xi^2}{d + 2 \xi \overline{\xi}} \overset{\eqref{ddef}}{=}
-\frac{d - 2 \xi \overline{\xi}}{2 \overline{\xi}^2} . \]
For clarity, we remark that in the formalism of the continuum limit (i.e., disregarding
errors of the form~\eqref{ap1}), these formulas
can be simplified with the help of the contraction rules~\eqref{eq52}--\eqref{eq54} to obtain
\begin{gather}
d = 2 \sqrt{ \frac{(z+\overline{z})^2}{4} - z^2 \overline{z}^2}
= 2 \sqrt{ \frac{(z-\overline{z})^2}{4} } = z-\overline{z}, \label{dval} \\
F_{-} \overline{\slashed{\xi}} = \left( \frac{1}{2} +\frac{\xi \overline{\xi} - \overline{\xi}^2}{d} \right) \slashed{\xi}
= \left( \frac{1}{2} +\frac{z + \overline{z} - 2 \overline{z}}{2 (z-\overline{z})} \right) \slashed{\xi} = \slashed{\xi}, \nonumber\\
F_{-} \slashed{\xi} = \left( \frac{1}{2} -\frac{\xi \overline{\xi} - \xi^2}{d} \right) \slashed{\xi}
= \Big( \frac{1}{2} -\frac{z + \overline{z} - 2 z}{2 (z-\overline{z})} \Big) \slashed{\xi} = \slashed{\xi}, \nonumber\\
F_{+} \slashed{\xi} = \left( \frac{1}{2} + \frac{\xi \overline{\xi} - \xi^2}{d} \right) \slashed{\xi}
= \left( \frac{1}{2} + \frac{z+\overline{z} - 2 z}{2 (z-\overline{z})} \right) \slashed{\xi} = 0 ,\nonumber
\end{gather}
giving agreement with the computations in~\cite[Section~2.6.1]{cfs}.

Similar as in the proof of Lemma~\ref{lemmadel2}, we consider the different contributions
after each other:
\begin{itemize}[leftmargin=1.5em]\itemsep=0pt
\item[(i)] Second order perturbation of the eigenvalues:
We again use the formula~\eqref{lambda2} for the
second order perturbation of the eigenvalue~$\lambda^{xy}_{nL+}$ to obtain
\begin{gather*}
\Delta \big| \lambda^{xy}_{L+} \big|^2 \asymp 2 \sum_{c=L,R} \re \left( \frac{\overline{\lambda^{xy}_{L+}}}
{\lambda^{xy}_{L+} - \lambda^{xy}_{c-}}
 \Tr\big( F^{xy}_{L+} \nabla A_{xy} F^{xy}_{c-} \nabla A_{xy} \big) \right) \\
\hphantom{\Delta \big| \lambda^{xy}_{L+} \big|^2}{}\! \overset{(*)}{=} 2 \re \left( \frac{\overline{z}}
{z - \overline{z}}
 \Tr\big( F^{xy}_{L+} \nabla A_{xy} F^{xy}_{-} \nabla A_{xy} \big) \right) .
\end{gather*}
Since we saw after~\eqref{2trace} that the trace vanishes in the continuum limit,
it gives rise to a~factor~$\varepsilon/t$. Therefore, the other factors can be computed
in the formalism of the continuum limit. Thus in~$(*)$ we could use
the explicit form of the eigenvalues in the continuum limit (see~\cite[Section~3.6.1]{cfs}).
Adding the similar formula for the right-handed eigenvalue, we obtain
\begin{gather*}
 \sum_{c=L,R} \Delta \big| \lambda^{xy}_{c+} \big|^2 \asymp
2 \re \left( \frac{\overline{z}}
{z - \overline{z}}
 \Tr\big( F^{xy}_{+} \nabla A_{xy} F^{xy}_{-} \nabla A_{xy} \big) \right) \\
\hphantom{\sum_{c=L,R} \Delta \big| \lambda^{xy}_{c+} \big|^2}{}
= 2 \re \left( \frac{\overline{z}}
{z - \overline{z}} \right) \Tr\big( F^{xy}_{+} \nabla A_{xy} F^{xy}_{-} \nabla A_{xy} \big)
= - \Tr\big( F^{xy}_{+} \nabla A_{xy} F^{xy}_{-} \nabla A_{xy} \big) \\
\hphantom{\sum_{c=L,R} \Delta \big| \lambda^{xy}_{c+} \big|^2}{}
= -\Tr\big( F^{xy}_{+} (\nabla P) P^* F^{xy}_{-} P (\nabla P^*) \big)
-\Tr\big( F^{xy}_{+} P (\nabla P^*) F^{xy}_{-} (\nabla P) P^* \big) \\
\hphantom{\sum_{c=L,R} \Delta \big| \lambda^{xy}_{c+} \big|^2=}{} - \Tr\big( F^{xy}_{+} (\nabla P) P^* F^{xy}_{-} (\nabla P) P^* \big)
- \Tr\big( F^{xy}_{+} P (\nabla P^*) F^{xy}_{-} P (\nabla P^*) \big) .
\end{gather*}

\item[(ii)] First order perturbation of the eigenvalues:
Exactly as explained in the proof of Lemma~\ref{lemmadel2},
we only need to consider the contribution~\eqref{foDelta}.
It affects the absolute square of the eigenvalue according to
\begin{gather*}
 \Delta \big|\lambda^{xy}_{L+} \big|^2 \asymp 2 \re \big( \Delta \lambda^{xy}_{L+} \overline{\lambda^{xy}_{L+} } \big)
 = \Tr \big( F^{xy}_{L+} (\nabla P) (\nabla P^*) \big) \overline{\lambda^{xy}_{L+} }
+ \Tr \big( F^{xy}_{R-} (\nabla P) (\nabla P^*) \big) \lambda^{xy}_{L+} \\
\hphantom{\Delta \big|\lambda^{xy}_{L+} \big|^2}{}
= \Tr \big( F^{xy}_{L+} P^* P (\nabla P) (\nabla P^*) \big)
+ \Tr \big( F^{xy}_{R-} P^* P (\nabla P) (\nabla P^*) \big) .
\end{gather*}
Adding the corresponding right-handed contribution, we obtain
\[ \sum_{c=L,R} \Delta \big|\lambda^{xy}_{c+} \big|^2 \asymp
\Tr \big( P^* P (\nabla P) (\nabla P^*) \big) . \]
\end{itemize}
Collecting all the contributions gives the result.
\end{proof}

For the computation of the resulting contributions to the Lagrangian,
we shall make essential use of the following symmetry argument:
\begin{Lemma} \label{lemmasymm} Let~$S$ be an expression
linear in~$P$ and~$P^*$ which may contain factors of~$\xi$.
Assume that the expression is
even under the replacements~$P \leftrightarrow P^*$. Then the combination of
derivatives
\begin{gather} \label{dercombi}
( \nabla_{1, \u} - \nabla_{2, \u} )( \nabla_{1, \v} + \nabla_{2, \v} ) S
\end{gather}
is odd $($even$)$ under the replacements~$x \leftrightarrow y$ if
the number of factors~$\xi$ is even $($respectively odd$)$.\end{Lemma}
\begin{proof} Since~$P(x,y)^*=P(y,x)$, the expression~$S$ is
even (odd) under the replacements~$x \leftrightarrow y$
if the number of factor~$\xi$ is odd (even).
Clearly, the combination of derivatives in~\eqref{dercombi}
reverses the symmetry from odd to even and vice versa.
This gives the result.
\end{proof}

\begin{Prp} \label{prpfermiformnew}
The contributions~${\sim} \delta^{-4} \cdot J^2 \varepsilon/t$ to the Lagrangian have the form
\begin{gather} \label{fermiform}
 ( \nabla_{1, \u} - \nabla_{2, \u} ) ( \nabla_{1, \v} + \nabla_{2, \v} ) \L(x,y)
= J^{kl}(x,y) \frac{1}{\delta^4} \frac{\xi_l \xi_k}{t^2} {\rm i} K_0(\xi) ,
\end{gather}
where the tensor~$J^{lk}(x,y)$ has the components
\begin{gather}
J^{00}(x,y) \simeq \sum_{c=L,R} \im \big( ( \nabla_{1, \u} - \nabla_{2, \u} ) \Sl \psi^\u(x) |
\chi_c \psi^\u(y) \Sr \nonumber\\
\qquad\hphantom{J^{00}(x,y) \simeq}{}\times
 ( \nabla_{1, \v} + \nabla_{2, \v} ) \Sl \psi^\v(x) | \chi_{\bar{c}} \psi^\v(y) \Sr \big) \label{Jscalar} \\
\hphantom{J^{00}(x,y) \simeq}{} -\sum_{c=L,R} \im \big(( \nabla_{1, \u} - \nabla_{2, \u} ) \Sl \psi^\u(x) | \gamma^\alpha
\chi_c \psi^\u(y) \Sr \notag \\
\qquad \hphantom{J^{00}(x,y) \simeq}{} \times ( \nabla_{1, \v} + \nabla_{2, \v} ) \Sl \psi^\v(x) | \gamma_\alpha
\chi_c \psi^\v(y) \Sr \big), \label{Jjj} \\
J^{0\alpha}(x,y) \simeq \re \Tr \big(
 ( \nabla_{1, \u} - \nabla_{2, \u} ) | \psi^\u(y) \Sr \Sl \psi^\u(x) | \notag \\
\qquad \hphantom{J^{0\alpha}(x,y) \simeq}{} \times ( \nabla_{1, \v} + \nabla_{2, \v} ) |\sigma^{0\alpha}
\psi^\v(x) \Sr \Sl \psi^\v(y) | \big) \label{J0a1} \\
\hphantom{J^{0\alpha}(x,y) \simeq}{} - \re \Tr \big(
 ( \nabla_{1, \u} - \nabla_{2, \u} ) | \sigma^{0\alpha} \psi^\u(y) \Sr \Sl \psi^\u(x) | \notag \\
\qquad \hphantom{J^{0\alpha}(x,y) \simeq}{} \times ( \nabla_{1, \v} + \nabla_{2, \v} ) | \psi^\v(x) \Sr \Sl \psi^\v(y) | \big),
\label{J0a2} \\
J^{\alpha \beta}(x,y) \simeq \sum_{c=L,R} \im \big( ( \nabla_{1, \u} - \nabla_{2, \u} ) \Sl \psi^\u(x) | \sigma^{0\alpha} \chi_c \psi^\u(y) \Sr \notag \\
\qquad\hphantom{J^{\alpha \beta}(x,y) \simeq}{} \times ( \nabla_{1, \v} + \nabla_{2, \v} )
\Sl \psi^\v(x) | \sigma^{0\beta} \chi_{\bar{c}} \psi^\v(y) \Sr \big) \label{Jab1} \\
\hphantom{J^{\alpha \beta}(x,y) \simeq}{} - \sum_{c=L,R} \im \big( ( \nabla_{1, \u} - \nabla_{2, \u} ) \Sl \psi^\u(x) | \gamma^\alpha
\chi_c \psi^\u(y) \Sr \notag \\
\qquad\hphantom{J^{\alpha \beta}(x,y) \simeq}{} \times ( \nabla_{1, \v} + \nabla_{2, \v} )
\Sl \psi^\v(x) | \gamma^\beta \chi_c \psi^\v(y) \Sr \big) , \label{Jab2}
\end{gather}
where~$\bar{L}=R$ and~$\bar{R}=L$
$($moreover, the indices~$1$ and~$2$ again refer to the derivatives with respect to~$x$ and~$y$, respectively$)$.
\end{Prp}
\begin{proof} We first decompose the trace in~\eqref{s1} as
\begin{gather} \label{sums}
\Tr\big( F^{xy}_{+} (\nabla P) P^* F^{xy}_{-} (\nabla P) P^* \big)
= \sum_{c,c'=L,R} \Tr\big( F^{xy}_{c+} (\nabla P) P^* F^{xy}_{c'-} (\nabla P) P^* \big) .
\end{gather}
The contributions with~$c\neq c'$ depend only on the even components of~$\nabla P$.
For example for~$c=L$, the term can be rewritten as
\begin{gather*}
\Tr \big( F^{xy}_{L+} (\nabla P) P^* F^{xy}_{R-} (\nabla P) P^* \big) \\
\qquad{} = \Tr\big( F^{xy}_{L+} (\nabla P) F^{xy}_{L+} P^* (\nabla P) P^* \big)
\overset{(*)}{=} \Tr\big( F^{xy}_{L+} (\nabla P) \big) \Tr \big( F^{xy}_{L+} P^* (\nabla P) P^* \big) \\
\qquad{} = \Tr\big( F^{xy}_{L+} (\nabla P) \big) \Tr \big( F^{xy}_{R-} (\nabla P) P^* P^* \big)
= \Tr\big( F^{xy}_{L+} (\nabla P) \big) \Tr \big( F^{xy}_{R-} (\nabla P) \big) (P^* )^2 .
\end{gather*}
In~$(*)$, the fact that the operator~$F^{xy}_{L+}$ has a one-dimensional image made it possible
to factorize the trace (for details on this method see~\cite[Appendix~G.2]{pfp}).
Since the factor~$(P^*)^2$ already gives the scaling factor~$\varepsilon/t$,
the resulting traces can be computed in the formalism of the continuum limit. This gives
(see~\eqref{ddef} and~\eqref{dval} or~\cite[equation~(2.6.16)]{cfs})
\begin{gather}
- \sum_{c=L,R} \Tr \big( F^{xy}_{c+} (\nabla P) P^* F^{xy}_{\bar{c}-} (\nabla P) P^* \big) \notag \\
\qquad {} = -\frac{1}{2} \Tr ( \chi_L \nabla P ) \Tr ( \chi_R \nabla P ) (P^* )^2
\label{gives27} \\
 \qquad\quad{} + \frac{1}{2} \Tr\left( \chi_L \frac{[\slashed{\xi}, \overline{\slashed{\xi}}]}{z-\overline{z}} \nabla P \right)
\Tr\left( \chi_R \frac{[\slashed{\xi}, \overline{\slashed{\xi}}]}{z-\overline{z}} \nabla P \right)
 (P^* )^2 . \label{gives31}
\end{gather}

In the summands in~\eqref{sums} with~$c=c'$, on the other hand, only the odd component of~$\nabla P$
enters. Denoting the left- and right-handed components of~$\nabla P$ with an index,
\[ \nabla P \asymp \chi_L \nabla P_L + \chi_R \nabla P_R \]
(where~$\nabla P_{L/R}$ are vectorial), we can make use of the fact that
the anti-commutator $\{\nabla P_L, P^*\}$ is a multiple of the identity matrix.
We thus obtain
\begin{gather*}
-\Tr \big( F^{xy}_{L+} (\nabla P_L) P^* F^{xy}_{L-} (\nabla P_L) P^* \big) \\
\qquad{} = \Tr\big( F^{xy}_{L+} (\nabla P_L) P^* F^{xy}_{L-} P^* (\nabla P_L) \big)
= \Tr\big( F^{xy}_{L+} (\nabla P_L) F^{xy}_{R+} (\nabla P_L) \big) (P^* )^2 .
\end{gather*}
Again using the explicit form of the spectral projectors~$F_\pm$ in the continuum
limit (see~\eqref{ddef} and~\eqref{dval} or~\cite[equation~(2.6.16)]{cfs}), we obtain
\begin{gather}
-\Tr \big( F^{xy}_{L+} (\nabla P) P^* F^{xy}_{L-} (\nabla P) P^* \big) \nonumber\\
\qquad{} =\frac{1}{2} \Tr ( (\nabla P_L) (\nabla P_L) ) (P^* )^2
 +\Tr\left( \frac{[\slashed{\xi}, \overline{\slashed{\xi}}]}{z-\overline{z}}
(\nabla P_L) (\nabla P_L) \right) (P^* )^2 \nonumber \\
\qquad \quad {} +\frac{1}{2} \Tr\left(\frac{[\slashed{\xi}, \overline{\slashed{\xi}}]}{z-\overline{z}}
(\nabla P_L) \frac{[\slashed{\xi}, \overline{\slashed{\xi}}]}{z-\overline{z}} (\nabla P_L) \right) (P^* )^2 \notag \\
\qquad{} = \frac{1}{2} \Tr ( (\nabla P_L) (\nabla P_L) ) (P^* )^2 \label{gives28} \\
\qquad\quad{} + \frac{1}{2 (z-\overline{z})^2} \Tr\big( [\slashed{\xi}, \overline{\slashed{\xi}}]
(\nabla P_L) [\slashed{\xi}, \overline{\slashed{\xi}}] (\nabla P_L) \big) (P^* )^2 . \label{gives32}
\end{gather}
Finally, we add the resulting formula for the right-handed component.

It remains to consider the summands~\eqref{s3}--\eqref{s5} involving
one factor~$\nabla P$ and one factor~$\nabla P^*$. Rewriting~\eqref{s3} as
\begin{gather*}
\eqref{s3} = -\Tr\big( F^{xy}_{+} (\nabla P) F^{xy}_{+} P^* P (\nabla P^*) \big) \\
\hphantom{\eqref{s3}}{} = -\Tr\big( F^{xy}_{+} (\nabla P) F^{xy}_{+} (\nabla P^*) \big)
 \Tr \big( F^{xy}_{L+} P^* P \big) ,
\end{gather*}
one sees that~\eqref{s3} is symmetric under the replacement~$\nabla P \leftrightarrow \nabla P^*$.
As a consequence, the symmetry argument of Lemma~\ref{lemmasymm}
applies. The counting of the number of factors~$\xi$ is already taken into account
by our expansion in powers of~$\varepsilon/t$.
Representing the contribution in the form~\eqref{fermiform}, the expression~$J^{kl}(x,y)$
is odd under the replacements~$x \leftrightarrow y$.
Since the factor~$K_0(\xi)$ in~\eqref{fermiform} is also odd, we conclude that the
contribution on the right of~\eqref{fermiform} is necessarily even.
But the left side of~\eqref{fermiform} is odd. Therefore, these contributions must
vanish in~\eqref{fermiform}.

The term~\eqref{s4} can be treated similarly after rewriting it according to
\begin{gather*}
\eqref{s4} = -\Tr\big( F^{xy}_{-} P^* P (\nabla P^*) F^{xy}_{-} (\nabla P) \big) \\
\hphantom{\eqref{s4}}{} = -\Tr\big( F^{xy}_{-} P^* P \big) \Tr \big(F^{xy}_{R-} (\nabla P^*) F^{xy}_{-} (\nabla P) \big) .
\end{gather*}

It remains to consider the summand~\eqref{s5}.
Since contributions symmetric under the replacement~$\nabla P \leftrightarrow \nabla P^*$
again vanish, it suffices to consider the contribution anti-symmetrized under this replacement,
\begin{gather}
 \frac{1}{4} \Tr \big(P^* P [\nabla P, \nabla P^* ] \big)
 =\frac{1}{4} \Tr (P^* P (\nabla P) (\nabla P^*) )
- \frac{1}{4} \Tr \big((\nabla P) P^* P (\nabla P^*) \big) \notag \\
\hphantom{\frac{1}{4} \Tr \big(P^* P [\nabla P, \nabla P^* ] \big)}{}
=\frac{1}{8} \Tr \big( [P^*, P] (\nabla P) (\nabla P^*) \big)
- \frac{1}{8} \Tr \big((\nabla P) [P^*, P] (\nabla P^*) \big) , \label{gives30}
\end{gather}
where in the last step we used that~$P$ and~$P^*$ are vectorial, so
that their anti-commutator commutes with~$\nabla P$ and~$\nabla P^*$.

Finally, we specialize the above formulas for a spherically symmetric regularization. Then
\[ \frac{[\slashed{\xi}, \overline{\slashed{\xi}}]}{z-\overline{z}} \simeq \gamma^0 \gamma^\alpha \frac{\xi_\alpha}{t}
\qquad \text{and} \qquad
[P^*, P] \simeq {\rm i} \varepsilon \gamma^0 \gamma^\alpha \xi_\alpha \big| T^{(-1)} \big|^2
= \gamma^0 \gamma^\alpha t \xi_\alpha \frac{{\rm i} \varepsilon}{t} \big| T^{(-1)} \big|^2 .
\]
Collecting all the results and expressing~$\nabla P$ and its adjoint
according to~\eqref{nabP1} and~\eqref{nabP2} in terms of the fermionic wave functions gives the result.
More precisely, the contribution~\eqref{gives27} gives~\eqref{Jscalar}, and~\eqref{gives31} gives~\eqref{Jab1}.
Computing the tensor contractions in~\eqref{gives28} and~\eqref{gives32} yields~\eqref{Jjj} and~\eqref{Jab2}.
Finally, the contributions~\eqref{gives30} give~\eqref{J0a1} and~\eqref{J0a2}.
\end{proof}

\subsection{Convolution lemmas} \label{secconv}
We now derive lemmas which are needed for computing convolution integrals
(similar convolution lemmas were obtained in a different context in~\cite[Section~5]{vacstab}).

\begin{Lemma} \label{lemmaintconv}
For any test function~$h \in C^\infty_0\big(\R^4\big)$ $($or in our applications a spinorial
test function~$h \in C^\infty_0\big(\R^4, \C^4\big))$ and for
any momentum~$q$ with~$q^2>0$ and any mass parameter~$m>0$,
\begin{gather*}
\int \frac{{\rm d}^4p}{(2 \pi)^4} \hat{K}_0(p) \delta\big( (p-q)^2 -m^2 \big) h(p-q) \nonumber\\
\qquad{} = \frac{1}{32 \pi^3} \frac{q^2-m^2}{q^2} \epsilon\big( q^0 \big) h(-q)
\big(1+\O \big(q^2-m^2\big) \big) .
\end{gather*}
\end{Lemma}
\begin{proof} Due to Lorentz symmetry we may assume that~$q=(\Omega, \vec{0})$ with~$\Omega \neq 0$.
Moreover, we first consider the situation that~$h(p-q)$ is a constant~$\chi$.
Then, using spherical symmetry of the resulting integrals,
\begin{gather}
 \int \frac{{\rm d}^4p}{(2 \pi)^4} \hat{K}_0(p) \delta\big( (p-q)^2 -m^2 \big) \chi \notag \\
\qquad{}= \frac{4 \pi}{(2 \pi)^4} \chi \int_{-\infty}^\infty {\rm d}\omega \int_0^\infty k^2 \,{\rm d}k\,
\delta \big(\omega^2-k^2 \big) \epsilon(\omega) \delta\big( (\omega-\Omega)^2 - k^2 -m^2 \big) \notag \\
\qquad{}= \frac{1}{4 \pi^3} \chi \int_{-\infty}^\infty {\rm d}\omega\, \frac{|\omega|}{2}
\epsilon(\omega) \delta\big( (\omega-\Omega)^2 - \omega^2 -m^2 \big) \notag \\
\qquad{}= \frac{1}{8 \pi^3} \chi \int_{-\infty}^\infty {\rm d}\omega\, \omega
\delta\big( -2 \omega \Omega + \Omega^2 -m^2 \big)
= \frac{1}{8 \pi^3} \chi \frac{1}{2 |\Omega|} \frac{\Omega^2-m^2}{2 \Omega} . \label{zwies}
\end{gather}
This gives the result if~$h$ is a constant.
Moreover, the above computation shows that if~$q^2 \approx m^2$, one only gets a contribution
for~$p \approx 0$ (as one also sees graphically by varying~$q$ on the right in Fig.~\ref{figsupport}).
Therefore, we may expand~$h(p-q)$ in a Taylor polynomial around~$p=0$ to obtain the result.
\end{proof}

\begin{Lemma} \label{lemmaintconv2}
For any test function~$h$ and any momentum~$q$ inside the upper mass cone $\{q^2>0$ and~$q^0>0\}$,
\begin{gather}
\int \frac{{\rm d}^4p}{(2 \pi)^4} \frac{\Theta\big( p^2 \big) }{|\vec{p}|} \delta\big( (p-q)^2 -m^2 \big)
\Theta \big( q^0 - p^0 \big) h(p-q) \nonumber\\
\qquad{} = \left\{ \frac{\ell_{\max}}{16 \pi^3} +
\frac{m^2}{32 \pi^3} \frac{1}{|\vec{q}|} \log
\bigg( \frac{q^0 - |\vec{q}| - \ell}{q^0 + |\vec{q}| - \ell} \bigg) \bigg|_{\ell=0}^{\ell=\ell_{\max}} \right\}
h(-q)
\big(1+\O \big(q^2-m^2\big) \big)\label{intconv2}
\end{gather}
with
\[ \ell_{\max} := q^0 - \sqrt{|\vec{q}|^2 + m^2} . \]
Moreover, the curly brackets in~\eqref{intconv2} vanish quadratically on the mass cone, i.e.,
\[ \{ \cdots \} = \O \big( \big(q^2-m^2\big)^2 \big) . \]
\end{Lemma}
\begin{proof} In the case~$q^2 \geq m^2$, in~\eqref{intconv2} one integrates only over~$p$ inside the upper mass cone.
Likewise, if~$q^2 < m^2$ one integrates over~$p$ inside the lower mass cone.
We consider these two cases after each other and begin with the case~$q^2 \geq m^2$.
If in~\eqref{zwies} one replaces the factor~$\hat{K}_0$ by a $\delta$-distribution supported on the boundary of the upper
mass cone, one gets for any~$r \in \R^4$,
\begin{gather} \label{intrel1}
\int \frac{{\rm d}^4p}{(2 \pi)^4} \delta\big(p^2\big) \Theta\big(p^0\big) \delta\big( (p-r)^2 -m^2 \big)
= \frac{1}{32 \pi^3} \frac{r^2-m^2}{r^2} \Theta \big(r^2-m^2 \big) \Theta\big(r^0\big) .
\end{gather}
Next, we rewrite the Heaviside function in the integrand in~\eqref{intconv2}
as an integral over a $\delta$-distribution. Namely, for~$p=(\omega, \vec{p})$ with~$\omega>0$,
\[ \frac{\Theta\big( p^2 \big)}{|\vec{p}|} = 2 \int_0^\infty {\rm d}\ell\, \delta\big( (\omega-\ell)^2 - |\vec{p}|^2 \big)
 \Theta\big( \omega-\ell \big) =
2 \int_0^\infty {\rm d}\ell\, \delta\big( (p - l)^2 \big) \Theta\big( (p-l)^0 \big) , \]
where in the last step we set~$l=(\ell, \vec{0}) \in \R^4$. Substituting this relation into the
integral in~\eqref{intconv2} and using~\eqref{intrel1}, we obtain
\begin{gather}
\int \frac{{\rm d}^4p}{(2 \pi)^4} \frac{\Theta\big( p^2 \big) }{|\vec{p}|} \delta\big( (p-q)^2 -m^2 \big) \notag \\
\qquad{} = 2 \int_0^\infty {\rm d}\ell \int \frac{{\rm d}^4p}{(2 \pi)^4} \delta\big( (p - l)^2 \big) \Theta\big( (p-l)^0 \big)
 \delta\big( (p-q)^2 -m^2 \big) \notag \\
 \qquad{} = 2 \int_0^\infty {\rm d}\ell \int \frac{{\rm d}^4p}{(2 \pi)^4} \delta\big( p^2 \big) \Theta\big( p^0 \big)
 \delta\big( (p+l-q)^2 -m^2 \big) \notag \\
 \qquad{} \!\!\!\overset{\eqref{intrel1}}{=} \frac{1}{16 \pi^3} \int_0^\infty \frac{(q-l)^2-m^2}{(q-l)^2} \Theta \big( (q-l)^2-m^2 \big) \Theta\big(q^0-l^0\big) \, {\rm d}\ell \notag \\
\qquad{} = \frac{1}{16 \pi^3} \int_0^{\ell_{\max}} \frac{(q-l)^2-m^2}{(q-l)^2}\, {\rm d}\ell . \label{zwies2}
\end{gather}
Computing the $\ell$-integral gives the result in the case~$q^2 \geq m^2$.

In the remaining case~$q^2<m^2$, we proceed similarly for the Heaviside function
and the $\delta$-distribution in the lower mass cone,
\begin{gather*}
 \int \frac{{\rm d}^4p}{(2 \pi)^4} \delta\big(p^2\big) \Theta\big({-}p^0\big) \delta\big( (p-r)^2 -m^2 \big)
= \frac{1}{32 \pi^3} \frac{r^2-m^2}{r^2} \Theta\big(m^2-r^2\big) ,
\end{gather*}
valid for all~$r$ inside the upper mass cone,
\[ r^2>0 \qquad \text{and} \qquad r^0>0 . \]
Moreover, for~$p=(\omega, \vec{p})$ with~$\omega<0$,
\[ \frac{\Theta\big( p^2 \big)}{|\vec{p}|}
= 2 \int_{-\infty}^0 {\rm d}\ell\, \delta\big( (p - l)^2 \big) \Theta\big( (l-p)^0 \big) , \]
and thus
\begin{gather*}
\int \frac{{\rm d}^4p}{(2 \pi)^4} \frac{\Theta\big( p^2 \big) }{|\vec{p}|} \delta\big( (p-q)^2 -m^2 \big) \\
\qquad{} = 2 \int_{-\infty}^0 {\rm d}\ell \int \frac{{\rm d}^4p}{(2 \pi)^4} \delta\big( (p - l)^2 \big) \Theta\big( (l-p)^0 \big)
 \delta\big( (p-q)^2 -m^2 \big) \\
\qquad{} = \frac{1}{16 \pi^3} \int_{-\ell_{\max}}^0 \frac{(q-l)^2-m^2}{(q-l)^2} \, {\rm d}\ell .
\end{gather*}
Again carrying out the $\ell$-integral completes the proof.
\end{proof}

In the next lemma, we show that inserting a factor~$\omega$ in the integrand in~\eqref{intconv2}
makes the contribution smaller by one order on the mass cone:
\begin{Lemma} \label{lemmaintconv3}
For any momentum~$q$ inside the upper mass cone $\big\{q^2>0$ and~$q^0>0\big\}$,
\[
\int \frac{{\rm d}^4p}{(2 \pi)^4} \Theta\big( p^2 \big) \frac{\omega}{|\vec{p}|} \delta\big( (p-q)^2 -m^2 \big)
\Theta \big( q^0 - p^0 \big) h(p-q) = \O \big( \big(q^2-m^2 \big)^3 \big) . \]
\end{Lemma}
\begin{proof} We proceed as in the proof of Lemma~\ref{lemmaintconv2}
and estimate the resulting contributions. First, inserting a factor~$\omega$ in~\eqref{intrel1},
a calculation similar to~\eqref{zwies} shows that
\[
\int \frac{{\rm d}^4p}{(2 \pi)^4} \omega \delta\big(p^2\big) \Theta\big(p^0\big) \delta\big( (p-r)^2 -m^2 \big)
= \O \big( \big( r^2-m^2 \big)^2 \big) \Theta \big(r^2-m^2 \big) \Theta\big(r^0\big) . \]
In the case~$q^2 \geq m^2$, a computation similar to~\eqref{zwies2} yields
\begin{gather*}
\int \frac{{\rm d}^4p}{(2 \pi)^4} \Theta\big( p^2 \big) \frac{\omega}{|\vec{p}|} \delta\big( (p-q)^2 -m^2 \big) \\
\qquad{} = \frac{1}{16 \pi^3} \int_0^{\ell_{\max}}
\O \big( \big( (q-l)^2-m^2 \big)^2 \big) \, {\rm d}\ell = \O \big( \big( q^2-m^2 \big)^3 \big) .
\end{gather*}
The case~$q^2<m^2$ can be treated similarly.
\end{proof}

\subsection{Conservation of surface layer integrals} \label{secfermiconserve}
According to Proposition~\ref{prpfermiformnew}, the integrand of the surface layer integral is of the form
\begin{gather} \label{JklK}
J^{kl}(x,y) {\mathfrak{K}}(x,y)
\qquad \text{with} \qquad
{\mathfrak{K}}(x,y) := \frac{1}{\delta^4} \frac{\xi_k \xi_l}{t^2} {\rm i} K_0(\xi) ,
\end{gather}
where~$J^{kl}(x,y)$ is real-valued and symmetric, i.e.,
\[ J^{kl}(x,y) = J^{lk}(y,x) . \]
As we shall see, the surface layer integrals are conserved only under
certain assumptions on the fermionic jets (see~\eqref{conscond}
and the explanations in Section~\ref{secconservefinal}).
In the applications, these assumptions pose conditions on the nature of microscopic mixing.
This is the main result of this section:

\begin{Thm} \label{thmfermiconserve}
Assume that all fermionic jets satisfy for all~$x,y \in \scrM$ and~$\alpha=1,2,3$
the conditions
\begin{gather} \label{conscond}
\Sl \psi(x) | \big(\1\pm i \gamma^0\big) \gamma^\alpha \delta \psi(y) \Sr = 0
= \Sl \psi(x) | \gamma^5 \big(\1\pm \gamma^0\big) \gamma^\alpha \delta \psi(y) \Sr
\end{gather}
$($where in each equation we can choose the plus or the minus sign independently$)$.
Then the surface layer integrals are conserved.
\end{Thm}

\begin{figure}[t]\centering
\includegraphics{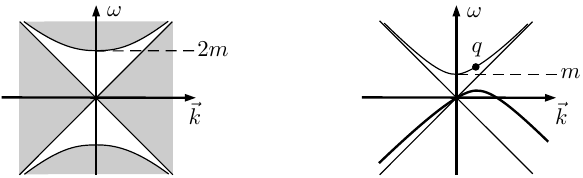}
\caption{Support properties used in the computation of fermionic surface layer integrals.}\label{figsupport}
\end{figure}

The significance of the condition~\eqref{conscond} will be explained and discussed
in detail in Section~\ref{secconservefinal} below.
We now enter the proof of this theorem, which will be completed
at the end of Section~\ref{secconservefinal}.
It is useful to write the $y$-dependence of~$J^{kl}(x,y)$ symbolically as
\begin{gather} \label{psiphi}
\psi(y) \cdot \overline{\phi(y)} ,
\end{gather}
where~$\psi$ and~$\phi$ are two solutions of the Dirac equation.
Since the spinorial character of the wave functions is of no relevance for what follows,
it is preferable to consider~$\psi$ and~$\phi$ as solutions of the {\em{Klein--Gordon equation}}
of mass~$m>0$.
Thus in momentum space, both $\hat{\psi}$ and~$\hat{\phi}$ are supported on the
mass shell. Rewriting the integral~\eqref{pointwise} in momentum space,
our task is to analyze the integral
\begin{gather} \label{cint}
\int \frac{{\rm d}^4p}{(2 \pi)^4}
\hat{\mathfrak{K}}(p) \int \frac{{\rm d}^4q}{(2 \pi)^4} \hat{\psi}(q) \hat{\overline{\phi}}(p-q) .
\end{gather}
Using that both $\hat{\psi}$ and~$\hat{\phi}$ are supported on the
mass shell, their convolution is supported in the shaded region
on the left of Fig.~\ref{figsupport}.
The Fourier transform~$\hat{\mathfrak{K}}$ of the kernel~${\mathfrak{K}}$
in~\eqref{JklK}, on the other hand, is obtained by taking the second derivatives of the
distribution shown on the right of Fig.~\ref{figK12}.
This shows in particular that~$\hat{\mathfrak{K}}(p)$ vanishes inside the upper and lower mass cone.
As a consequence, the integral~\eqref{cint} vanishes if
the arguments of~$\hat{\psi}$ and~$\hat{\overline{\phi}}$ are both on the upper
or both on the lower mass shell.
Hence in what follows, it suffices to consider the case that one momentum is on the upper
and the other on the lower mass shell (as shown on the right of Fig.~\ref{figsupport}).

For the following analysis, it is convenient to decompose~$\hat{\mathfrak{K}}$
into the so-called {\em{equal time}} and {\em{zero momentum contributions}},
\[ \hat{\mathfrak{K}} = \hat{\mathfrak{K}}_\text{et} + \hat{\mathfrak{K}}_\text{zm} , \]
which are defined by
\begin{gather}
\hat{\mathfrak{K}}_\text{et}(p) := \frac{1}{\delta^4} \frac{\partial^2}{\partial p^j p^k} \frac{{\rm i} \omega}{k}, \label{Ket} \\
\hat{\mathfrak{K}}_\text{zm}(p) := -\frac{1}{\delta^4} \frac{\partial^2}{\partial p^j p^k} \left( \frac{{\rm i} \omega}{k} \Theta\big(p^2\big) - {\rm i} \epsilon\big(p^0\big) \Theta\big(p^2\big) \right) . \label{Kzm}
\end{gather}
Note that~$\hat{\mathfrak{K}}_\text{et}(p)$ is a polynomial in~$\omega$, implying that
its Fourier transform~${\mathfrak{K}}_\text{et}(x,y)$ gives a~contribution only
for equal times~$x^0=y^0$. In view of the right of Fig.~\ref{figsupport}, $\hat{\mathfrak{K}}_\text{zm}$
gives a contribution only if~$p=(\omega, \vec{k})=0$, explaining the name
zero momentum contribution.

\subsubsection{The equal time contribution}
We first compute the condition for conservation~\eqref{integrated}
for the equal time contribution without the assumptions in Theorem~\ref{thmfermiconserve}:
\begin{Lemma} \label{lemmaet}
The contribution by~${\mathfrak{K}}_\text{\rm{et}}$ to the surface layer integrals is
given by
\begin{gather}
 \int_{\R^3} {\rm d}^3x \int_\scrM {\rm d}^4y \, J^{jk}(x,y) \delta^4 {\mathfrak{K}}_\text{\rm{et}}(x,y) \Big|_{x^0=0} \notag \\
\qquad {} =-\frac{1}{2} \frac{\partial}{\partial t}
\int_{\R^3} {\rm d}^3x \int_{\R^3} {\rm d}^3y\, J^{\alpha \beta}\big((t,\vec{x}), (t,\vec{y}) \big)\big|_{t=0}
\left( \frac{\xi_\alpha \xi_\beta}{|\vec{\xi}|^2} - \frac{\delta_{\alpha \beta}}{3} \right) .
\label{noconserve}
\end{gather}
\end{Lemma}
\begin{proof} Transforming~\eqref{Ket} back to position space, one finds
\[ {\mathfrak{K}}_\text{et}(x,y) \simeq \frac{1}{\delta^4} \xi_j \xi_k \frac{\delta'\big(\xi^0 \big)}{|\vec{\xi}|^2} \]
(where again~$\xi = y-x$). Hence
\begin{gather*}
 \int_\scrM J^{jk}(x,y) \delta^4 {\mathfrak{K}}_\text{et}(x,y) \big|_{x^0=0} \, {\rm d}^4y
\simeq \int_\scrM {\rm d}^4y \, J^{jk}\big((0,\vec{x}), (t',\vec{y}) \big) \xi_j \xi_k \frac{\delta'(t') }{|\vec{\xi}|^2} \, {\rm d}^4y \\
 \qquad{} = \int_\scrM \big(J^{0\alpha}+ J^{\alpha0}\big)\big((0,\vec{x}), (t',\vec{y}) \big) t' \xi_\alpha
 \frac{\delta'(t')}{|\vec{\xi}|^2} \,{\rm d}^4y \\
 \qquad\quad {}+\int_\scrM J^{\alpha \beta}\big((0,\vec{x}), (t',\vec{y}) \big) \xi_\alpha \xi_\beta
 \frac{\delta'(t')}{|\vec{\xi}|^2} \, {\rm d}^4y\\
 \qquad{}= -\int_{\R^3} \big(J^{0\alpha}+ J^{\alpha0}\big)\big((0,\vec{x}), (0,\vec{y}) \big) \xi_\alpha
 \frac{1}{|\vec{\xi}|^2}\, {\rm d}^3y \\
 \qquad\quad {}-\int_{\R^3} \partial_{t'} J^{\alpha \beta}\big((0,\vec{x}), (t',\vec{y}) \big)\big|_{t'=0} \xi_\alpha \xi_\beta
 \frac{1}{|\vec{\xi}|^2} \, {\rm d}^3y .
\end{gather*}
It follows that
\begin{gather}
 \int_{\R^3} {\rm d}^3x \int_\scrM {\rm d}^4y\, J^{jk}(x,y) \delta^4 {\mathfrak{K}}_\text{\rm{et}}(x,y) \big|_{x^0=0} \notag \\
\qquad{} = -\int_{\R^3} {\rm d}^3x \int_{\R^3} {\rm d}^3y\, \big(J^{0\beta}+ J^{\beta0}\big)\big((0,\vec{x}), (0,\vec{y}) \big)
 \frac{\xi_\beta}{|\vec{\xi}|^2} \notag \\
\qquad\quad {}-\int_{\R^3} {\rm d}^3x \int_{\R^3} {\rm d}^3y \, \partial_{t'} J^{\alpha \beta}\big((0,\vec{x}), (t',\vec{y}) \big)\big|_{t'=0} \frac{\xi_\alpha \xi_\beta}{|\vec{\xi}|^2} \notag \\
\qquad{} \!\overset{(*)}{=}
-\int_{\R^3} {\rm d}^3x \int_{\R^3} {\rm d}^3y \, \partial_{t'} J^{\alpha \beta}\big((0,\vec{x}), (t',\vec{y}) \big)\big|_{t'=0}
\frac{\xi_\alpha \xi_\beta}{|\vec{\xi}|^2} \label{exA} \\
\qquad{} =-\frac{1}{2} \frac{\partial}{\partial t}
\int_{\R^3} {\rm d}^3x \int_{\R^3} {\rm d}^3y\, J^{\alpha \beta}\big((t,\vec{x}), (t,\vec{y}) \big)\big|_{t=0}
\frac{\xi_\alpha \xi_\beta}{|\vec{\xi}|^2} ,\nonumber
\end{gather}
where in~$(*)$ we used the symmetry of~$J$.

Finally, we rewrite the term involving~$\xi$ in the integrand as
\[ \frac{\xi_\alpha \xi_\beta}{|\vec{\xi}|^2} = \left( \frac{\xi_\alpha \xi_\beta}{|\vec{\xi}|^2}
- \frac{\delta_{\alpha \beta}}{3} \right) + \frac{\delta_{\alpha \beta}}{3} . \]
The first summand gives~\eqref{noconserve}. In the contribution by the first summand to the integrals,
\begin{gather} \label{endproof}
\int_{\R^3} {\rm d}^3x \int_{\R^3} {\rm d}^3y \, \delta_{\alpha \beta} J^{\alpha \beta}\big((t,\vec{x}), (t,\vec{y}) \big)\big|_{t=0} ,
\end{gather}
on the other hand,
the integration over~$\vec{y}$ implies that the wave functions~$\psi(y)$ and~$\overline{\phi(y)}$
in~\eqref{psiphi} have opposite momenta. Since they are both on-shell and their frequencies
have opposite signs (as explained after~\eqref{cint}), their frequencies coincide except for a sign.
Therefore, their product~\eqref{psiphi} is time independent.
As a consequence, the integral over their $t'$-derivative in~\eqref{exA} vanishes.
This concludes the proof.
\end{proof}

\subsubsection{The zero momentum contribution}

\begin{Lemma} The contribution by the kernel~${\mathfrak{K}}_\text{zm}$ to the
surface layer integrals is conserved in the sense that
\[
\int_\scrM J^{kl}(x,y) {\mathfrak{K}}_\text{\rm{zm}}(x,y) \, {\rm d}^4y = 0 . \]
\end{Lemma}
\begin{proof} We again write the $y$-dependence of~$J^{kl}$ according to~\eqref{psiphi} as the product
of two Dirac solutions. Moreover, we again work in momentum space,
\begin{gather}
\int_{\scrM} {\mathfrak{K}}_\text{zm}(x,y) \psi(y) \overline{\phi}(y) \,{\rm d}^4y
= \int \frac{{\rm d}^4q}{(2 \pi)^4} \int \frac{{\rm d}^4p}{(2 \pi)^4}
\hat{\mathfrak{K}}_\text{zm}(p) \hat{\psi}(q) {\rm e}^{-{\rm i} p x} \hat{\overline{\phi}}(p-q) \notag \\
\hphantom{\int_{\scrM} {\mathfrak{K}}_\text{zm}(x,y) \psi(y) \overline{\phi}(y) \,{\rm d}^4y}{} = \int \frac{{\rm d}^4q}{(2 \pi)^4} {\rm e}^{-{\rm i}qx} \hat{\psi}(q) \int \frac{{\rm d}^4p}{(2 \pi)^4}
\hat{\mathfrak{K}}_\text{zm}(p) {\rm e}^{-{\rm i} (p-q) x} \hat{\overline{\phi}}(p-q) .\label{intprod}
\end{gather}
As explained after~\eqref{cint}, it suffices to consider the case that
momenta~$q$ and~$q-p$ lie both on the upper or both on the lower mass shell.

We consider the two summands in~\eqref{Kzm} after each other.
In the second summand, we can carry out one derivative,
\[ \delta^4 \hat{\mathfrak{K}}_\text{zm}(p) \asymp \frac{\partial^2}{\partial p^j p^k} \big( {-} {\rm i} \epsilon\big(p^0\big) \Theta\big(p^2\big) \big)
= \frac{\partial}{\partial p^j} \big( - {\rm i} \epsilon\big(p^0\big) \delta\big(p^2\big) 2 p_k \big)
\simeq \frac{\partial}{\partial p^j} \big( p_k {\rm i}K_0(p) \big) . \]
Using this formula in~\eqref{intprod}, integrating by parts and using that
all the other terms in the integrand depend on~$p-q$, the inner integral in~\eqref{intprod}
can be written as
\[ \int \frac{{\rm d}^4p}{(2 \pi)^4} \delta^4 \hat{\mathfrak{K}}_\text{zm}(p) {\rm e}^{-{\rm i} (p-q) x} \hat{\overline{\phi}}(p-q)
\asymp -\frac{\partial}{\partial q^j}
\int \frac{{\rm d}^4p}{(2 \pi)^4} p_k {\rm i}K_0(p) {\rm e}^{-{\rm i} (p-q) x} \hat{\overline{\phi}}(p-q) . \]
Since~$\hat{\overline{\phi}}$ is supported on the mass shell, we can write the integrand as
\begin{gather} \label{hdef}
{\rm e}^{-{\rm i} (p-q) x} \hat{\overline{\phi}}(p-q) = \delta \big( (p-q)^2 - m^2 \big) h(p-q) .
\end{gather}
Then the resulting integral can be computed with the help of Lemma~\ref{lemmaintconv}.
Due to the factor~$p_k$, the integrals grows at least quadratically in
a Taylor expansion in~$q$ around a momentum on the mass shell.
Thus the first $q$-derivative vanishes, and we get zero.

It remains to consider the first summand in~\eqref{Kzm}.
The resulting contribution to the inner integral in~\eqref{intprod}
can be simplified using integration by parts to
\begin{gather*}
 \int \frac{{\rm d}^4p}{(2 \pi)^4} \big({-}\delta^4\big) \hat{\mathfrak{K}}_\text{zm}(p) {\rm e}^{-{\rm i} (p-q) x} \hat{\overline{\phi}}(p-q) \\
\qquad{} \asymp \int \frac{{\rm d}^4p}{(2 \pi)^4}
\left( \frac{\partial^2}{\partial p^j p^k} \left( \frac{{\rm i} \omega}{k} \Theta\big(p^2\big) \right) \right)
 {\rm e}^{-{\rm i} (p-q) x} \hat{\overline{\phi}}(p-q) \\
\qquad{} = \frac{\partial^2}{\partial q^j q^k}\int \frac{{\rm d}^4p}{(2 \pi)^4}
\frac{{\rm i} \omega}{k} \Theta\big(p^2\big)
 {\rm e}^{-{\rm i} (p-q) x} \hat{\overline{\phi}}(p-q) \\
\qquad{} = \frac{\partial^2}{\partial q^j q^k}\int \frac{{\rm d}^4p}{(2 \pi)^4}
\frac{{\rm i} \omega}{k} \Theta\big(p^2\big) \delta \big( (p-q)^2 - m^2 \big) h(p-q) ,
\end{gather*}
where in the last step we again used the notation~\eqref{hdef}.
According to Lemma~\ref{lemmaintconv3}, the integral vanishes
cubically on the mass cone. Therefore, its second $q$-derivatives
vanish for~$q^2=m^2$. This concludes the proof.
\end{proof}

\subsubsection{The conditions for conservation} \label{secconservefinal}
The computations of the previous lemmas showed that the only contribution
to the integrals in~\eqref{integrated} is given by~\eqref{noconserve}.
It is remarkable that the conservation laws
are not satisfied automatically, but they hold only if~\eqref{noconserve} vanishes,
leading to the condition
\begin{gather} \label{noconszero}
\frac{\partial}{\partial t}
\int_{\R^3} {\rm d}^3x \int_{\R^3} {\rm d}^3y\, J^{\alpha \beta}\big((t,\vec{x}), (t,\vec{y}) \big)\big|_{t=0}
\left( \frac{\xi_\alpha \xi_\beta}{|\vec{\xi}|^2} - \frac{\delta_{\alpha \beta}}{3} \right) = 0 .
\end{gather}
Since the conservation laws follow from the causal action principle
(see~\cite{jet} and~\cite{osi}), one can say that~\eqref{noconszero}
gives information on the interaction described by the causal action principle.
Of course, it is an important consistency check that the condition~\eqref{noconszero}
can be satisfied in physically relevant situations.
Indeed, the condition~\eqref{noconszero} can be arranged in various ways, as we now discuss.

First, one should keep in mind that we get a contribution only
if both Dirac solutions~$\psi, \phi$ describing the~$y$-dependence
of~$J^{kl}(x,y)$ in~\eqref{psiphi} are Dirac solutions which are supported both on the upper mass shell
or both on the lower mass shell.
Consequently, we only need to take into account the contributions in~\eqref{Jab1} and~\eqref{Jab2}
where the two wave functions at~$y$ (and similarly at~$x$) have frequencies of opposite signs.
Moreover, due to momentum conservation, we only get a contribution if the spatial momenta add up to
zero. For example, in the contribution
\[ \Sl \delta \psi^\u(x) | \sigma^{0\alpha} \psi^\u(y) \Sr
\Sl \psi^\v(x) | \sigma^{0\beta} \delta \psi^\v(y) \Sr , \]
the corresponding momenta must satisfy the relation
\begin{gather} \label{momconserve}
\vec{p}_{\delta \psi^\u} - \vec{p}_{\psi^\u} = \vec{p}_{\delta \psi^\v} - \vec{p}_{\psi^\v} .
\end{gather}
Time independence of this contribution
means that also the frequencies must cancel each other, i.e.,
\begin{gather} \label{freqconserve}
\omega_{\delta \psi^\u} - \omega_{\psi^\u} = \omega_{\delta \psi^\v} - \omega_{\psi^\v} .
\end{gather}
If the left and right side of~\eqref{momconserve} vanish separately, then~\eqref{freqconserve}
is also satisfied (exactly as explained in the proof of Lemma~\ref{lemmaet}
after~\eqref{endproof}).
However, if~$\vec{p}_{\delta \psi^\u} - \vec{p}_{\psi^\u} \neq 0$,
then~\eqref{freqconserve} will in general be violated.
This is the reason why~\eqref{noconszero} imposes conditions on the
fermionic jets.

One way of satisfying~\eqref{noconszero} is to ensure the implication
\begin{gather}
\eqref{momconserve} \quad \Longrightarrow \quad \eqref{freqconserve} .
\label{implication}
\end{gather}
For example, one can assume that there is a function~$\Omega\colon \R^3 \rightarrow \R$ with
\[ \omega_{\delta \psi^\u} - \omega_{\psi^\u} = \Omega \big( \vec{p}_{\delta \psi^\u} - \vec{p}_{\psi^\u} \big) . \]
This leaves a lot of freedom to choose the fermionic jets~$(\delta \psi^\u, \psi^\u)$.
More specifically, one could consider the situation that the momenta~$\vec{p}_{\delta \psi^\u}$
and~$\vec{p}_{\psi^\u}$ are related to each other by a linear transformation~$A\colon \R^3 \rightarrow \R^3$,
\[ \vec{p}_{\delta \psi^\u} = A \vec{p}_{\psi^\u} . \]
But clearly, there are many other ways of choosing the jets such that
the implication~\eqref{implication} holds.

Another strategy for satisfying~\eqref{noconszero} is to arrange that the
contributions in~\eqref{Jab1} and~\eqref{Jab2} vanish or cancel each other.
This has the advantage that~$J^{\alpha \beta}(x,y)$ can be arranged
to vanish {\em pointwise}, making it unnecessary to argue with momentum conservation.
In particular, we not only satisfy the conservation law for the surface layer integral~\eqref{integrated},
but even the pointwise relation~\eqref{pointwise} which should hold
in order to comply with the EL equations. In Theorem~\ref{thmfermiconserve},
such pointwise conditions are implemented, as is shown in the following lemma:
\begin{Lemma} If the conditions~\eqref{conscond} hold, then the
components~$J^{\alpha \beta}$ $($see~\eqref{Jab1} and~\eqref{Jab2}$)$ vanish.
\end{Lemma}
\begin{proof} We rewrite the sums over the chiral index in~\eqref{Jab1} and~\eqref{Jab2} as
\begin{gather*}
\sum_{c=L,R} \Tr \big( \chi_c \sigma^{0\alpha} \nabla P \big)
\Tr \big( \chi_{\bar{c}} \sigma^{0\beta} \nabla P \big) \xi_\alpha \xi_\beta \\
\qquad{} = \frac{1}{2} \Tr \big( \sigma^{0\alpha} \nabla P \big)
\Tr \big( \sigma^{0\beta} \nabla P \big) \xi_\alpha \xi_\beta
 -\frac{1}{2} \Tr \big( \gamma^5 \sigma^{0\alpha} \nabla P \big)
\Tr \big( \gamma^5 \sigma^{0\beta} \nabla P \big) \xi_\alpha \xi_\beta, \\
-\sum_{c=L,R} \Tr \big( \chi_c \gamma^\alpha \nabla P \big)
\Tr \big( \chi_c \gamma^\beta \nabla P \big) \xi_\alpha \xi_\beta \\
\qquad{} = -\frac{1}{2} \Tr \big( \gamma^\alpha \nabla P \big)
\Tr \big(\gamma^\beta \nabla P \big) \xi_\alpha \xi_\beta
 -\frac{1}{2} \Tr \big( \gamma^5 \gamma^\alpha \nabla P \big)
\Tr \big(\gamma^5 \gamma^\beta \nabla P \big) \xi_\alpha \xi_\beta .
\end{gather*}
The condition~$J^{\alpha \beta}=0$ gives rise to quadratic equations.
Since all the above traces are in general complex (note that~$\nabla P$ may involve
arbitrary phase factors), the only sensible method for satisfying these quadratic equations
by linear relations seems to impose that pairs of these traces are multiples of each other,
\[ \Tr \big( \sigma^{0\alpha} \nabla P \big) = \pm \Tr \big( \gamma^\alpha \nabla P \big)
\qquad \text{and} \qquad
\Tr \big( \gamma^5 \sigma^{0\alpha} \nabla P \big) = \pm {\rm i} \Tr \big( \gamma^5 \gamma^\alpha \nabla P \big) . \]
Expressing~$\nabla P$ according to~\eqref{nabP1} and~\eqref{nabP2}
gives precisely the conditions in~\eqref{conscond}.
\end{proof}

This lemma also completes the proof of Theorem~\ref{thmfermiconserve}.

We finally remark that if the fragmentation of the universal measure is taken into account
(see~\cite[Section~5]{jet} or the mechanism of microscopic mixing in~\cite{qft}),
then~\eqref{noconszero} must be satisfied only after summing over the subsystems.
This gives many more possibilities to satisfy the conditions for conservation.
For example, one could arrange that, after summing over the subsystems,
the tensor~$J^{\alpha \beta}$ is spherically symmetric,
implying that its contractions in~\eqref{noconszero} vanish.

\subsection{Computation of the conserved surface layer integral} \label{secfermicompute}
Knowing that the surface layer integral is conserved, we may compute it
using again the formula of Lemma~\ref{lemmaconservedform}.
This gives the following result:
\begin{Thm} \label{thmfermi} Under the assumptions of Theorem~{\rm \ref{thmfermiconserve}},
for the contribution~\eqref{fermiform} the surface layer integral is computed by
\begin{gather}
\int_{-\infty}^{t_0} {\rm d}t \int_{\R^3} {\rm d}^3x \int_{t_0}^\infty {\rm d}t' \int_{\R^3} {\rm d}^3y\,
 ( \nabla_{1, \u} - \nabla_{2, \u} )
 ( \nabla_{1, \v} + \nabla_{2, \v} ) \L(t,\vec{x}; t', \vec{y}) \notag \\
\qquad{} \simeq \frac{1}{\delta^4} \int_{\R^3} {\rm d}^3x \int_{\R^3} {\rm d}^3y \sum_{s,s'=\pm1}
\frac{1}{m^2} \big( G^{s,s'}+H^{s,s'} \big)\big((t,\vec{x}), (t,\vec{y}) \big)\big|_{t=0} , \label{osires}
\end{gather}
where
\begin{gather}
G^{s,s'}(x,y) \simeq \sum_{c=L,R} \im \big( ( \nabla_{1, \u} - \nabla_{2, \u} ) \Sl \Pi_s \psi^\u(x) |
\chi_c \Pi_{s'} \partial_0^2 \psi^\u(y) \Sr \notag \\
\hphantom{G^{s,s'}(x,y) \simeq}{} \times ( \nabla_{1, \v} + \nabla_{2, \v} ) \Sl \Pi_{-s} \psi^\v(x) | \chi_{\bar{c}}
\Pi_{-s'} \psi^\v(y) \Sr \big) \label{Gscalar} \\
\hphantom{G^{s,s'}(x,y) \simeq}{} -\sum_{c=L,R}
\im \big( ( \nabla_{1, \u} - \nabla_{2, \u} \big) \Sl \Pi_s \psi^\u(x) | \gamma^\alpha \chi_c \Pi_{s'}
\partial_0^2 \psi^\u(y) \Sr \notag \\
\hphantom{G^{s,s'}(x,y) \simeq}{} \times ( \nabla_{1, \v} + \nabla_{2, \v} ) \Sl
\Pi_{-s} \psi^\v(x) | \gamma_\alpha \chi_c \Pi_{-s'} \psi^\v(y) \Sr \big), \label{G0} \\
H^{s,s'}(x,y) \simeq \re \Tr \big(
 ( \nabla_{1, \u} - \nabla_{2, \u} ) | \partial_{0 \alpha} \Pi_{s'} \psi^\u(y) \Sr \Sl \Pi_s \psi^\u(x) | \notag \\
\hphantom{H^{s,s'}(x,y) \simeq}{}\times \big( \nabla_{1, \v} + \nabla_{2, \v} \big) |\sigma^{0\alpha} \Pi_s \psi^\v(x) \Sr
\Sl \Pi_{s'} \psi^\v(y) | \big) \label{H1} \\
\hphantom{H^{s,s'}(x,y) \simeq}{} - \re \Tr \big(
 ( \nabla_{1, \u} - \nabla_{2, \u} ) | \sigma^{0\alpha}
\partial_{0 \alpha} \Pi_{s'} \psi^\u(y) \Sr \Sl \Pi_s \psi^\u(x) | \notag \\
\hphantom{H^{s,s'}(x,y) \simeq}{}\times ( \nabla_{1, \v} + \nabla_{2, \v} ) |\Pi_s \psi^\v(x) \Sr
\Sl \Pi_{s'} \psi^\v(y) | \big) . \label{H2}
\end{gather}
Here~$\Pi_+$ and~$\Pi_-$ denote the projections in~$\Hil_m$ on the solutions of positive
and negative frequency, respectively.
\end{Thm}
For the proof, we need to compute the expression~\eqref{osival}.
The factor~$(y^0-t)$ gives rise to a~factor~$t$ in~\eqref{JklK}, so that we need to consider
the kernel
\[
{\mathfrak{K}}(\xi) = \frac{1}{\delta^4} \frac{\xi_k \xi_l}{t} K_0(\xi) . \]
We again work in momentum space and decompose~$\hat{\mathfrak{K}}$
similar to~\eqref{Ket} and~\eqref{Kzm} into the equal time and
zero momentum contributions,
\begin{gather}
\hat{\mathfrak{K}}_\text{et}(p) := \frac{1}{\delta^4} \frac{\partial^2}{\partial p^j p^k} \frac{1}{k}, \label{Ketc} \\
\hat{\mathfrak{K}}_\text{zm}(p) := -\frac{1}{\delta^4} \frac{\partial^2}{\partial p^j p^k} \left( \frac{1}{k} \Theta\big(p^2\big) \right) . \label{Kzmc}
\end{gather}
We again analyze these contributions after each other.

\subsubsection{The equal time contribution}
\begin{Lemma} Under the assumptions of Theorem~{\rm \ref{thmfermiconserve}},
the equal time contribution to the surface layer integrals vanishes.
\end{Lemma}
\begin{proof} Transforming~\eqref{Ketc} back to position space, one finds
\[ {\mathfrak{K}}_\text{et}(x,y) \simeq \frac{1}{\delta^4} \xi_j \xi_k \frac{\delta\big(\xi^0\big) }{|\vec{\xi}|^2} . \]
Hence
\[ \int_\scrM J^{jk}(x,y) {\mathfrak{K}}_\text{et}(x,y) \big|_{x^0=0}\, {\rm d}^4y
= \frac{1}{\delta^4} \int_{\R^3} J^{\alpha \beta}\big((0,\vec{x}), (0,\vec{y}) \big) \frac{\xi_\alpha \xi_\beta}{|\vec{\xi}|^2}\, {\rm d}^3y . \]
Under the assumptions~\eqref{conscond}, the terms~\eqref{Jab1} and~\eqref{Jab2} cancel each other,
so that~$J^{\alpha \beta}$ is zero. This gives the result.
\end{proof}

\subsubsection{The zero momentum contribution}
\begin{Lemma} \label{lemmazmc}
The contribution by the kernel~${\mathfrak{K}}_\text{zm}$ to the
conserved surface layer integrals is given by~\eqref{osires}
with~$G^{s,s'}$ and~$H^{s,s'}$ as in~\eqref{Gscalar}--\eqref{H2}.
\end{Lemma}
\begin{proof} We again write the $y$-dependence of~$J^{kl}$ according to~\eqref{psiphi} as the product
of two Dirac solutions. Moreover, we work again in momentum space~\eqref{intprod}.
Exactly as explained after~\eqref{cint}, it again suffices to consider the case that the
momenta~$q$ and~$q-p$ lie both on the upper or both on the lower mass shell.

Using~\eqref{Kzmc} and integrating by parts, we obtain
for the inner integral in~\eqref{intprod}
\begin{gather*}
\int \frac{{\rm d}^4p}{(2 \pi)^4} \big({-}\delta^4\big) \hat{\mathfrak{K}}_\text{zm}(p) {\rm e}^{-{\rm i} (p-q) x} \hat{\overline{\phi}}(p-q) \\
\qquad{}\asymp \int \frac{{\rm d}^4p}{(2 \pi)^4}
\left( \frac{\partial^2}{\partial p^j p^k} \left( \frac{1}{k} \Theta\big(p^2\big) \right) \right)
 {\rm e}^{-{\rm i} (p-q) x} \hat{\overline{\phi}}(p-q) \\
\qquad{}= \frac{\partial^2}{\partial q^j q^k}\int \frac{{\rm d}^4p}{(2 \pi)^4}
\frac{1}{k} \Theta\big(p^2\big)
 {\rm e}^{-{\rm i} (p-q) x} \hat{\overline{\phi}}(p-q) \\
\qquad{}= \frac{\partial^2}{\partial q^j q^k}\int \frac{{\rm d}^4p}{(2 \pi)^4}
\frac{1}{k} \Theta\big(p^2\big) \delta \big( (p-q)^2 - m^2 \big) h(p-q) ,
\end{gather*}
where in the last step we again used the notation~\eqref{hdef}.
Using the assumption~\eqref{conscond}, the spatial components~$J^{\alpha \beta}$
vanish. Therefore, at least one tensor index is zero. Setting for example the index~$k$ to zero,
the corresponding derivative can be carried out and integrated by parts,
\begin{gather*}
 \frac{\partial^2}{\partial q^j q^0}\int \frac{{\rm d}^4p}{(2 \pi)^4}
\frac{1}{k} \Theta\big(p^2\big) \delta \big( (p-q)^2 - m^2 \big) h(p-q) \\
\qquad{}= \frac{\partial}{\partial q^j}\int \frac{{\rm d}^4p}{(2 \pi)^4}
\left( \frac{\partial}{\partial p^0} \frac{1}{k} \Theta\big(p^2\big) \right) \delta \big( (p-q)^2 - m^2 \big) h(p-q) \\
\qquad{} = \frac{\partial}{\partial q^j}\int \frac{{\rm d}^4p}{(2 \pi)^4}
\frac{2 p^0}{k} \delta\big(p^2\big) \delta \big( (p-q)^2 - m^2 \big) h(p-q) \\
\qquad{} = 2 \frac{\partial}{\partial q^j}\int \frac{{\rm d}^4p}{(2 \pi)^4}
K_0(p) \delta \big( (p-q)^2 - m^2 \big) h(p-q)
\end{gather*}
(in the last step we used that on the mass cone, $p^0/k=\epsilon\big(p^0\big)$).
Computing the last integral with the help of Lemma~\ref{lemmaintconv}, we obtain
(again in the case that the tensor index~$k=0$)
\begin{gather}
\int \frac{{\rm d}^4p}{(2 \pi)^4} \delta^4 \hat{\mathfrak{K}}_\text{zm}(p) {\rm e}^{-{\rm i} (p-q) x} \hat{\overline{\phi}}(p-q)
\simeq \frac{q_j}{m^2} \epsilon\big( q^0 \big) h(-q) . \label{innerint}
\end{gather}

In order to relate~$h(-q)$ back to the Dirac wave function~$\phi$,
we evaluate~\eqref{hdef} at~$p=0$, multiply by~${\rm e}^{-{\rm i}qx}$ and integrate over the frequency.
We thus obtain
\begin{gather*}
\int_{-\infty}^\infty \hat{\overline{\phi}}(-q)\big|_{q=(\omega, \vec{q})} \,{\rm d}\omega
 = \int_{-\infty}^\infty {\rm e}^{-{\rm i} q x} \delta \big( q^2 - m^2 \big) h(-q) \big|_{q=(\omega, \vec{q})} \,{\rm d}\omega \\
\hphantom{\int_{-\infty}^\infty \hat{\overline{\phi}}(-q)\big|_{q=(\omega, \vec{q})} \,{\rm d}\omega}{}
= \frac{1}{2 \omega(\vec{q})} \sum_\pm {\rm e}^{-{\rm i} r x} h(-r) \big|_{r=(\pm \omega(\vec{q}), \vec{q})} ,
\end{gather*}
where~$\omega(\vec{q}) := \sqrt{|\vec{q}|^2+m^2}$.
Combining this equation with~\eqref{intprod} and~\eqref{innerint}, we obtain
\begin{gather*}
\int_{\scrM} \delta^4 {\mathfrak{K}}_\text{zm}(x,y) \psi(y) \overline{\phi}(y)\, {\rm d}^4y
\simeq \int \frac{{\rm d}^4q}{(2 \pi)^4} {\rm e}^{-{\rm i}qx} \hat{\psi}(q)
\frac{q_j}{m^2} \epsilon\big( q^0 \big) h(-q) \\
 \hphantom{\int_{\scrM} \delta^4 {\mathfrak{K}}_\text{zm}(x,y) \psi(y) \overline{\phi}(y)\, {\rm d}^4y}{} \overset{(*)}{=}\int \frac{{\rm d}^4q}{(2 \pi)^4} \hat{\psi}(q)
\frac{q_j}{m^2} \epsilon\big( q^0 \big) \sum_{\pm} {\rm e}^{-{\rm i}rx} h(-r) \Big|_{r=(\pm \omega(\vec{q}), \vec{q})} \\
\hphantom{\int_{\scrM} \delta^4 {\mathfrak{K}}_\text{zm}(x,y) \psi(y) \overline{\phi}(y)\, {\rm d}^4y}{}
=\int \frac{{\rm d}^4q}{(2 \pi)^4} \hat{\psi}(q)
\frac{q_j q_0}{m^2} \frac{1}{\omega(\vec{q})} \sum_{\pm} {\rm e}^{-{\rm i}rx} h(-r) \Big|_{r=(\pm \omega(\vec{q}), \vec{q})} \\
\hphantom{\int_{\scrM} \delta^4 {\mathfrak{K}}_\text{zm}(x,y) \psi(y) \overline{\phi}(y)\, {\rm d}^4y}{}
\simeq \int \frac{{\rm d}^4q}{(2 \pi)^4} \hat{\psi}(q)
\frac{q_j q_0}{m^2} \int_{-\infty}^\infty \hat{\overline{\phi}} (-\omega, \vec{q} ) \, {\rm d}\omega\\
\hphantom{\int_{\scrM} \delta^4 {\mathfrak{K}}_\text{zm}(x,y) \psi(y) \overline{\phi}(y)\, {\rm d}^4y}{}
\simeq \frac{1}{m^2} \int_{\R^3} \big(\partial_{0l} \psi(0,\vec{y}) \big) \overline{\phi}(0, \vec{y}) \,{\rm d}^3y ,
\end{gather*}
where in~$(*)$ we made use of the assumption that the frequencies of~$\hat{\psi}$
and~$\overline{\hat{\phi}}$ have opposite signs.
Collecting all the contributions gives the result.
\end{proof}

This concludes the proof of Theorem~\ref{thmfermi}.

\subsubsection{Building in a chiral symmetry}
Since the result of Theorem~\ref{thmfermi} is rather complicated, it is helpful
to simplify the setting by imposing additional assumptions on the form of the
fermionic jets.
Since we restrict attention to Maxwell fields which
do not break the chirality (in other words, the corresponding gauge potential~$A_j \gamma^j$
is vectorial but not axial),
we should consequently only consider the perturbations by the Dirac wave functions
which preserve the chiral degeneracy of the eigenvalues of the closed chain.
To this end, it would be natural to assume that there is a space-like unit vector~$u$ such that the
imaginary vector~$v={\rm i}u$ describes a symmetry of the fermionic projector in the sense that
\begin{gather} \label{chiralsymm}
\left\{ \begin{matrix} \text{odd} \\ \text{even} \end{matrix} \right\}
\text{contributions to~$P$ and~$\nabla P$}
 \left\{ \begin{matrix} \text{anti-commute} \\ \text{commute} \end{matrix} \right\}
 \text{with~$\slashed{v}$}
\end{gather}
(here we use the same notation as introduced in~\cite[Section~2.6.2]{cfs}).
Indeed, here it suffices to impose a weaker condition, as we now explain.
In preparation, we note that the conditions in~\eqref{conscond} imply that~$\nabla P$ can be written as
\begin{gather} \label{Pansatz}
\nabla P = \gamma^0 g + \big(\1 \mp {\rm i} \gamma^0 \big) \vec{a} \vec{\gamma} + \alpha \1
+\gamma^5 \gamma^0 h + \gamma^5 \big( \1 \mp \gamma^0 \big) \vec{b} \vec{\gamma} + \beta {\rm i} \gamma^5
\end{gather}
with complex-valued coefficients~$g,h \in \C$, $\vec{a}, \vec{b} \in \C^3$ and~$\alpha, \beta \in \C$
This ansatz indeed satisfies~\eqref{conscond} because
\begin{gather*}
\Tr \big( \big(\1 \pm {\rm i} \gamma^0 \big) \gamma^\alpha \nabla P \big)
= \Tr \big( \big(\1 \pm {\rm i} \gamma^0 \big) \gamma^\alpha \big(\1 \mp {\rm i} \gamma^0 \big) \vec{a} \vec{\gamma} \big) \\
\hphantom{\Tr \big( \big(\1 \pm {\rm i} \gamma^0 \big) \gamma^\alpha \nabla P \big)}{}
= \Tr \big( \big(\1 \pm {\rm i} \gamma^0 \big) \big(\1 \pm {\rm i} \gamma^0 \big) \gamma^\alpha \vec{a} \vec{\gamma} \big)
= \Tr \big( \big( \1 \pm 2 {\rm i} \gamma^0 - \1 \big) \gamma^\alpha \vec{a} \vec{\gamma} \big) = 0, \\
\Tr \big( \gamma^5 \big(\1 \pm \gamma^0\big) \gamma^\alpha \nabla P \big)
= \Tr \big( \gamma^5 \big(\1\pm \gamma^0\big) \gamma^\alpha
\gamma^5 \big( \1 \mp \gamma^0 \big) \vec{b} \vec{\gamma} \big) \\
\hphantom{\Tr \big( \gamma^5 \big(\1 \pm \gamma^0\big) \gamma^\alpha \nabla P \big)}{}
= -\Tr \big( \big(\1 \mp \gamma^0\big) \big( \1 \pm \gamma^0 \big) \gamma^\alpha \vec{b} \vec{\gamma} \big) = 0 .
\end{gather*}
Moreover, the fact that~\eqref{Pansatz} is the most general ansatz satisfying~\eqref{conscond}
can be verified by counting dimensions (\eqref{Pansatz} has $10$ degrees of freedom, which
is consistent with the fact that~\eqref{conscond} gives $6$ constraints).

The condition~\eqref{chiralsymm} would imply that all the contributions involving~$\gamma^5$ must vanish, i.e.,
\[ \nabla P = \gamma^0 g + \big(\1 \mp {\rm i} \gamma^0 \big) \vec{a} \vec{\gamma} + \alpha \1 \]
(and then~$v$ in~\eqref{chiralsymm} can be chosen as~$(0, {\rm i} \vec{u})$, where~$\vec{u}$
is a unit vector orthogonal to~$\vec{a}$ and~$\vec{\xi}$).
For our arguments, however, it suffices to merely assume that the vector~$\vec{b}$ vanishes:
\begin{Def} \label{defchiralsymm}
The {\it fermionic jet}~$\nabla P$ {\it is chirally symmetric} if it can be written as
\begin{gather} \label{weakchiral}
\nabla P = \gamma^0 g + \big(\1 \mp {\rm i} \gamma^0 \big) \vec{a} \vec{\gamma} + \alpha \1
+\gamma^5 \gamma^0 h + \beta {\rm i} \gamma^5
\end{gather}
with complex-valued coefficients~$g,h \in \C$, $\vec{a} \in \C^3$ and~$\alpha, \beta \in \C$.
\end{Def}
Under this assumption, we can simplify the result of Theorem~\ref{thmfermi} as follows.
\begin{Prp} \label{prpfermisimp} If the fermionic jets preserve the chirality, then
the formula for the function~$H^{s,s'}$ given in~\eqref{H1} and~\eqref{H2} can be written alternatively as
\begin{gather}
H^{s,s'}(x,y) \simeq \mp \re \Tr \big(
 ( \nabla_{1, \u} - \nabla_{2, \u} ) | \partial_{0 \alpha} \Pi_{s'} \psi^\u(y) \Sr \Sl \Pi_s \psi^\u(x) | \notag \\
\hphantom{H^{s,s'}(x,y) \simeq}{}\times ( \nabla_{1, \v} + \nabla_{2, \v} ) |\gamma^\alpha \Pi_s \psi^\v(x) \Sr
\Sl \Pi_{s'} \psi^\v(y) | \big) \label{Hf1} \\
\hphantom{H^{s,s'}(x,y) \simeq}{} \mp \re \Tr \big(
 ( \nabla_{1, \u} - \nabla_{2, \u} ) | \gamma^\alpha
\partial_{0 \alpha} \Pi_{s'} \psi^\u(y) \Sr \Sl \Pi_s \psi^\u(x) | \notag \\
\hphantom{H^{s,s'}(x,y) \simeq}{}\times ( \nabla_{1, \v} + \nabla_{2, \v} ) |\Pi_s \psi^\v(x) \Sr
\Sl \Pi_{s'} \psi^\v(y) | \big) \label{Hf2}
\end{gather}
$($the sign~$\pm$ in~\eqref{Hf1} and~\eqref{Hf2} is to be chosen in agreement with the left equation in~\eqref{conscond}$)$.
\end{Prp}

The main point of this proposition is that the matrices~$\sigma^{0\alpha}$
in~\eqref{H1} and~\eqref{H2} have been replaced by Dirac matrices~$\gamma^\alpha$.
This will be crucial for the analysis of the positivity properties of the surface layer
inner product in Section~\ref{secfermipositive}.

In preparation for the proof, we observe that the term~\eqref{H2} vanishes as a consequence of the Dirac equation:
\begin{Lemma} \label{lemmabilzero}
For any solutions~$\psi$, $\phi$ of the Dirac equation
and every~$s \in \{\pm 1\}$,
\[ \int_{\R^3} \Sl \Pi_s \phi | \sigma^{0 \alpha} \partial_{0 \alpha} \Pi_s \psi \Sr \,{\rm d}^3y = 0 . \]
\end{Lemma}
\begin{proof} First,
\begin{gather}
\int_{\R^3} \Sl \phi | \sigma^{0 \alpha} \partial_{0 \alpha} \psi \Sr\, {\rm d}^3y
 = \int_{\R^3} \Sl \phi | \gamma^0 \partial_0 \big({\rm i} \gamma^\alpha \partial_{\alpha} \psi \big) \Sr \,{\rm d}^3y
 = \int_{\R^3} \Sl \phi | \gamma^0 \partial_0 \big( m - {\rm i} \gamma^0 \partial_0 \psi \big) \Sr \,{\rm d}^3y \notag \\
\hphantom{\int_{\R^3} \Sl \phi | \sigma^{0 \alpha} \partial_{0 \alpha} \psi \Sr \, {\rm d}^3y}{}
= m \int_{\R^3} \Sl \phi | \gamma^0 \partial_0 \psi \Sr \,{\rm d}^3y
- {\rm i} \int_{\R^3} \Sl \phi | \partial^2_0 \psi \,{\rm d}^3y . \label{t1}
\end{gather}
On the other hand, integration by parts gives
\begin{gather}
\int_{\R^3} \Sl \phi | \sigma^{0 \alpha} \partial_{0 \alpha} \psi \Sr \, {\rm d}^3y
 = -\int_{\R^3} \Sl \gamma^\alpha \phi | \gamma^0 \partial_0 \big( {\rm i} \partial_{\alpha} \psi \Sr \big) \,{\rm d}^3y
 = -\int_{\R^3} \Sl {\rm i} \gamma^\alpha \partial_\alpha \phi | \gamma^0 \partial_0 \psi \Sr \,{\rm d} ^3y \notag \\
\hphantom{\int_{\R^3} \Sl \phi | \sigma^{0 \alpha} \partial_{0 \alpha} \psi \Sr \, {\rm d}^3y}{}
 = -\int_{\R^3} \Sl \big( m - {\rm i} \gamma^0 \partial_0 \big) \phi | \gamma^0 \partial_0 \psi \Sr \,{\rm d}^3y \notag \\
\hphantom{\int_{\R^3} \Sl \phi | \sigma^{0 \alpha} \partial_{0 \alpha} \psi \Sr \, {\rm d}^3y}{}
 = -m \int_{\R^3} \Sl \phi | \gamma^0 \partial_0 \psi \Sr \,{\rm d}^3y
-{\rm i} \int_{\R^3} \Sl \partial_0 \phi | \partial_0 \psi \Sr \,{\rm d}^3y . \label{t2}
\end{gather}
Taking the mean of~\eqref{t1} and~\eqref{t2} gives
\begin{gather} \label{dtcombi}
\int_{\R^3} \Sl \phi | \sigma^{0 \alpha} \partial_{0 \alpha} \psi \Sr \,{\rm d}^3y
= -\frac{{\rm i}}{2} \int_{\R^3} \big( \Sl \phi | \partial^2_0 \psi \Sr
+ \Sl \partial_0 \phi | \partial_0 \psi \Sr \big) \, {\rm d}^3y .
\end{gather}
Due to momentum conservation,
we only get a contribution if the momenta of~$\hat{\phi}$ and~$\hat{\psi}$ coincide.
Thus the corresponding frequencies coincide up to signs.
The combination in~\eqref{dtcombi} vanishes if the frequencies have the same sign.
This gives the result.
\end{proof}

\begin{proof}[Proof of Proposition~\ref{prpfermisimp}]
It remains to simplify the term~\eqref{H1}. We first note that this term can be written as
\begin{gather*}
 \int_{\R^3} {\rm d}^3x \int_{\R^3} {\rm d}^3y \sum_{s,s'=\pm1}
\frac{1}{m^2} H^{s,s'}\big((t,\vec{x}), (t,\vec{y}) \big)\Big|_{t=0} \notag \\
\qquad{} = \int_\scrM \mathfrak{K}(x,y)
 ( \nabla_{1, \u} - \nabla_{2, \u} ) ( \nabla_{1, \v} + \nabla_{2, \v} )
\Tr \big( \sigma^{0 \alpha} P P^* \big) \, {\rm d}^4y . 
\end{gather*}
For ease in notation, we write the contributions as
\begin{gather}
\int_\scrM \mathfrak{K}(x,y)
\Tr \big( \sigma^{0 \alpha} \nabla P \nabla P^* \big) \,{\rm d}^4y , \label{intertwo}
\end{gather}
where~$\nabla$ stands for any of the combinations of derivatives in~\eqref{H1}.
Carrying out the $y$-integration as in the proof of Lemma~\ref{lemmazmc},
we only get a contribution if the two wave functions at~$y$ have frequencies of opposite
signs, and in this case, we get additional derivatives~$\partial_{0i}$
acting on one of the wave functions at~$y$.
Applying again Lemma~\ref{lemmabilzero}, we find that the following integral vanishes,
\[ \int_\scrM \mathfrak{K}(x,y)
\Tr \big( \nabla P \sigma^{0 \alpha} \nabla P^* \big) \, {\rm d}^4y = 0 . \]
Adding this integral to~\eqref{intertwo}, we obtain
\begin{gather}
\eqref{intertwo} = \int_\scrM \mathfrak{K}(x,y)
\Tr \big( \sigma^{0 \alpha} \nabla P \nabla P^*
+ \nabla P \sigma^{0 \alpha} \nabla P^* \big) \, {\rm d}^4y \notag \\
\hphantom{\eqref{intertwo}}{} = \int_\scrM \mathfrak{K}(x,y)
\Tr \big( \sigma^{0 \alpha} \big\{ \nabla P, \nabla P^* \big\} \big) \,{\rm d}^4y . \label{inter3}
\end{gather}

We next compute the anti-commutator of~$\nabla P$ and~$\nabla P^*$
using the ansatz~\eqref{weakchiral}. Taking its adjoint, we obtain
\[ \nabla P^* = \gamma^0 \overline{g} + \big(\1 \mp {\rm i} \gamma^0 \big) \overline{\vec{a}} \vec{\gamma} +
\overline{\alpha} \1
+\gamma^5 \gamma^0 \overline{h} + \overline{\beta} {\rm i} \gamma^5 . \]
Thus the contribution~${\sim} \sigma^{0 \alpha}$ is computed by
\[ \big\{\nabla P, \nabla P^* \} \asymp
+ \big\{ {\mp} {\rm i} \gamma^0 \vec{a} \vec{\gamma}, \overline{\alpha} \big\} + \big\{\alpha , \mp {\rm i} \gamma^0 \overline{\vec{a}} \vec{\gamma} \big\}
= \mp 2 {\rm i} \gamma^0 \big( \vec{a} \vec{\gamma} \overline{\alpha} + \overline{\vec{a}} \vec{\gamma} \alpha \big) . \]
Likewise, the contribution~${\sim} \gamma^\alpha$ to the anti-commutator is given by
\[ \{\nabla P, \nabla P^* \} \asymp \big\{ \vec{a} \vec{\gamma}, \overline{\alpha}\big\}
+ \big\{\alpha , \overline{\vec{a}} \vec{\gamma} \big\} = 2 \vec{a} \vec{\gamma} \overline{\alpha} + 2 \overline{\vec{a}} \vec{\gamma} \alpha . \]
It follows that
\begin{gather*}\begin{split}
&\Tr \big( \sigma^{0 \alpha} \{ \nabla P, \nabla P^* \} \big)
 = \pm 2 \Tr \big( \gamma^0 \gamma^\alpha
\gamma^0 \big( \vec{a} \vec{\gamma} \overline{\alpha} + \overline{\vec{a}} \vec{\gamma} \alpha \big) \big) \\
&\hphantom{\Tr \big( \sigma^{0 \alpha} \{ \nabla P, \nabla P^* \} \big)}{}= \mp 2 \Tr \big( \gamma^\alpha
\big( \vec{a} \vec{\gamma} \overline{\alpha} + \overline{\vec{a}} \vec{\gamma} \alpha \big) \big)
= \mp \Tr \big( \gamma^\alpha \{ \nabla P, \nabla P^* \} \big) .\end{split}
\end{gather*}
Thus we may replace the factor~$\sigma^{0 \alpha}$ in~\eqref{inter3} by~$\mp \gamma^\alpha$,
\begin{gather*}
 \int_{\R^3} {\rm d}^3x \int_{\R^3} {\rm d}^3y\, \sum_{s,s'=\pm1}
\frac{1}{m^2} H^{s,s'}\big((t,\vec{x}), (t,\vec{y}) \big)\Big|_{t=0} \notag \\
 \qquad{} = \mp \int_\scrM \mathfrak{K}(x,y)
\Tr \big( \gamma^\alpha \{ \nabla P, \nabla P^* \} \big) \, {\rm d}^4y .
\end{gather*}
This gives the formula for~$H^{s,s'}$ in~\eqref{Hf1} and~\eqref{Hf2},
concluding the proof.
\end{proof}

\subsection[Exciting the Dirac sea and positivity of~$(\cdot,\cdot)$]{Exciting the Dirac sea and positivity of~$\boldsymbol{(\cdot,\cdot)}$} \label{secfermipositive}
We now consider the special case that~$\psi^\u$ and~$\psi^\v$ have negative frequencies,
whereas~$\delta \psi^\u$ and~$\delta \psi^\v$ have positive frequencies.
Then the result of Theorem~\ref{thmfermi} simplifies to
 \begin{gather}
\int_{-\infty}^{t_0}{\rm d} t \int_{\R^3} {\rm d}^3x \int_{t_0}^\infty {\rm d}t' \int_{\R^3} {\rm d}^3y\,
 ( \nabla_{1, \u} - \nabla_{2, \u} )
 ( \nabla_{1, \v} + \nabla_{2, \v} ) \L(t,\vec{x}; t', \vec{y})\nonumber \\
\qquad{} \simeq \frac{1}{\delta^4} \int_{\R^3} {\rm d}^3x \int_{\R^3} {\rm d}^3y\,
\frac{1}{m^2} \big( G+H \big)\big((t,\vec{x}), (t,\vec{y}) \big)\Big|_{t=0} ,
 \label{osiex}
\end{gather}
where
\begin{gather}
G(x,y) \simeq \sum_{c=L,R} \im \big( \Sl \delta \psi^\u(x) |
\chi_c \partial_0^2 \psi^\u(y) \Sr \Sl \psi^\v(x) | \chi_{\bar{c}} \delta \psi^\v(y) \Sr \label{G1} \\
\hphantom{G(x,y) \simeq}{} - \Sl \psi^\u(x) |
\chi_c \partial_0^2 \delta \psi^\u(y) \Sr \Sl \delta \psi^\v(x) | \chi_{\bar{c}} \psi^\v(y) \Sr \big) \nonumber\\
\hphantom{G(x,y) \simeq}{} -\sum_{c=L,R} \im \big( \Sl \delta \psi^\u(x) | \gamma^\alpha \chi_c
\partial_0^2 \psi^\u(y) \Sr \Sl \psi^\v(x) | \gamma_\alpha \chi_c \delta \psi^\v(y) \Sr \nonumber\\
\hphantom{G(x,y) \simeq}{} + \Sl \psi^\u(x) | \gamma^\alpha \chi_c
\partial_0^2 \delta \psi^\u(y) \Sr \Sl \delta \psi^\v(x) | \gamma_\alpha \chi_c \psi^\v(y) \Sr \big),\label{G4} \\
H(x,y) \simeq \mp \re \big(
\Sl \delta \psi^\u(x) |\gamma^\alpha \delta \psi^\v(x) \Sr \Sl \psi^\v(y) | \partial_{0 \alpha} \psi^\u(y) \Sr \nonumber\\
\hphantom{H(x,y) \simeq}{} -\Sl \psi^\u(x) | \gamma^\alpha \psi^\v(x) \Sr \Sl \delta \psi^\v(y) | \partial_{0 \alpha} \delta \psi^\u(y) \Sr \big) \nonumber\\
\hphantom{H(x,y) \simeq}{} \mp \re \big(
\Sl \delta \psi^\u(x) | \delta \psi^\v(x) \Sr \Sl \psi^\v(y) | \gamma^\alpha \partial_{0 \alpha} \psi^\u(y) \Sr \nonumber\\
\hphantom{H(x,y) \simeq}{} -\Sl \psi^\u(x) |\gamma^\alpha \psi^\v(x) \Sr \Sl \delta \psi^\v(y) | \partial_{0 \alpha} \delta \psi^\u(y) \Sr \big) .\nonumber
\end{gather}
Due to the integration over~$\vec{y}$, the momenta of~$\psi^\u$ and~$\delta \psi^\v$
in~\eqref{G1} coincide. Hence their frequencies coincide except for a sign, implying that
the derivatives~$\partial_0^2$ can act just as well on the function~$\delta \psi^\v$.
Arguing similarly for the other summands, one sees that~$G$ is anti-symmetric in~$\u$ and~$\v$,
whereas~$H$ is symmetric. Therefore, $G$ and~$H$ give rise to the symplectic form and the
surface layer inner product, respectively. Rewriting the spatial integrals in momentum space
gives the following result.
\begin{Prp} \label{prpex}
Assume that~$\psi^\u$ and~$\psi^\v$ are Dirac solutions on the lower mass shell,
where\-as~$\delta \psi^\u$ and~$\delta \psi^\v$ are solutions on the upper mass shell. Then
the symplectic form~$\sigma(\cdot,\cdot )$ and the surface layer inner product~$(\cdot,\cdot )$ are given by
\begin{gather}
\sigma(\u, \v) \simeq \frac{1}{\delta^4} \int_{\R^3} \frac{{\rm d}^3k}{(2 \pi)^3} \int_{\R^3} \frac{{\rm d}^3q}{(2 \pi)^3}
 \frac{1}{m^2} \big( \omega(\vec{q})^2 + \omega(\vec{k})^2 \big) \notag \\
\hphantom{\sigma(\u, \v) \simeq}{} \times \sum_{c=L,R} \im \big( \Sl \delta \hat{\psi}^\u(\vec{k}) | \chi_c \hat{\psi}^\u(-\vec{q}) \Sr
\Sl \hat{\psi}^\v(-\vec{k}) | \chi_{\bar{c}} \delta \hat{\psi}^\v(\vec{q}) \Sr \notag \\
\hphantom{\sigma(\u, \v) \simeq}{}
- \Sl \delta \hat{\psi}^\u(\vec{k}) | \gamma^\alpha \chi_c \hat{\psi}^\u(-\vec{q}) \Sr
\Sl \hat{\psi}^\v(-\vec{k}) | \gamma_\alpha \chi_c \delta \hat{\psi}^\v(\vec{q}) \Sr\big), \label{sympex} \\
(\u, \v) \simeq \mp \frac{1}{\delta^4} \int_{\R^3} \frac{{\rm d}^3k}{(2 \pi)^3} \int_{\R^3} \frac{{\rm d}^3q}{(2 \pi)^3}
\re \big( \Sl \delta \hat{\psi}^\u(\vec{k}) | \delta \hat{\psi}^\v(\vec{k}) \Sr \Sl \hat{\psi}^\v(\vec{q}) | \hat{\psi}^\u(\vec{q}) \Sr \big) \notag \\
\hphantom{(\u, \v) \simeq}{} \times \frac{1}{m^3} \big\{ \vec{k} \cdot \vec{q} \big( \omega(\vec{q}) + \omega(\vec{k}) \big) + |\vec{q}|^2 \omega(\vec{q}) + |\vec{k}|^2 \omega(\vec{k}) \big\} , \label{osispex}
\end{gather}
where~$\omega(\vec{k}) := \sqrt{|\vec{k}|^2+m^2}$
$($and all functions are evaluated at~$t=0)$.
\end{Prp}
\begin{proof} The formula for the symplectic form~\eqref{sympex} follows immediately
by rewriting the spatial integrals in~\eqref{osiex} as integrals over momentum space
and using~\eqref{G1}--\eqref{G4}. The formula for the inner product~\eqref{osispex}
follows similarly if one removes the Dirac matrices inside the spinorial expectation values
using Lemma~\ref{lemmanogamma} below. This concludes the proof.
\end{proof}

\begin{Lemma} \label{lemmanogamma}
For any solutions~$\psi$, $\phi$ of the Dirac equation and every~$s \in \{\pm 1\}$,
\[ \int_{\R^3} \Sl \Pi_s \phi | \gamma_\alpha \Pi_s \psi \Sr \, {\rm d}^3x
= \frac{1}{m} \int_{\R^3} \Sl \Pi_s \phi | {\rm i} \partial_{\alpha} \Pi_s \psi \Sr \,{\rm d}^3x . \]
\end{Lemma}
\begin{proof} We begin with the expression involving a derivative on the right hand side,
\begin{gather*}
\int_{\R^3} \Sl \phi | \partial_{\alpha} \psi \Sr \,{\rm d}^3x
= \frac{1}{2} \int_{\R^3} \Sl \phi | \big\{ \gamma_\alpha, \gamma^\beta \big\}
 \partial_{\beta} \psi \Sr \,{\rm d}^3x \\
\hphantom{\int_{\R^3} \Sl \phi | \partial_{\alpha} \psi \Sr \,{\rm d}^3x}{}
 = -\frac{{\rm i}}{2} \int_{\R^3} \Sl \big(\phi | \gamma_\alpha {\rm i} \gamma^\beta \partial_{\beta} \psi \big) \Sr \, {\rm d}^3x
- \frac{{\rm i}}{2} \int_{\R^3} \Sl {\rm i} \gamma^\beta \partial_{\beta} \phi | \gamma_\alpha \psi \Sr \,{\rm d}^3x \\
\hphantom{\int_{\R^3} \Sl \phi | \partial_{\alpha} \psi \Sr \,{\rm d}^3x}{}
 = -\frac{{\rm i}}{2} \int_{\R^3} \Sl\big( \phi | \gamma_\alpha \big(m - {\rm i} \gamma^0 \partial_0 \big)
\psi \big) \Sr \, {\rm d}^3x \\
\hphantom{\int_{\R^3} \Sl \phi | \partial_{\alpha} \psi \Sr \,{\rm d}^3x=}{}
-\frac{{\rm i}}{2} \int_{\R^3} \Sl \big(\big(m - {\rm i} \gamma^0 \partial_0 \big) \phi | \gamma_\alpha \psi \big) \Sr \,{\rm d}^3x \\
\hphantom{\int_{\R^3} \Sl \phi | \partial_{\alpha} \psi \Sr \,{\rm d}^3x}{}
 = -{\rm i} m \int_{\R^3} \Sl (\phi | \gamma_\alpha \psi ) \Sr \, {\rm d}^3x \\
\hphantom{\int_{\R^3} \Sl \phi | \partial_{\alpha} \psi \Sr \,{\rm d}^3x=}{}
 -\frac{1}{2} \int_{\R^3} \Sl \big(\phi | \gamma_\alpha \gamma^0 \partial_0 \psi \big) \Sr \,{\rm d}^3x
+ \frac{1}{2} \int_{\R^3} \Sl \big(\gamma^0 \partial_0 \phi | \gamma_\alpha \psi \big) \Sr \,{\rm d}^3x \\
\hphantom{\int_{\R^3} \Sl \phi | \partial_{\alpha} \psi \Sr \,{\rm d}^3x}{}
 = -{\rm i} m \int_{\R^3} \Sl (\phi | \gamma_\alpha \psi ) \Sr \, {\rm d}^3x \\
\hphantom{\int_{\R^3} \Sl \phi | \partial_{\alpha} \psi \Sr \,{\rm d}^3x=}{}
 -\frac{1}{2} \int_{\R^3} \Sl \big(\phi | \gamma_\alpha \gamma^0 \partial_0 \psi \big) \Sr \,{\rm d}^3x
- \frac{1}{2} \int_{\R^3} \Sl \big(\partial_0 \phi | \gamma_\alpha \gamma^0 \psi \big) \Sr \,{\rm d}^3x .
\end{gather*}
The last line vanishes if~$\hat{\phi}$ and~$\hat{\psi}$ have the same frequency, giving the result.
\end{proof}

\begin{Prp}[definiteness of the surface layer inner product] \label{prpdefinite}
Under the assumptions of Proposition~{\rm \ref{prpex}},
depending on the sign in~\eqref{conscond}, the surface layer inner product is
either positive or negative semi-definite. It vanishes if and only if
\[ \vec{q} = - \vec{k} \qquad \text{for all~$\vec{q}$, $\vec{k}$ with~$\delta \hat{\psi}^u(\vec{k})$
\ and \ $\hat{\psi}^u(\vec{q}) \neq 0$} . \]
\end{Prp}
\begin{proof} We evaluate~\eqref{osispex} for~$\u = \v$.
Since~$\psi$ and~$\delta \psi$ are supported on the lower and upper mass shell,
respectively, the real part in~\eqref{osispex} is necessarily negative.
In order to show that the curly brackets in~\eqref{osispex} are non-negative,
we consider the worst possible case that~$\vec{k} \cdot \vec{q} = - |\vec{k}| |\vec{q}|$.
In this case, the curly brackets in~\eqref{osispex} simplify to
\[ |\vec{q}|^2 \omega(\vec{q}) + |\vec{k}|^2 \omega(\vec{k})
- |\vec{k}| |\vec{q}| \big( \omega(\vec{q}) + \omega(\vec{k}) \big)
= \big[ |\vec{q}| - |\vec{k}| \big] \big[ |\vec{q}| \omega(\vec{q}) - |\vec{k}| \omega(\vec{k}) \big] . \]
Since the function~$\omega(\vec{k})$ is strictly monotone increasing in~$|\vec{k}|$,
the two square brackets always have the same sign, and they vanish if and only if~$|\vec{q}| = |\vec{k}|$.
This concludes the proof.
\end{proof}

This result shows that, by a suitable choice of the sign in the first equation in~\eqref{conscond},
one can arrange that the surface layer inner product is positive definite.

\subsection[The contributions ${\sim} \delta^{-4} \cdot j J$ and ${\sim} \delta^{-4} \cdot j^2$]{The contributions $\boldsymbol{{\sim} \delta^{-4} \cdot j J}$ and $\boldsymbol{{\sim} \delta^{-4} \cdot j^2}$} \label{secaddition}

We remark that there are also contributions~${\sim} \delta^{-4} \cdot j J$
which are linear in the Dirac and linear in the Maxwell current, as well as
contributions~${\sim} \delta^{-4} \cdot j^2$ quadratic in the Maxwell current
(as well as expansions of these contributions in powers of~$\varepsilon/t$).
Moreover, there are contributions involving~$\Box j$ or higher derivatives of
the field tensor. All these
contributions have a different structure than the contributions quadratic in the
Dirac current (mainly because the Maxwell current gives rise to unbounded line integrals
as explained in Section~\ref{secunbound} for the field tensor terms).
As a~consequence, it seems impossible that these contributions partially cancel
contributions~${\sim} J^2$. With this in mind, for brevity we shall not enter the
analysis of these contributions.

\section{Computation of a positive surface layer integral} \label{secpositive}
In~\cite{positive} positive functionals in spacetime were derived.
We shall now verify that in Minkowski space, these functionals are indeed positive.
There are the two positive functionals involving volume integrals
(see~\cite[Theorem~1.1]{positive})
\begin{gather}
\int_M \nabla^2 \ell|_x(\u,\u) \,{\rm d}\rho(x) \geq 0, \label{Fpos1} \\
\int_M {\rm d}\rho(x) \int_M {\rm d}\rho(y)\, \nabla_{1,\u} \nabla_{2,\u} \L(x,y)
+ \int_M \nabla^2 \ell|_x(\u,\u) \, {\rm d}\rho(x) \geq 0 \label{Fpos2}
\end{gather}
as well as the positive surface layer integral (see~\eqref{osipos} and~\cite[Proposition~7.1]{positive})
\begin{gather} \label{osipos2}
-\int_\Omega {\rm d}\rho(x) \int_{M \setminus \Omega} {\rm d}\rho(y) \nabla_{1,\v} \nabla_{2,\v} \L(x,y) \geq 0 .
\end{gather}
Since in the formalism of the continuum limit, the functions~$\ell(x)$ and~$\L(x,y)$
vanish in the Minkowski vacuum, the corresponding measure~$\rho$ clearly is a minimizer.
Therefore, the inequali\-ties~\eqref{Fpos1} and~\eqref{Fpos2} obviously hold.

The positivity of the surface layer integral~\eqref{osipos2} is less obvious.
Therefore, it will be instructive to compute this surface layer integral in Minkowski space and to verify that it
is indeed positive. We begin with the contributions by the Maxwell current
(as we shall see below, these are indeed the dominant contributions).
The corresponding contribution to the fermionic projector is given by (see~\cite[equation~(D.0.7)]{cfs})
\begin{gather} \label{jint}
\frac{1}{4} \Tr \left( \slashed{\xi} \Delta P(x,y) \right)
\asymp -2 \int_x^y \big(\alpha-\alpha^2\big) \xi_k j^k T^{(1)}_{[0]}(x,y) .
\end{gather}
The resulting contribution to the perturbation of the eigenvalues of the closed chain
is computed by replacing the Maxwell current~$j$ in~\eqref{lamj} by the
line integral in~\eqref{jint},
\[
\Delta \lambda^{xy}_{c+} = {\rm i}g^2 \Big(
2 T^{(1)}_{[0]} \overline{T^{(-1)}_{[0]}} -T^{(0)}_{[0]} \overline{T^{(0)}_{[0]}} \Big)
\int_x^y \big(\alpha-\alpha^2\big) \xi_k j^k . \]
Hence the absolute values of the eigenvalues are perturbed by
\begin{gather*}
\Delta \big| \lambda^{xy}_{c+} \big|
 =\frac{1}{|\lambda^{xy}_{c+}|} \re \big( \big( \Delta \lambda^{xy}_{c+} \big) \overline{\lambda^{xy}_{c+}} \big) \\
\hphantom{\Delta \big| \lambda^{xy}_{c+} \big|}{} = {\rm i}g^2 \left( \left(
2 \frac{\big|T^{(-1)}_{[0]} \big|}{\big|T^{(0)}_{[0]} \big|}
 T^{(1)}_{[0]} \overline{T^{(0)}_{[0]}} -
\frac{\big|T^{(0)}_{[0]} \big|}{\big|T^{(-1)}_{[0]} \big|}
\overline{T^{(0)}_{[0]}} T^{(-1)}_{[0]} \right)- \text{c.c.} \right)
\int_x^y \big(\alpha-\alpha^2\big) \xi_k j^k \\
\hphantom{\Delta \big| \lambda^{xy}_{c+} \big|}{} \simeq \int_x^y \big(\alpha-\alpha^2\big) \xi_k j^k (\deg=2) .
\end{gather*}

If the potential is perturbed only at the point~$x$, we may apply Lemma~\ref{lemmaunbounded} to obtain
\[ \nabla_{1,\u} \big| \lambda^{xy}_{c+} \big| \simeq
\int_{-\infty}^\infty \epsilon(\alpha) \big(\alpha-\alpha^2\big) \xi_k j^k|_{z=\alpha y + (1-\alpha)x}\, {\rm d}\alpha (\deg=2) . \]
If~$y=x$, the unbounded line integral gives rise to a pole.
Therefore, in spherical symmetry we get
\begin{gather} \label{nab1j}
\nabla_{1,\u} \big| \lambda^{xy}_{c+} \big| \simeq
\left(\xi_k \big({\textstyle{\fint}} j^k \big)(x) \frac{1}{t^3} + \O\big( t^{-2} \big) \right) (\deg=2) ,
\end{gather}
where the symbol~$\fint$ is a short notation for the integral
\[ {\textstyle{\fint}} j^k = \int_{\infty}^\infty \alpha^2 \epsilon(\alpha) \xi_k j^k\left(x^0+\alpha, \vec{x} +
\alpha \frac{\vec{y}-\vec{x}}{\big| \vec{y}-\vec{x} \big|}\right) {\rm d}\alpha . \]
Likewise, exchanging the roles of~$x$ and~$y$, we obtain
\[ \nabla_{2,\u} \big| \lambda^{xy}_{c+} \big| \simeq
-\left(\xi_k \big({\textstyle{\fint}} j^k \big)(x) \frac{1}{t^3} + \O\big( t^{-2} \big) \right) (\deg=2) \]
(where the minus sign comes about because the factor~$t^{-3}$ in~\eqref{nab1j} flips its sign).
Using these formulas in~\eqref{DeltaLag} gives
\begin{gather}
\nabla_{1,\v} \nabla_{2,\v} \L(x,y)
 \simeq \nabla_{1,\u} \big| \lambda^{xy}_{c+} \big| \nabla_{2,\u} \big| \lambda^{xy}_{c+} \big|
 \simeq -\left(\big({\textstyle{\fint}} j_k \big)(x) \big({\textstyle{\fint}} j_l \big)(y) \xi^k \xi^l \frac{1}{t^6} + \O\big( t^{-5} \big) \right) (\deg=4) \notag \\
\hphantom{\nabla_{1,\v} \nabla_{2,\v} \L(x,y)}{} \simeq - \big({\textstyle{\fint}} j_k \big)(x) \big({\textstyle{\fint}} j_l \big)(y) \xi^k \xi^l \frac{1}{\varepsilon^3 t^{10}} \delta \big( |t|-r \big) . \label{Lpole}
\end{gather}
Using this formula in~\eqref{nab1j}, the $y$-integral diverges at~$y=x$.
Since this divergence is resolved on the regularization scale~$y-x \sim \varepsilon$,
we obtain a very large positive contribution.
More precisely, per spatial volume one obtains a contribution to the surface layer integral
with the scaling behavior
\[ 0 \leq -\int_{-\infty}^{t_0} {\rm d}t \int_{t_0}^\infty {\rm d} t' \int_{\R^3} {\rm d}^3y\,
\nabla_{1,\v} \nabla_{2,\v} \L\big((t,\vec{x}),(t',\vec{y}) \big)
\simeq \frac{\varepsilon^2}{\varepsilon^{10-1-4}} {\rm d}^3x =
\frac{1}{\varepsilon^6} . \]
This contribution clearly dominates the contributions involving the Dirac current,
because the Dirac current does not lead to unbounded line integrals, giving rise to
contributions to~$\L(x,y)$ which are less singular at~$y=x$.
We conclude that the surface layer integral in~\eqref{osipos2} is indeed positive.

We close with two remarks. We first point out that, since the currents have contributions
for space-like momenta (see Fig.~\ref{figsupport}), we cannot again
use the arguments in Sections~\ref{secosiconserve}
and~\ref{secreduce} to conclude that the unbounded line integrals drop out of
the surface layer integrals. It seems impossible to avoid the pole of order~$t^{-10}$
in~\eqref{Lpole}. We also remark that the contribution~\eqref{Lpole}
does not seem to have any direct physical significance. But it could nevertheless
be useful for analytic studies and estimates of the causal action principle.

\section{Remarks and outlook} \label{secdiscuss}
In this paper, we computed various contributions to the causal Lagrangian and
analyzed the corresponding the surface layer integrals.
We considered the contributions to decreasing degree on the light cone and in an expansion in
powers of~$\varepsilon/\delta$.
A subtle point of the analysis is the dependence on the ultraviolet
regularization of the fermionic projector. In physical terms, this ultraviolet regularization
describes the microscopic spacetime structure of the causal fermion system.
Since the detailed structure of spacetime on the Planck scale is unknown,
it is also not known how the physically correct regularization of the fermionic projector should look like.
Despite this general shortcoming, we saw that most results of this paper are independent
of the regularization. However, we also saw that certain contributions did depend on the
regularization, leading to additional assumptions and the discussion of different cases.
For clarity, in the next remark we recall which assumptions were made
at which point in the paper.

\begin{Remark}[assumptions on the regularization] \label{remreg}\quad
\begin{itemize}\itemsep=0pt
\item[$\blacktriangleright$] In the computation of the {\it bosonic conserved surface layer integral} in Section~\ref{secosib},
in order to make sense of the convolution integrals we had to make the assumptions~(a)--(c) on p.~\pageref{(a)}, (b$'$) on p.~\pageref{bp} as well as~(d) and~(e) on p.~\pageref{dp}.
These assumptions pose implicit conditions on the form of the regularized fermionic projector near the light cone.
\item[$\blacktriangleright$] In the computation of the {\it fermionic conserved surface layer integral} in Section~\ref{secosif},
we had to impose conditions on the regularization in order to ensure that the surface layer
integral is conserved. Necessary conditions are stated in~\eqref{conscond}, whereas
more general conditions are discussed in Section~\ref{secconservefinal}.
\item[$\blacktriangleright$] In Section~\ref{secdF21} we found bosonic contributions to the symplectic form
of the order \linebreak \mbox{${\sim} \delta^{-4} \cdot F^2 \varepsilon/t$}. These contributions are by
a scaling factor~$\delta/\varepsilon$ larger than the fermionic and bosonic contributions
stated in the introduction (see~\eqref{symplecticfermi} and~\eqref{symplecticbose}).
These contributions could vanish for specific regularizations, but they seem to be
non-zero in general. These contributions are discussed further in Remark~\ref{remsympdiv} below.
\end{itemize}
We also point out that our analysis is not exhaustive.
In Section~\ref{secaddition} we discuss those contributions which we do not analyze in detail.
\end{Remark}

We close with a few remarks. The first two remarks point to possible modifications
of our constructions which might be worth exploring in detail in the future.
\begin{Remark}[vanishing of contributions~$\big|\lambda^{xy}_{ncs}\big| \sim \delta^{-2}$]\label{remd4}
In Section~\ref{secdd} it was explained why it is natural and desirable to assume that in the vacuum,
the absolute values of the closed chain agree up to contributions of the order~$\delta^{-4}$
(see~\eqref{Dl4}). But we would like to point out that the argument leading to this assumption
was not compelling. We now outline how our constructions and results would have to be modified if
contributions
\begin{gather} \label{newcontri}
\Delta |\lambda_{ncs}| \sim \delta^{-2} (\deg=2)
\end{gather}
were present. The fact that this would no longer lead to physically sensible results can serve as
a further explanation why the contributions~\eqref{newcontri} must indeed be zero.

First, instead of the contributions~${\sim} \delta^{-4} \cdot F^2 \varepsilon/t$
in Section~\ref{secdF21}, one would have contributions~${\sim}\delta^{-2} \cdot F^2 \varepsilon/t$.
These contributions differ from the contributions~${\sim}\delta^{-4} \cdot F^2 \varepsilon^2/t^2$
in Section~\ref{secdF22} only by a constant prefactor~$\delta^2/\varepsilon^2$.
Therefore, the results of Theorem~\ref{thmbose} would remain valid except for this prefactor,
giving rise to the scalings
\begin{gather} \label{newscale}
\sigma(\u, \v), \big(\u, \v \big) \sim \frac{1}{\varepsilon^2 \delta^2} .
\end{gather}
The advantage of this procedure is that it becomes unnecessary to introduce a
regularization condition to arrange the conservation of the surface layer
integral~${\sim} \delta^{-4} \cdot F^2 \varepsilon/t$ (see the discussion after Proposition~\ref{prp414}).

The drawback of having contributions~$\Delta |\lambda_{ncs}| \sim \delta^{-2}$
is that the fermionic surface layer integrals would no longer have the correct scaling.
In particular, the contributions in Section~\ref{secd4JJ} would be of the order~${\sim}\delta^{-2}
\cdot J^2 \varepsilon/t$, being by a scaling factor~$\varepsilon/t$ smaller than the bosonic surface layer
integrals in~\eqref{newscale}. A further regularization expansion or
an expansion in powers of~$\delta^{-2}$ would not be helpful at this point,
because this would make the resulting contributions to the fermionic surface layer integrals even smaller.
We conclude that without assuming that the contributions~\eqref{newcontri} vanish,
the bosonic and fermionic components of the surface layer integral would necessarily have a
different scaling behavior in~$\delta/\varepsilon$. This seems to be an obstruction for getting
a scalar product on the jets having bosonic and fermionic components as needed for
getting the connection to quantum field theory~\cite{fockbosonic, fockfermionic}.
\end{Remark}

\begin{Remark}[shear and general surface states only in neutrino sector]\label{remshear}
Following the procedure in~\cite{cfs}, here we consider the shear and general surface states
only in the neutrino sector. This procedure seems natural
and has the advantage that it also breaks the chiral
symmetry in the neutrino sector, as is needed in order to explain why the neutrinos do not
take part in the strong and electromagnetic interactions.
However, one could introduce shear and general surface states also in the charged sectors,
provided that they preserve the chiral symmetry.
For the results in~\cite{cfs}, this procedure would not have any influence on the results
because in the analysis of the continuum limit, only the difference of the regularization effects
in the charged sectors and the neutrino sector comes into play.
This is the main reason why in~\cite{cfs} we could simply disregard shear and general surface states in the
charged sectors.

For the computations in this paper, having shear and general surface states in the charged sectors
would make a substantial difference. Namely, when expanding in powers of~$\delta^{-2}$,
these factors could also appear in the eigenvalues of the charged sectors. For example,
in addition to the contributions to the Lagrangian~${\sim}\delta^{-4} \cdot J^2$, there would also
be contributions~${\sim} \delta^{-4} J \cdot J$ (where the dot again refers to the notation~\eqref{notation}).
These additional contributions would not change the general structure of our computations,
but they would affect the quantitative details in such a way that it is difficult to predict
how our results would have to be modified.
\end{Remark}

We finally mention a question which our analysis did not answer:
\begin{Remark}[separate conservation of the bosonic symplectic form]\label{remsympdiv}
In Section~\ref{secdF21} a conservation law for the bosonic symplectic form was
derived (see Proposition~\ref{prpsympdiv}). We now discuss the significance of this conservation law.

The fact that the bosonic symplectic form appears several times to different degrees
on the light cone (cf.\ Proposition~\ref{prpsympdiv} and Theorem~\ref{thmbose})
can be understood similar as explained for the classical field equations in the introduction.
The expression in~\eqref{sympdiv} has the surprising feature that
it diverges if the infrared regularization is removed.
This does not pose any principal problem, because one can consider the system in finite spatial volume
and take the infinite volume limit.
But the infrared divergence in~\eqref{sympdiv} implies that the bosonic and fermionic
parts of the symplectic form must be conserved independently.
While this result seems physically sensible, we would like to point out that this result is not
compelling. Indeed, it is conceivable that for physical regularizations, the constant~$c$
in~\eqref{sympdiv} is zero. The main unknown at the present stage is why the
contributions of the order~$\delta^{-4} \cdot F^2 \varepsilon/t$ are conserved
(as discussed after Proposition~\ref{prp414}). In order to clarify the situation,
one would have to compute the effect of the terms without logarithms in more detail
and analyze the question if one really needs a~regularization condition in order to ensure the conservation
of the surface layer integral. If the answer is affirmative, one would have to analyze
which condition on the regularization is required. This would lead to the question if this
regularization condition also implies that the contribution~\eqref{sympdiv} vanishes.
\end{Remark}

Generally speaking, it would be desirable to have more detailed information
on the regularization of the physical vacuum. The main open question seems to be how
the regularized neutrino sector looks like (see Remarks~\ref{remd4} and~\ref{remshear}).
For example, our results might change considerably if the neutrino sector contained additional Dirac seas corresponding to
yet unobserved particles. Clearly, the uncertainties in the neutrino sector
are related to the fact that particles in the neutrino sector interact only very weakly.
Hopefully, future experimental input will clarify the situation.

\subsection*{Acknowledgments}
I would like to thank the referees for many helpful suggestions.

\LastPageEnding

\end{document}